\documentclass[apj]{emulateapj}
\usepackage{amsmath}
\usepackage{apjfonts}
\usepackage{natbib}

\newcommand{\be}{\begin{equation}}
\newcommand{\ee}{\end{equation}}
\newcommand{\bea}{\begin{eqnarray}}
\newcommand{\eea}{\end{eqnarray}}

\newcommand{\Msun}{~{\rm M}_\odot}
\newcommand{\ac}{a_{\rm c}}
\def\kms        {km$\;$s$^{-1}$}
\def\chandra    {\emph{Chandra}}
\def\xmm        {\emph{XMM}}
\def\hseventy   {$H_0=70$~km$\;$s$^{-1}\,$Mpc$^{-1}$}

\begin{document}

\title{The origin of cold fronts in the cores of relaxed galaxy clusters}

\author{Yago Ascasibar\altaffilmark{1} and Maxim
  Markevitch\altaffilmark{2}}

\affil{Harvard-Smithsonian Center for Astrophysics, 60 Garden St.,
  Cambridge, MA 02138}

\altaffiltext{1}{Presently at the Astrophysikalisches Institut Potsdam,
  An der Sternwarte 16, D-14482 Potsdam, Germany; yago@aip.de}

\altaffiltext{2}{Also Space Research Institute, Russian Acad.\ Sci.,
  Profsoyuznaya 84/32, Moscow 117997, Russia}

\setcounter{footnote}{3}

\shorttitle{COLD FRONTS IN RELAXED GALAXY CLUSTERS}
\shortauthors{ASCASIBAR AND MARKEVITCH}


\begin{abstract}

\emph{Chandra} X-ray observations revealed the presence of
cold fronts (sharp contact discontinuities between gas
regions with different temperatures and densities) in the
centers of many, if not most, relaxed clusters with cool
cores. We use high-resolution simulations of idealized
cluster mergers to show that they are due to sloshing of the
cool gas in the central gravitational potential, which is
easily set off by minor mergers and can persist for
gigayears. The only necessary condition is a steep entropy
profile, as observed in cooling flow clusters. Even if the
infalling subcluster has no gas during core passage, the
gravitational disturbance sets the main mass peak (gas and
dark matter together) in motion relative to the surrounding
gas. A rapid change in the direction of motion causes a
change in ram pressure, which pushes the cool gas away from
the dark matter peak and triggers sloshing. For nonzero
impact parameters, the cool gas acquires angular momentum,
resulting in a characteristic spiral pattern of cold
fronts. There is little visible disturbance outside the cool
core in such a merger. If the subcluster retains its gas
during core passage, the cool central gas of the main
cluster is more easily decoupled from the dark matter
peak. Subsequently, some of that gas, and often the cool gas
from the subcluster, falls back to the center and starts
sloshing. However, in such a merger, global disturbances are
readily visible in X-rays for a long time.  We conclude that
cold fronts at the centers of relaxed clusters, often spiral
or concentric-arc in shape, are probably caused by
encounters with small subhalos stripped of all their gas at
the early infall stages.

\end{abstract}


  \section{Introduction}
  \label{secIntro}

\begin{figure*}
\centering
\plottwo{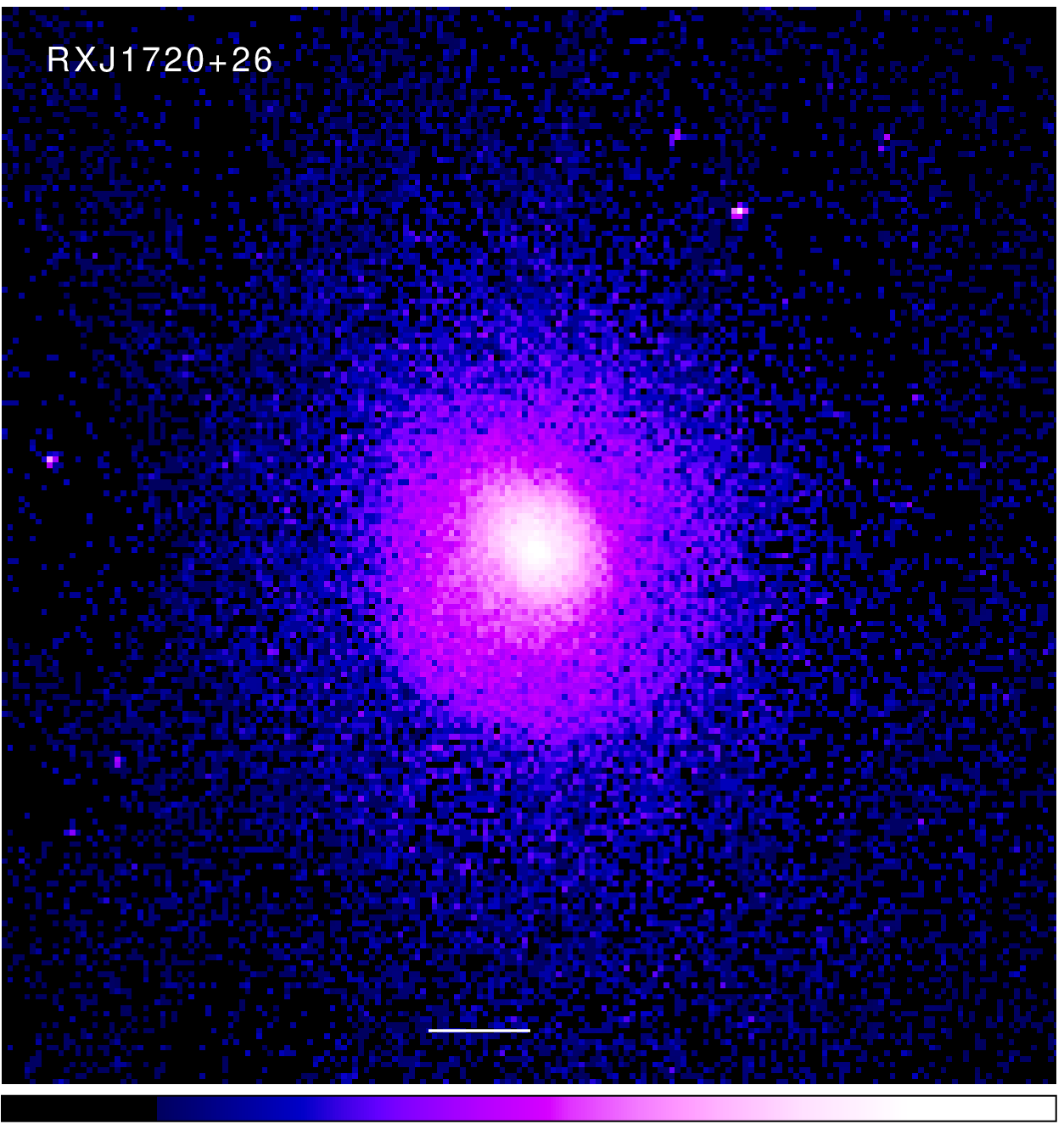}{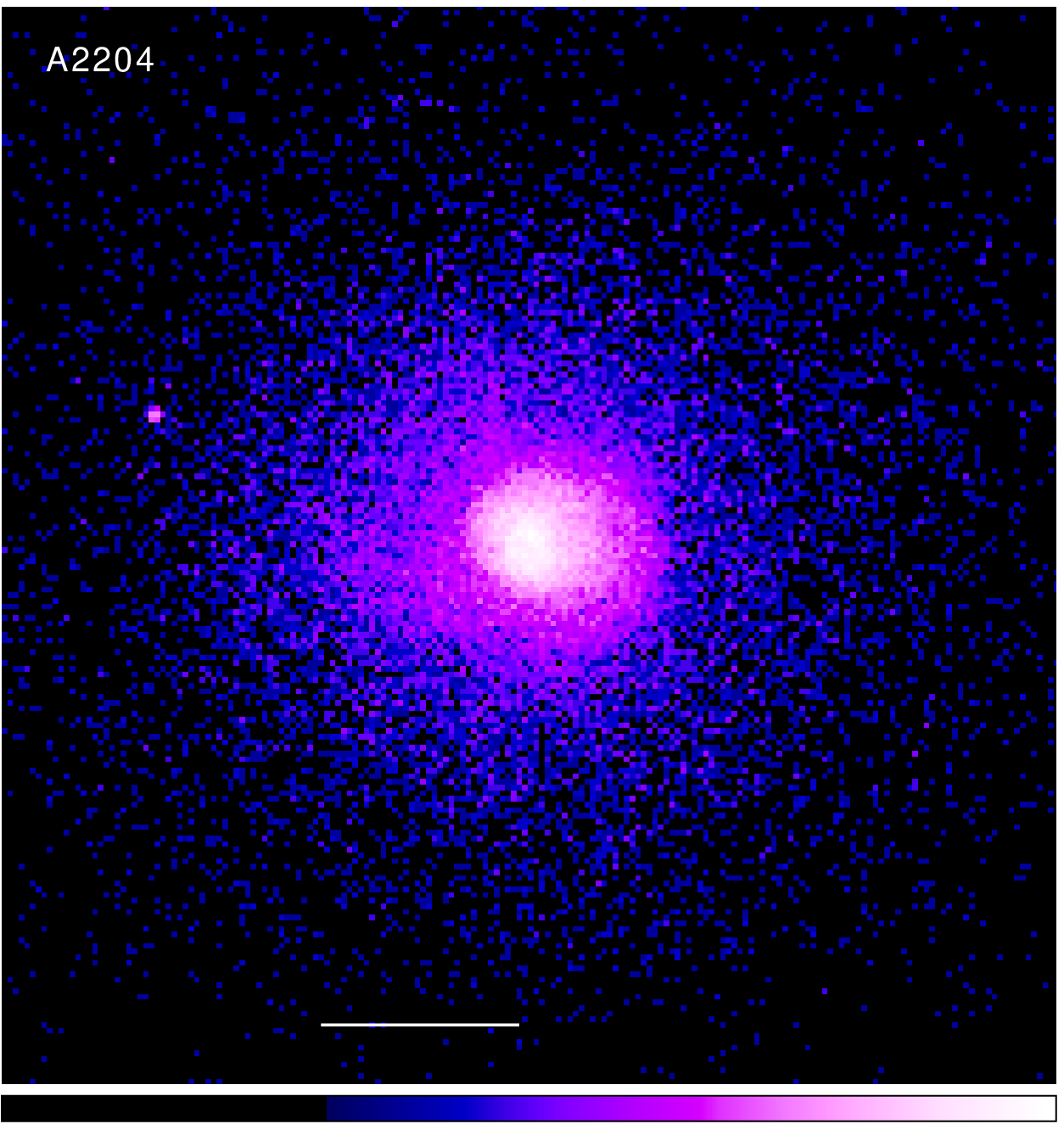}
\vspace{2mm}
\plottwo{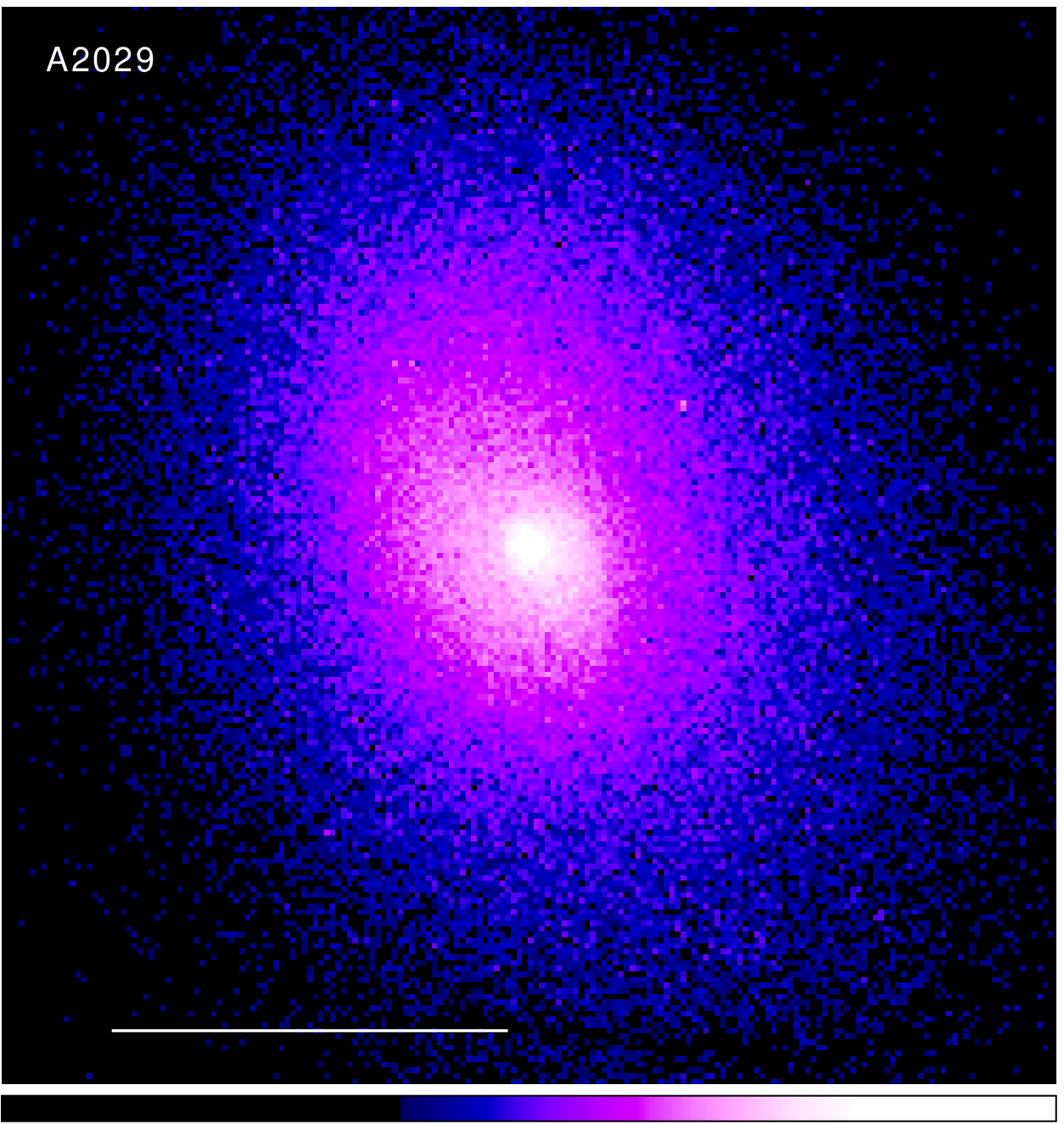}{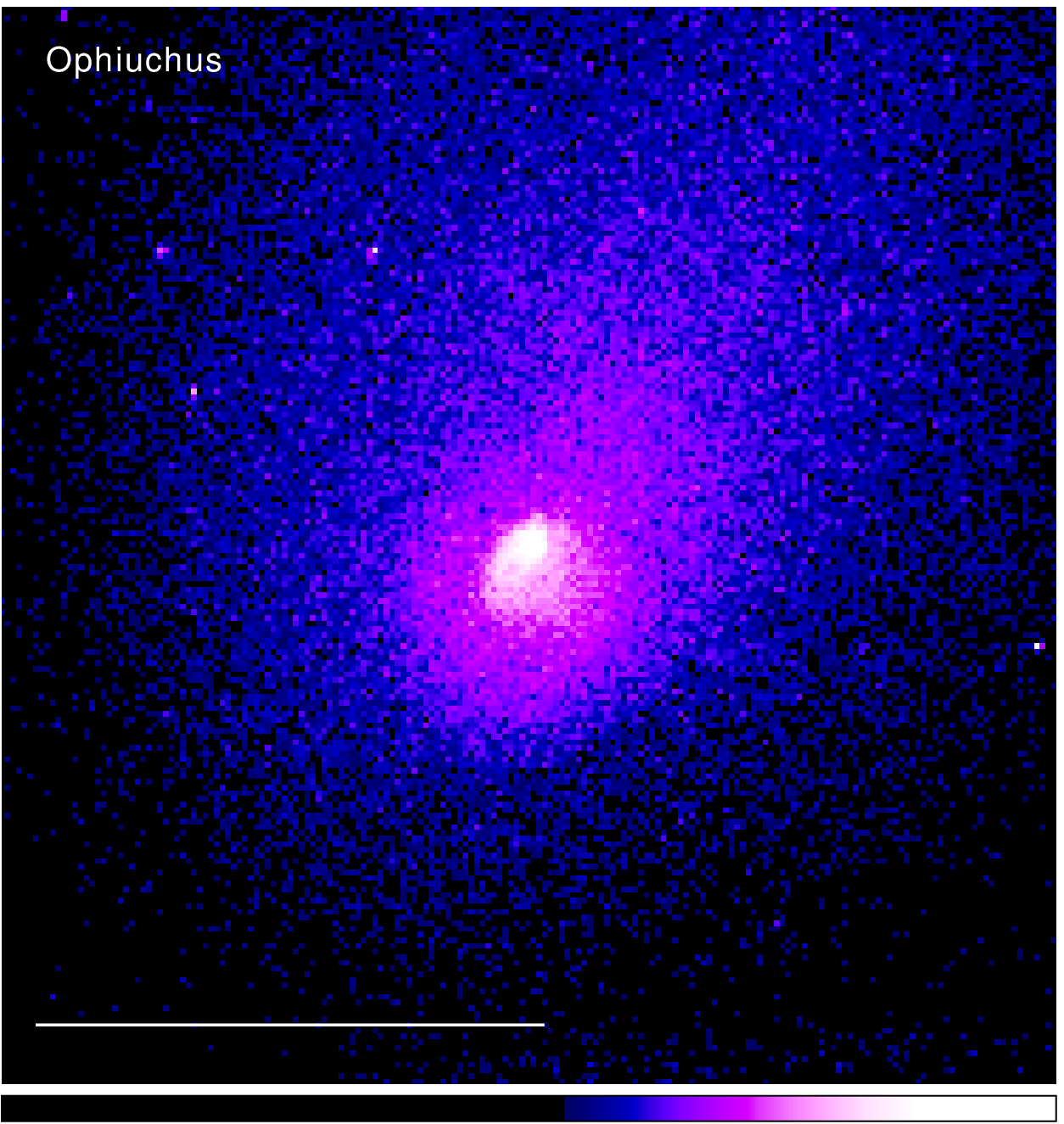}
\caption{Archival \emph{Chandra} X-ray images in a 0.5--4 keV energy band of
  examples of real clusters exhibiting cold fronts in their cool cores.
  Horizontal bars are 100 kpc.  A2029 (analyzed by Clarke et al.\ 2004) is
  among the most relaxed clusters known; it exhibits 2 edges in a spiral
  pattern at $r\sim 7$ and 20 kpc (between white and pink and between light
  and dark pink), and possibly another one at smaller scale.  RXJ1720+26
  (Mazzotta et al.\ 2001) and A2204 (Sanders et al.\ 2005) are relaxed on
  large scales; RXJ1720+26 exhibits a large, $r\approx 250$ kpc edge (pink -
  blue), while A2204 shows a spiral pattern consisting of least two edges at
  20 and 70 kpc (white - pink, pink - blue).  Ophiuchus has some evidence in
  the outskirts of a recent merger. It shows three edges on scales $r\sim 3$
  kpc, 8 kpc and 40 kpc (white - pink, white - darker pink, blue - darker
  blue).}
\label{figObs}
\end{figure*}

High-resolution X-ray observations of galaxy clusters with the
\emph{Chandra} satellite revealed a number of interesting and unexpected
features in the intracluster medium (ICM).  Many clusters were found to
exhibit sharp, arc-shaped, edge-like jumps in their gas density and
temperature.  Unlike in shock fronts, the gas on the dense side of these
``cold fronts'' is cooler, so the pressure is continuous across the front
(e.g., Markevitch et al.\ 2000; Vikhlinin et al.\ 2001).  Cold fronts are
much more ubiquitous than shock fronts; only two convincing examples of the
latter have been found so far (1E\,0657--56 and A520), while cold fronts are
observed in many if not most clusters.  In merging systems, they were
immediately interpreted as contact discontinuities between gases from
different subclusters (Markevitch et al.\ 2000). In some cases, such as the
cluster 1E\,0657--56 or the galaxy NGC\,1404, this is quite obvious from the
X-ray and optical images which show that the cold front is a boundary of the
cool gas belonging to the infalling subcluster (e.g., Markevitch et al.\ 
2002; Machacek et al.\ 2005).  However, cold fronts are also observed near
the centers of cooling flow%
\footnote{For the purpose of this work, we will use the term ``cooling
  flow'' simply to denote the observed centrally peaked gas density and
  temperature profiles, fully aware of the problems of the physical cooling
  flow model.}
clusters, many of which are relaxed and show little or no signs of recent
merging (e.g., Mazzotta et al.\ 2001; Markevitch et al.\ 2001, 2003;
Churazov et al.\ 2003; Dupke \& White 2003; Sanders et al.\ 2005). These
fronts are typically more subtle in terms of the density jump than those in
mergers.  Some of the examples are shown in Fig.\ 1; one of them is in
A2029, which on scales $r>100-200$ kpc is the most relaxed cluster known
(e.g., Buote \& Tsai 1996).  In these clusters, the moving gas clearly does
not belong to any infalling subcluster.  Markevitch et al.\ (2001) showed
that the gas forming the cold front in A1795, another very relaxed cluster,
is not in hydrostatic equilibrium and proposed that the front is caused by
subsonic ``sloshing'' of the cluster's own central gas in the gravitational
potential well as a result of a disturbance of the central potential by past
subcluster infall, or a gas disturbance from the central AGN activity. It is
important to establish the nature of these features, because they may have
significant effect on the energy balance in the cooling flows, as well as on
estimates of the cluster total mass based on the assumption of hydrostatic
equilibrium.

In this work, we wish to test whether gas sloshing in the cluster cores can
create cold fronts and investigate the possible origin of such sloshing.  We
use high-resolution numerical simulations of idealized cluster mergers to
explore the possibility that such features are a long-lived consequence of
the infall of small subclusters.  We will be particularly interested in
determining whether it is possible for a merger to create cold fronts in the
cool core, without significantly disturbing the ICM as a whole.

Several earlier works analyzed cluster mergers found in cosmological
simulations and demonstrated that cold fronts can indeed form as a result of
ram-pressure stripping and subsequent adiabatic expansion of the cold gas
belonging to merging subclusters \citep{Bialek02,NagaiKravtsov03,Mathis05}.
Detailed effects of ram-pressure stripping on a substructure moving through
the ICM have also been studied using idealized 3D or 2D merger simulations
\citep[e.g.][among others]{Heinz03,Acreman03,Asai04,Takizawa05}.  Closer to
the questions that we wish to address here, Churazov et al.\ (2003) and
Fujita et al.\ (2004) used 2D simulations to model a weak shock or acoustic
wave propagating toward the cluster center. They found that it can displace
the cool gas from the gravitational potential well and cause gas sloshing
and cold fronts. On the other hand, Tittley \& Henriksen (2005), using
mergers extracted from a cosmological simulation, suggested that cold fronts
in the cores can result from oscillations of the dark matter core caused by
gravitational disturbance from a merging subcluster. We will see that both
of these processes are at work.

The simulations presented in this paper improve on the earlier works in
several important respects.  First, we achieve much higher resolution, 5--10
kpc within the central $r<100$ kpc, sufficient to produce sharp cold fronts
and comparable to the resolution of the X-ray observations.  Second, we use
controlled mergers of idealized symmetric subclusters, spanning a range of
merger mass ratios and impact parameters, testing various dark matter and
gas profiles as well as mergers where the smaller subcluster has lost all
its gas. This makes it easier to identify physical effects at play and
separate them from the effects of random initial conditions in cosmological
simulations.  Most importantly, we use typical {\em observed}\/ cluster gas
density and temperature profiles as initial conditions, including the
observed temperature decline in cluster centers, and do not include
radiative cooling.  Current cosmological simulations, with or without
cooling, cannot reproduce central cool regions of real clusters, even
qualitatively. At the same time, the steep gas entropy profile in the center
turns out to be the most important condition for the creation of cold
fronts, so it is necessary to use realistic gas profiles for such
simulations, as we do.  We use a Smoothed Particle Hydrodynamics (SPH) code,
which lets us trace individual gas particles during the merger.  This opens
several possibilities that are uniquely interesting for our study --- for
example, we can see where the gas in the cold front comes from, whose gas
ends up in the center of the final merged cluster, and what is the exact
mechanism by which the initially continuous gas distribution evolves into a
cold front.

  \section{Numerical method}
  \label{secSims}

All our simulations have been accomplished with the parallel PMTree+SPH code
{\sc Gadget2} \citep{Gadget2}.  Cosmological expansion has not been
considered, and vacuum boundary conditions have been used.  The simulation
box size was 300 Mpc. The total mass of the system, consisting of a
$T\sim10$~keV galaxy cluster and an infalling group-sized substructure, is
$M=1.7\times10^{15}\Msun$ in all the experiments. The gas mass fraction has
been set to $\Omega_{\rm gas}/\Omega_{\rm dm}=0.04/0.3$, which is the
observed cluster value for \hseventy\ (e.g., Vikhlinin et al.\ 2005). This
corresponds to a total dark matter (DM) mass $M_{\rm dm}\equiv
M_0=1.5\times10^{15}\Msun$ and a total gas mass $M_{\rm
  gas}=2\times10^{14}\Msun$.  Our experiments involve $N=2\times10^7$
particles ($10^7$ DM particles and $10^7$ gas particles), which translates
into a mass resolution $m_{\rm dm}=1.5\times10^8\Msun$ and $m_{\rm
  gas}=2\times10^7\Msun$.  The gravitational softening length was set to 2
kpc, and SPH quantities are based on the nearest 64 particles.  The
resulting linear resolution (estimated as the distance to the 64-th neighbor)
for our default cluster density profile is $5-10$ kpc within $r=100$ kpc for
the main cluster ($3''-6''$ at $z=0.1$), widening to 25 kpc at $r=500$
kpc. As we will see below (\S\ref{secnum}), this is indeed the width of the
sharpest features produced by the simulations.  Snapshots of the simulations
have been saved every 10 Myr.

In order to set up the initial conditions, we model both objects, the main
cluster and the subhalo, at $t=0$ as spherically symmetric systems in
hydrostatic equilibrium.  For computational convenience, we use a
\citet{Hernquist90} profile for their DM distribution,
\begin{equation}
\rho_{\rm dm}(r)=\frac{M_0}{2\pi a^3}\frac{1}{r/a(1+r/a)^3},
\label{eqRhoH}
\end{equation}
where $M_0$ and $a$ are the mass and scale length of the DM halo.
For the gas temperature, we use a phenomenological formula
\begin{equation}
T(r)=\frac{T_0}{1+r/a}~\frac{c+r/\ac}{1+r/\ac},
\label{eqT}
\end{equation}
where $0<c<1$ is a free parameter that characterizes the depth of the
temperature drop in the cluster center and $a_{\rm c}$ is a characteristic
radius of that drop. The corresponding gas density can be derived by
imposing hydrostatic equilibrium,
\begin{equation}
\rho_{\rm gas}(r)=
        \rho_0 \left( 1+\frac{r}{\ac} \right)
        \left( 1+\frac{r/\ac}{c} \right)^\alpha
        \left( 1+\frac{r}{a}   \right)^\beta,
\label{eqRhog}
\end{equation}
with exponents
\begin{equation}
\alpha\equiv-1-n\frac{c-1}{c-a/\ac},\;\;\;\;\; 
\beta\equiv1-n\frac{1-a/\ac}{c-a/\ac}.
\end{equation}
We set $n=5$ in order to have a constant baryon fraction at large radii, and
compute the value of $\rho_0$ from the constraint $M_{\rm
  gas}/M_0=\Omega_{\rm gas}/\Omega_{\rm dm}$. We also will investigate
briefly a mass profile without a central cusp in~\S\ref{secFlat}.

\begin{figure}
\centering
\epsscale{.8}
\plotone{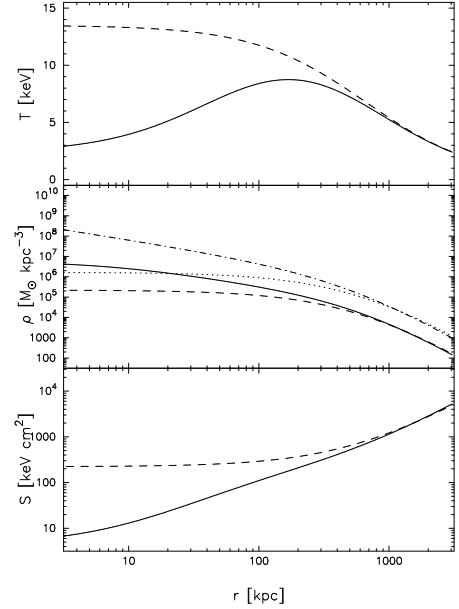}
\epsscale{1}
\caption{
  Initial conditions for our experiments (from top to bottom, gas
  temperature, gas and dark matter density and gas entropy).  Solid lines
  show the gas profiles for our `standard' model with
  $M_0=1.5\times10^{15}\Msun$, $a=600~$kpc, $c=0.17$ and $\ac=60~$ kpc,
  while dashed lines correspond to a cluster without a cool core (i.e.
  $c=1$).  The dark matter density of a \citet{Hernquist90} sphere is
  plotted as a dot-dashed line, while the dotted line shows a `cored' dark
  matter halo described by equation~(\ref{eqFlat}).  }
\label{figIC}
\end{figure}

\begin{figure*}
\centering
\includegraphics[width=5.4cm]{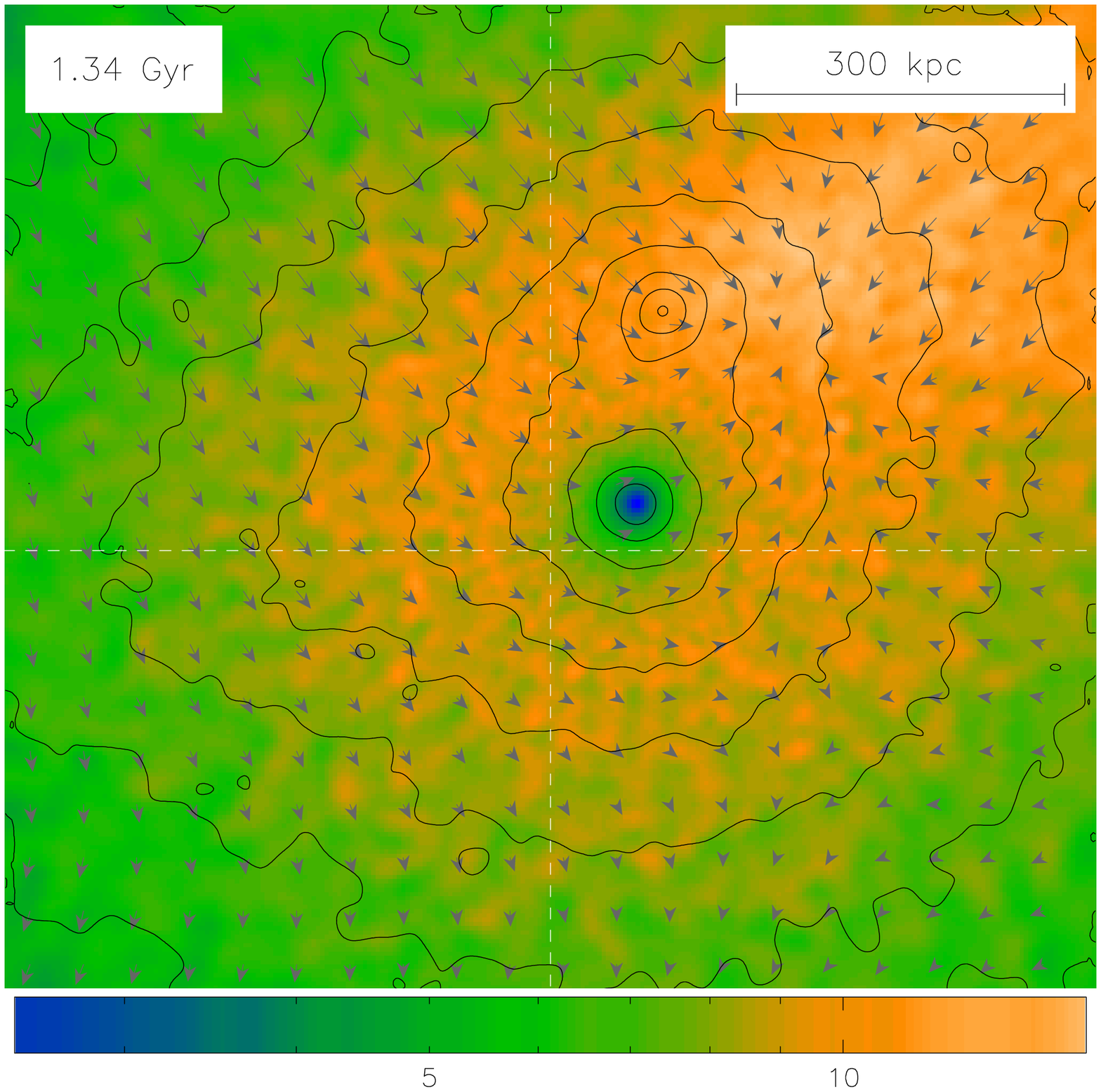}
\includegraphics[width=5.4cm]{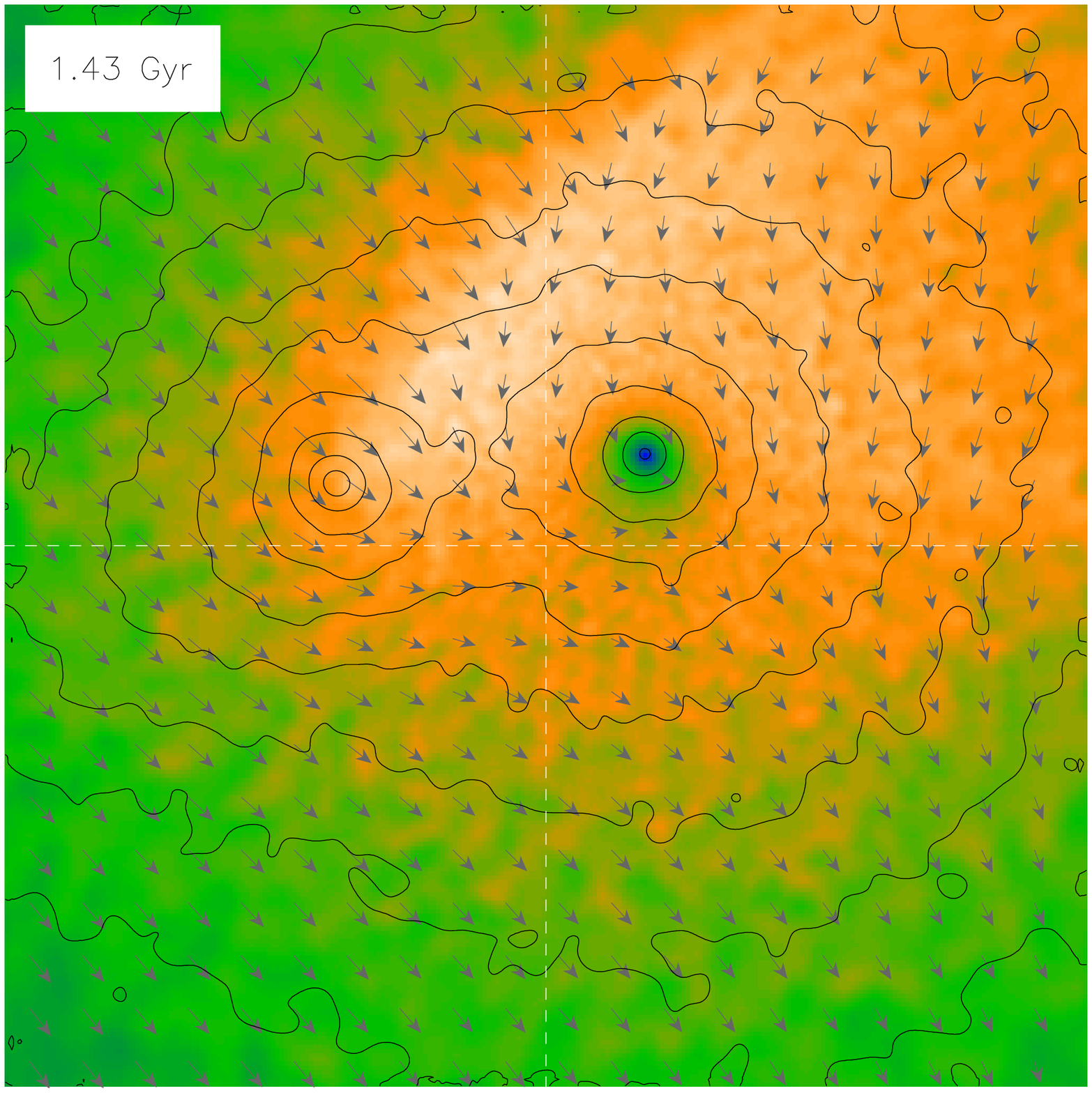}
\includegraphics[width=5.4cm]{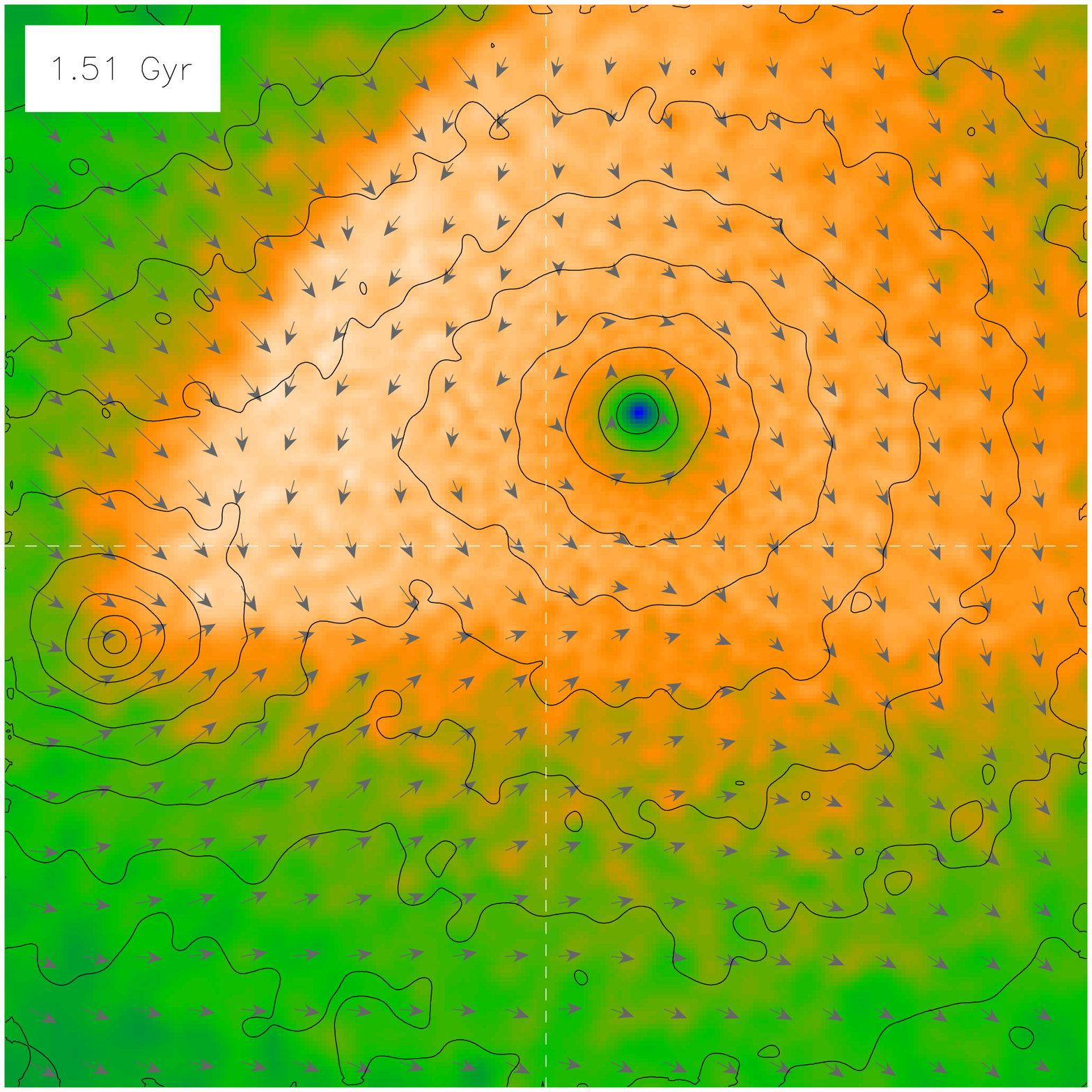}\\[1mm]
\includegraphics[width=5.4cm]{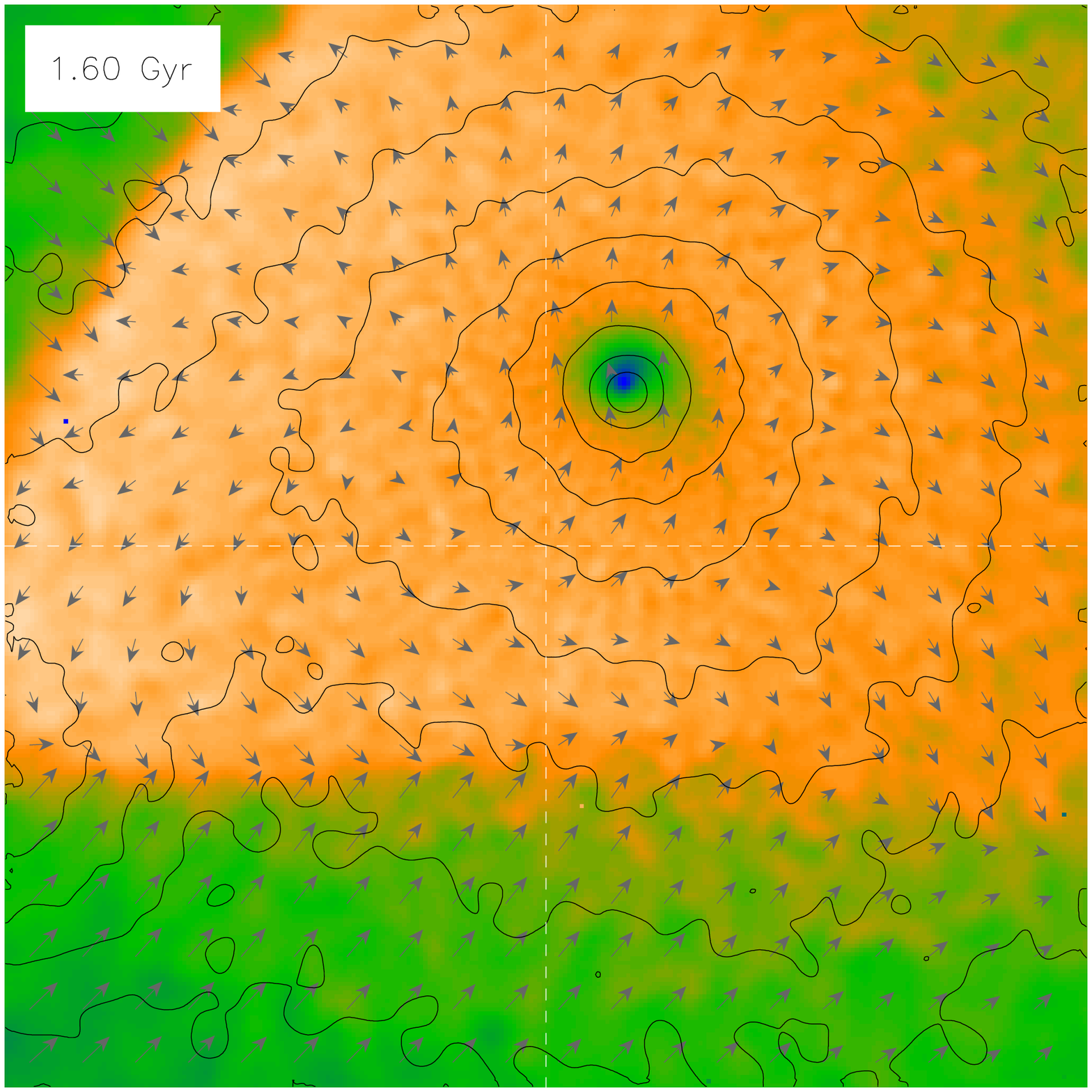}
\includegraphics[width=5.4cm]{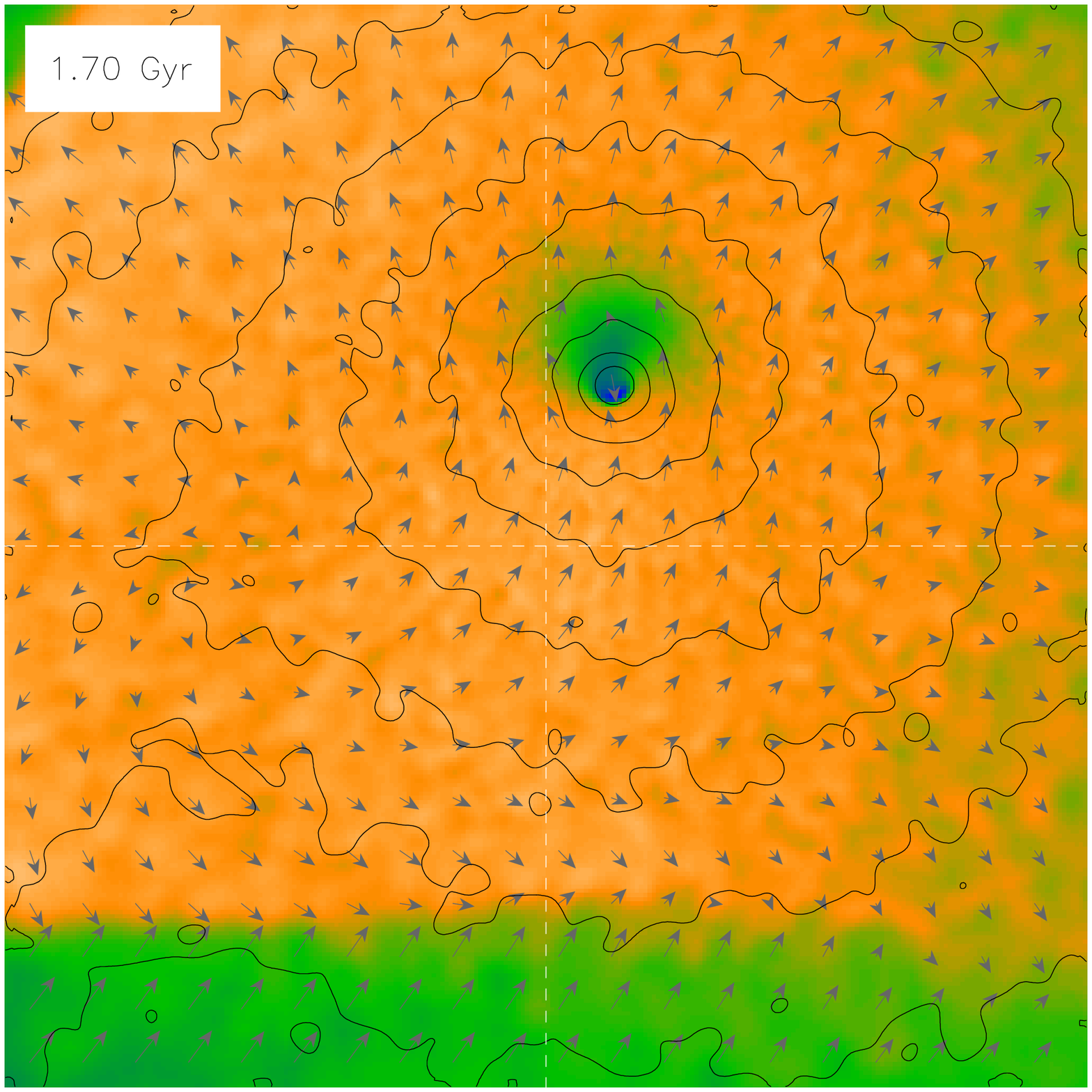}
\includegraphics[width=5.4cm]{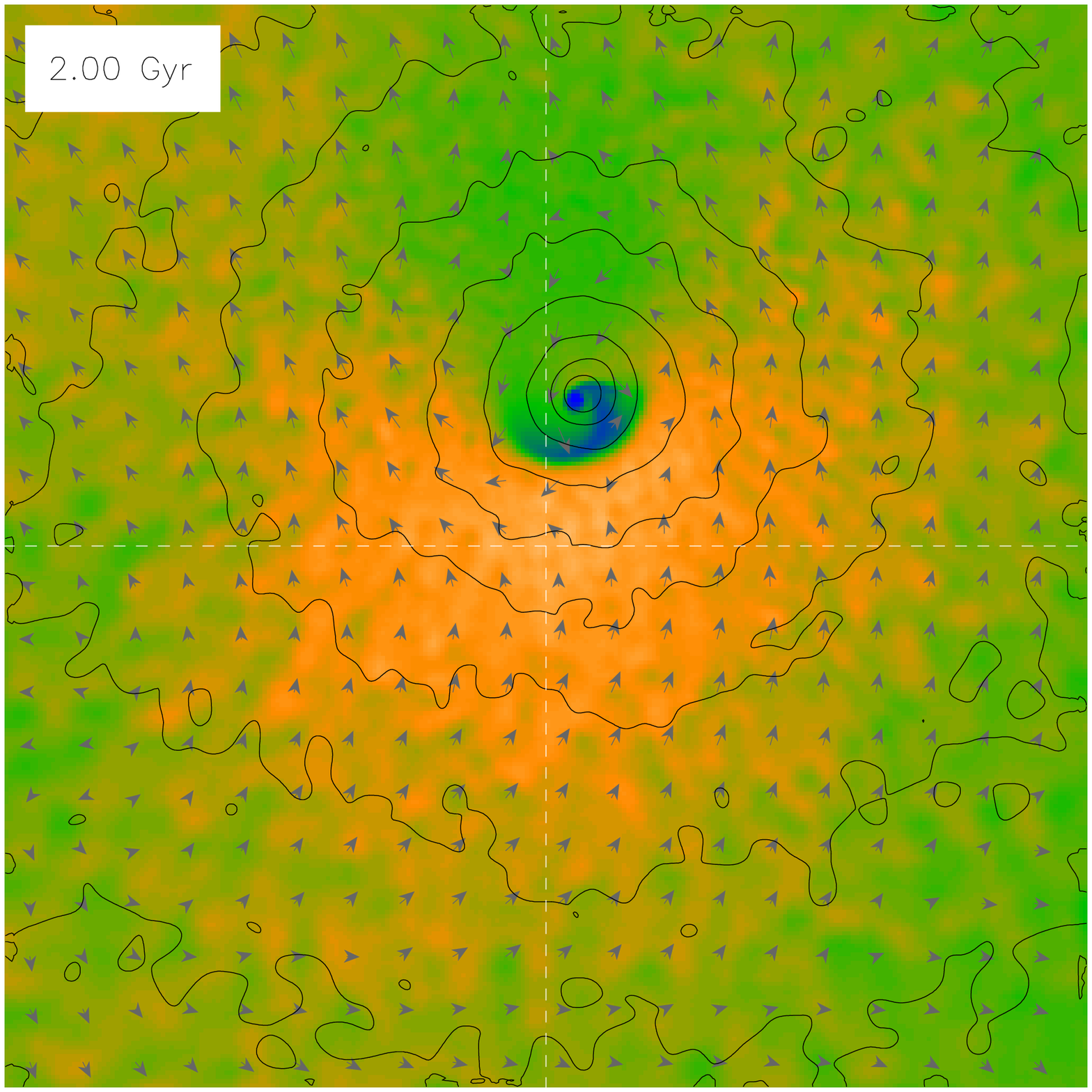}\\[1mm]
\includegraphics[width=5.4cm]{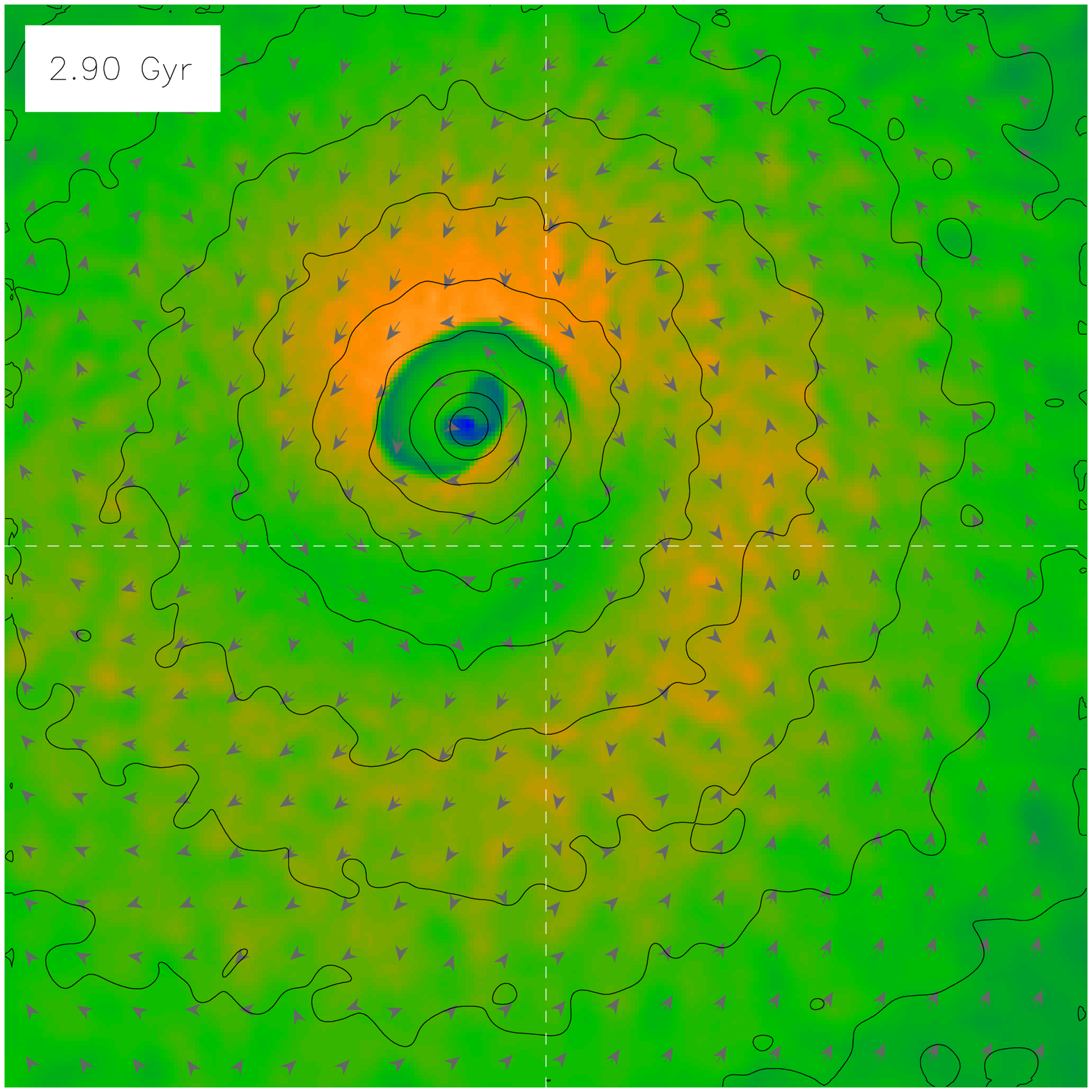}
\includegraphics[width=5.4cm]{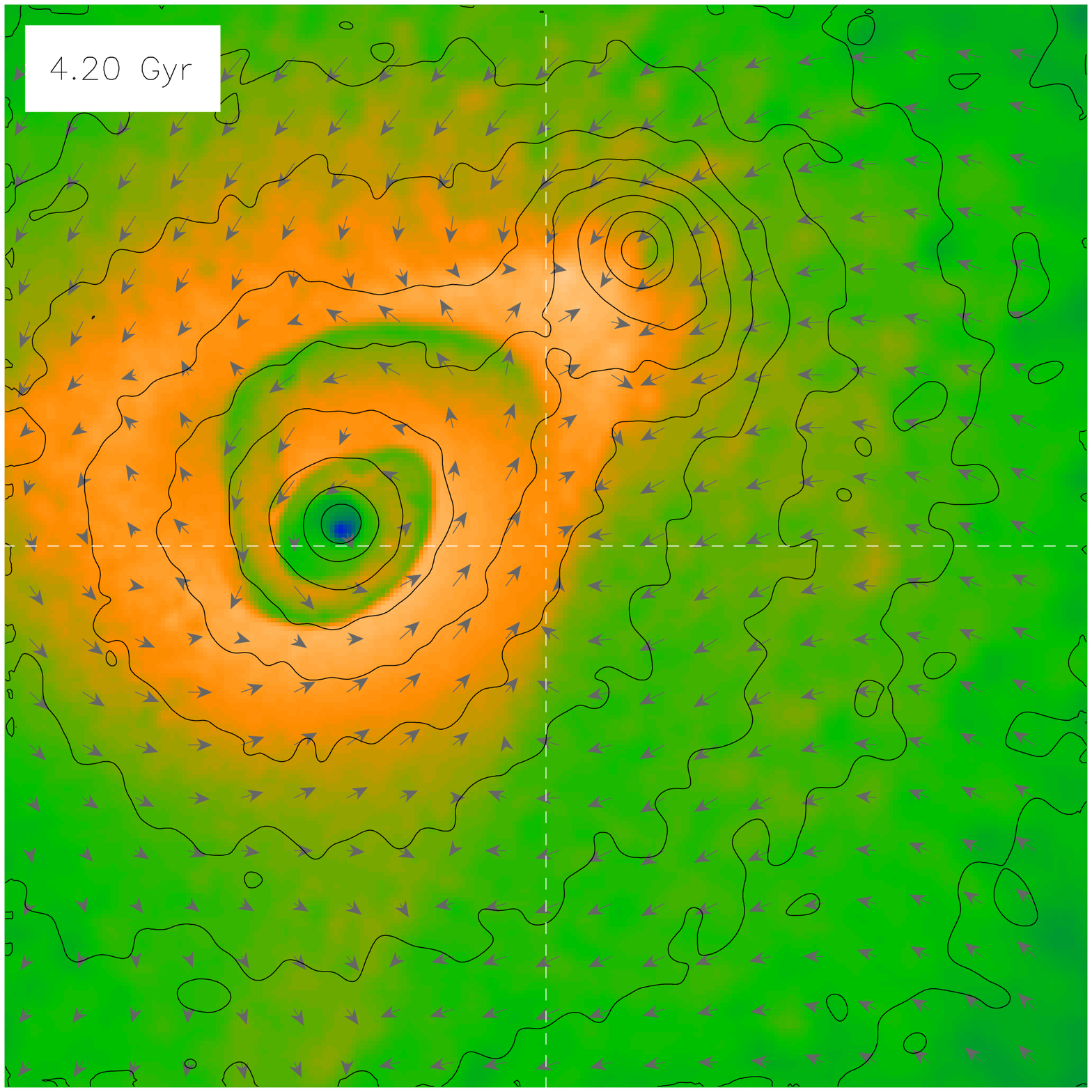}
\includegraphics[width=5.4cm]{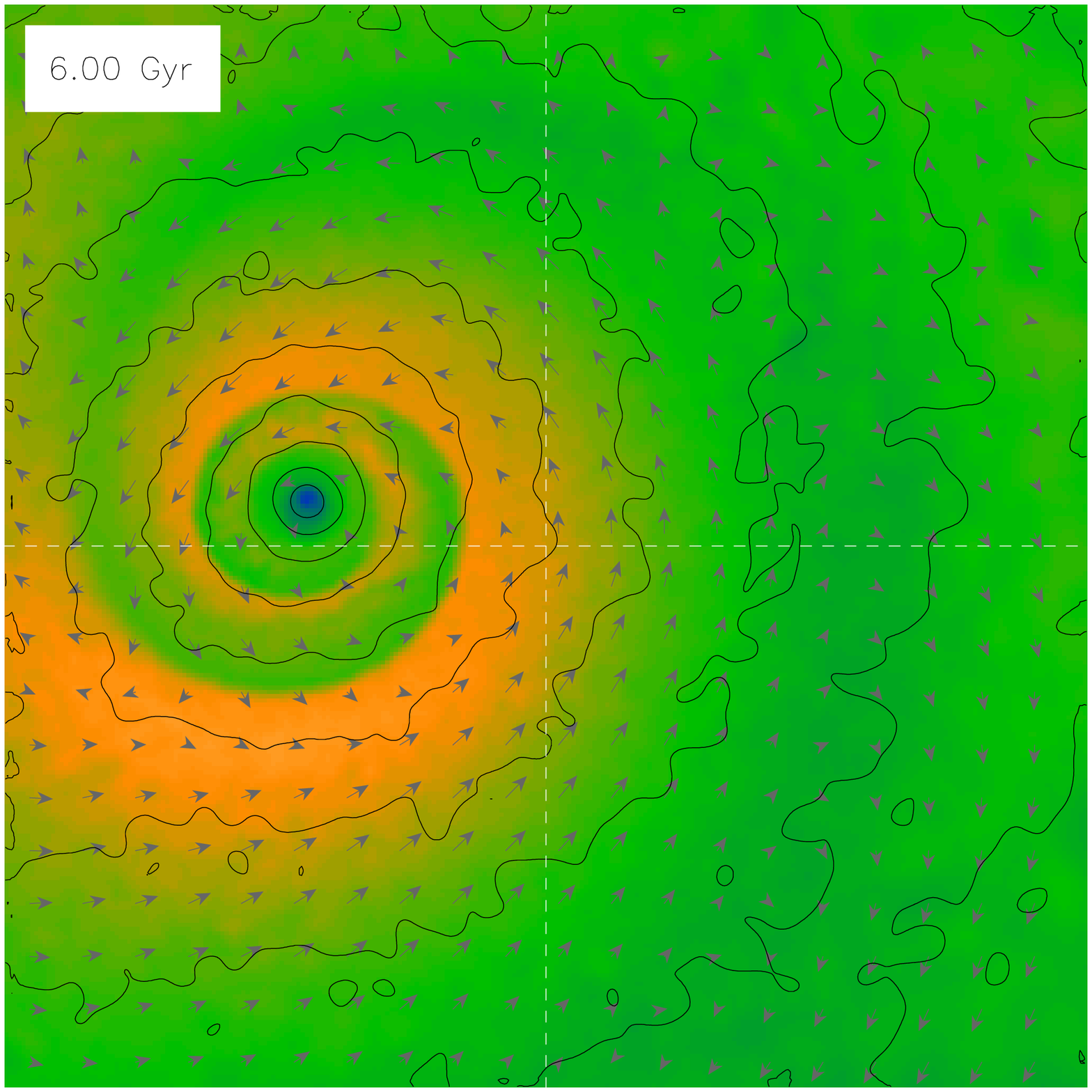}
\caption{
  Evolution of the cold front induced by a purely dark matter satellite.
  Parameters of the encounter are $R=5$ and $b=500~$kpc; the pericenter
  distance at the first core passage (which occurs at 1.37 Gyr) is $\sim
  150$ kpc.  Color maps show the gas temperature (in keV) in a slice in the
  orbital plane. The temperature scale shown in the top left panel (in keV)
  is the same for all panels.  Arrows represent the gas velocity field
  w.r.t.\ the main dark matter density peak (for clarity, the velocity scale
  is linear at low values, then saturates). Contours are drawn at increments
  of a factor of 2 in the local dark matter density. The white cross shows
  the center of mass for the main cluster DM particles (not for the whole
  system).  The panel size is 1 Mpc.}
\label{figDM}
\end{figure*}

As noted above, $M_0=M_1+M_2=1.5\times10^{15}\Msun$.  We chose $a=600~$kpc,
$c=0.17$ and $\ac=60~$kpc, so that the gas density and temperature profiles
of our objects resemble as close as possible those typically observed in
massive galaxy clusters at all observable radii, in particular, A2029 (e.g.,
Vikhlinin et al.\ 2005). These model profiles are shown in Fig.\ 
\ref{figIC}.  We will also try a gas profile for the main cluster without a
central temperature drop, as well as a subcluster without gas.

We enforce the gas density, temperature and velocity to be single-valued
(unlike collisionless dark matter, which may be multi-streaming).  The gas
density at any point is simply $\rho_{\rm gas}=\rho_1+\rho_2$.  Temperatures
and velocities have been weighted by the factor $w_i=\rho_i/\rho$, that is,
$T=w_1T_1+w_2T_2$ and $v=w_1v_1+w_2v_2$.  In the following, we will use
``entropy'' defined as $S\equiv T/n^{2/3}$, where $n=\rho_{\rm gas}/\mu
m_{\rm p}$.

Several experiments have been performed to investigate the effects of the
merger mass ratio, impact parameter and energy of the encounter.  The mass
ratio is defined as
\begin{equation}
R\equiv M_2 / M_1,
\end{equation}
so that $M_1=M_0/(1+R)$ and $M_2=M_0R/(1+R)$.  In the following, $M_1$ and
$M_2$ denote the masses of the infalling satellite and the main cluster,
respectively (i.e. $R\ge1$). In order to scale the initial profiles for
various mass ratios of the subclusters, we held $M/a^3$, $c$ and $\ac/a$
constant.

Both objects start at a separation $d=3$ Mpc, moving towards each other with
an initial impact parameter $b$.  The total kinetic energy of the system is
set to a fraction $0\le K\le1$ of its potential energy, approximating the
objects as point masses,
\begin{equation}
E \approx (K-1)\frac{GM_1M_2}{d}=(K-1)\frac{R}{{(1+R)}^2}\frac{GM_0^2}{d}.
\end{equation}

Thus, the initial velocities in the reference frame of the center of mass
are set to
\begin{equation}
v_1=\frac{R\sqrt{2K}}{1+R}\sqrt{\frac{GM_0}{d}}~~;~~
v_2=\frac{\sqrt{2K}}{1+R}\sqrt{\frac{GM_0}{d}}
\end{equation}
and the total angular momentum is
\begin{equation}
J \approx \frac{R\sqrt{2K}}{{(1+R)}^2}\,b\,M_0\sqrt{\frac{GM_0}{d}}
\end{equation}

Different mass ratios ($R=$2, 5, 20 and 100) and impact parameters ($b=$0,
500 and 1000 kpc) have been investigated.  The initial kinetic energy of the
merger has been set to $K=1/2$.

\begin{figure}
\centering
\plotone{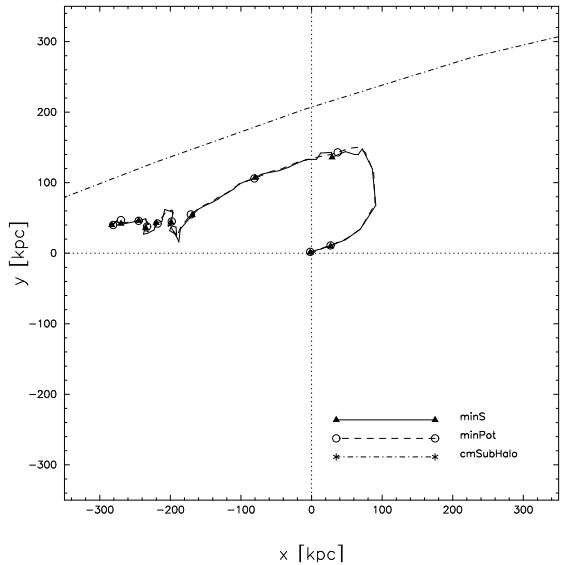}
\caption{Trajectories in the merger plane of the main cluster's
  DM peak (circles on dashed line), its gas peak (entropy minimum; triangles
  on solid line) and the subhalo DM centroid (dotted line) in the reference
  frame of the {\em main cluster}\/ mass, for the DM-only subcluster merger.
  Symbols are spaced by 1 Gyr; the subhalo flyby occurs between 1--2 Gyr.
  (The DM peak relaxes to a position offset from its initial position in
  part because the subcluster loses its mass predominantly on one side of
  the cluster, which shifts the center of mass of the main cluster particles
  from that of the whole system.)}
\label{figtraj}
\end{figure}

\begin{figure*}
\centering
\plotone{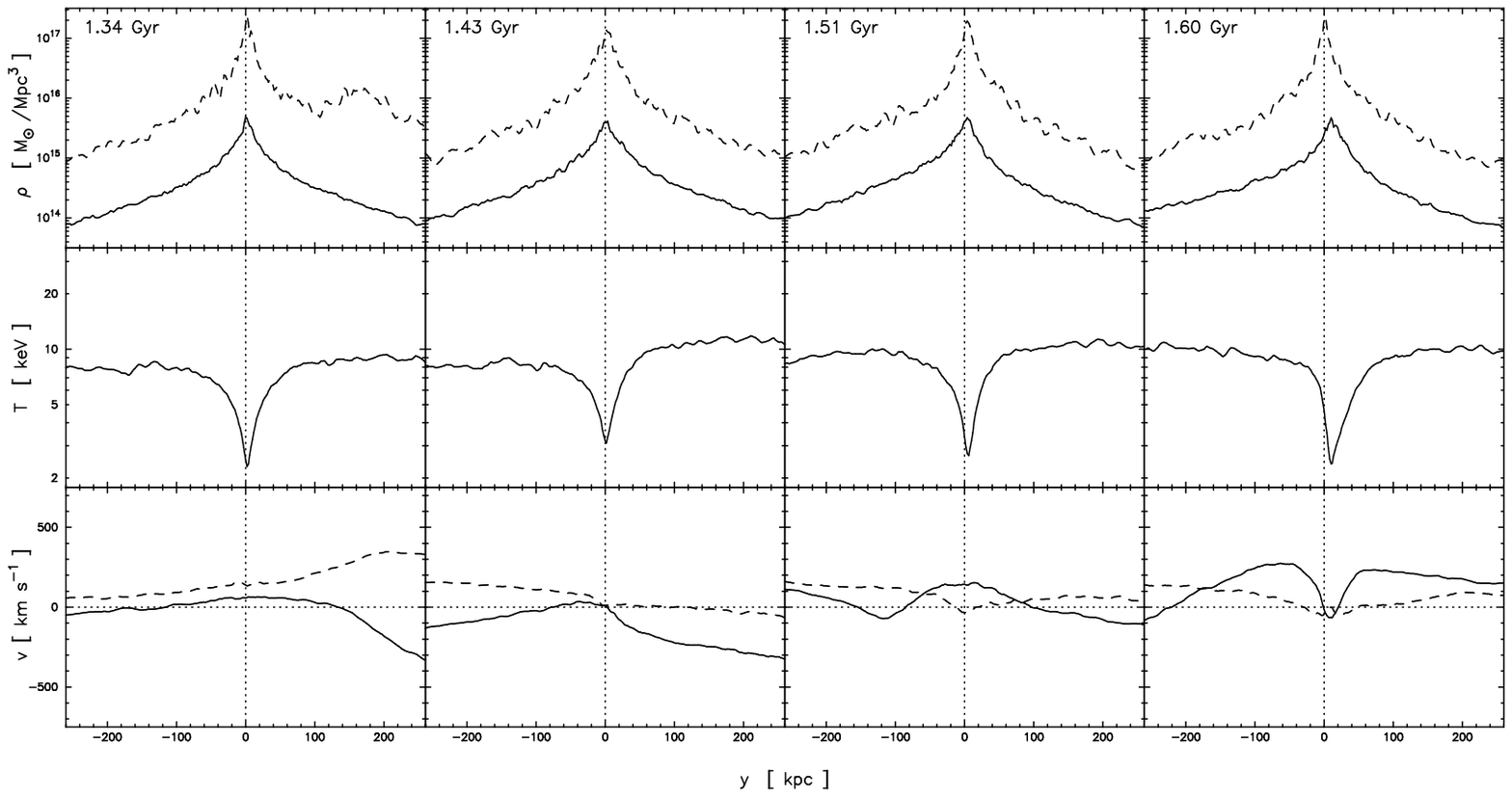}
\plotone{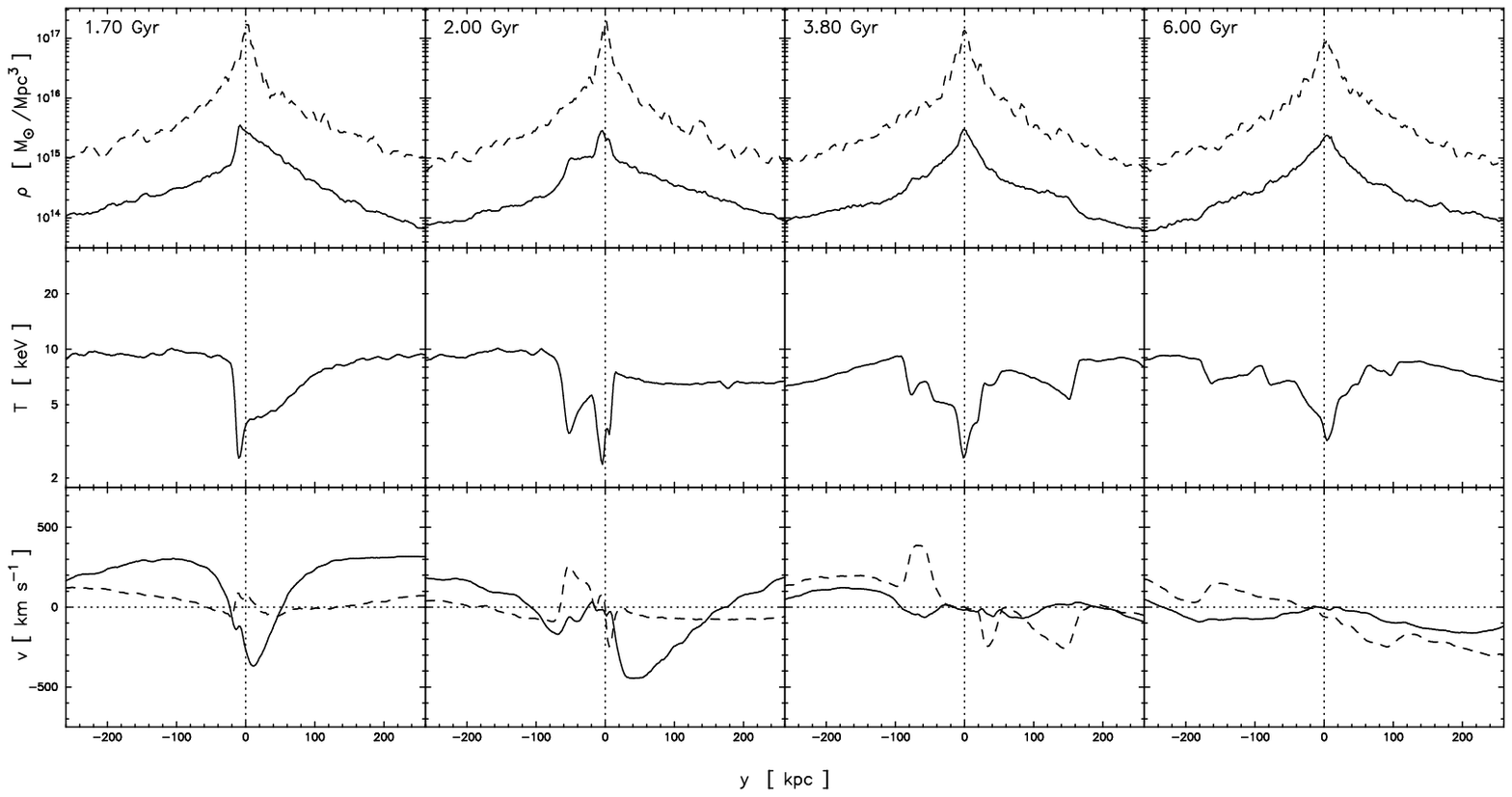}
\caption{
  Profiles of the DM and gas density, gas temperature, and gas velocity, for
  the DM-only subcluster run (see Figs.\ \ref{figDM} and \ref{figDMzoom}).
  The quantities have been computed along a vertical line through the DM
  peak of the main cluster, which is also the frame of reference ($y=0$,
  $v=0$).  Solid and dashed lines in the top panels represent gas and DM
  density, respectively.
  Solid lines in the velocity panels show the gas velocity along the $y$-axis,
  while dashed lines are used to plot the horizontal component.}
\label{figJumpDM}
\end{figure*}

\begin{figure*}
\centering
\plotone{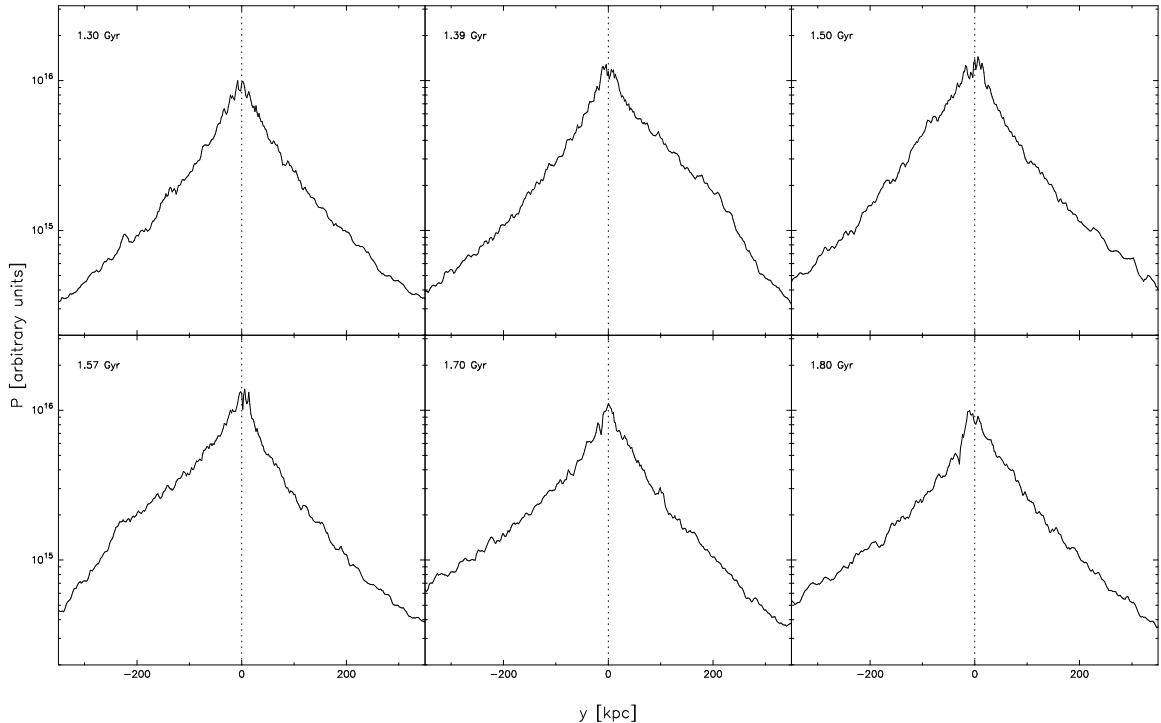}
\caption{Gas pressure along a vertical line through the DM peak, for the
  DM-only subcluster run, showing the passage of a wake created by the subcluster.}
\label{figP}
\end{figure*}

\begin{figure*}
\centering
\includegraphics[width=6.9cm]{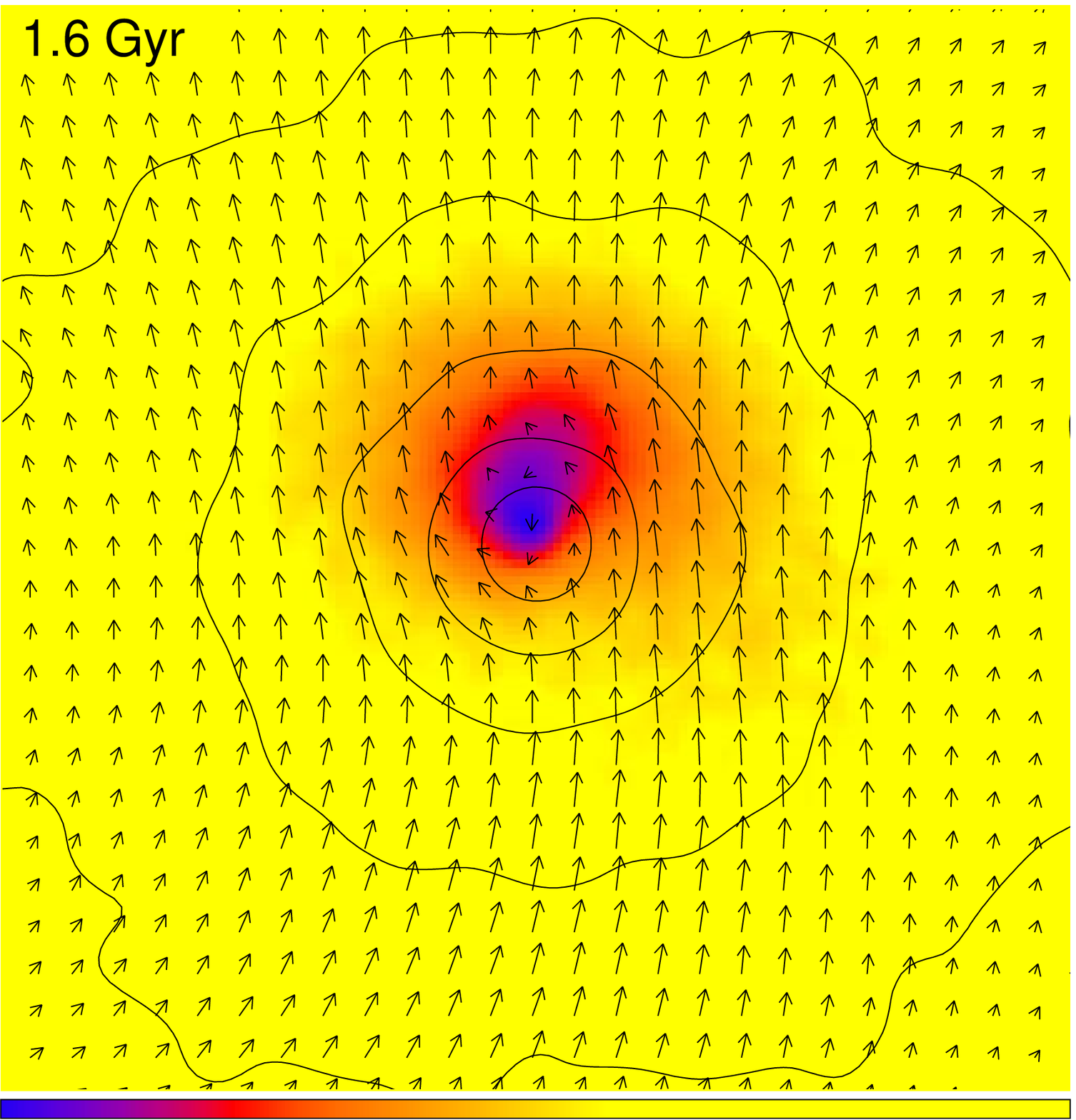}
\includegraphics[width=6.9cm]{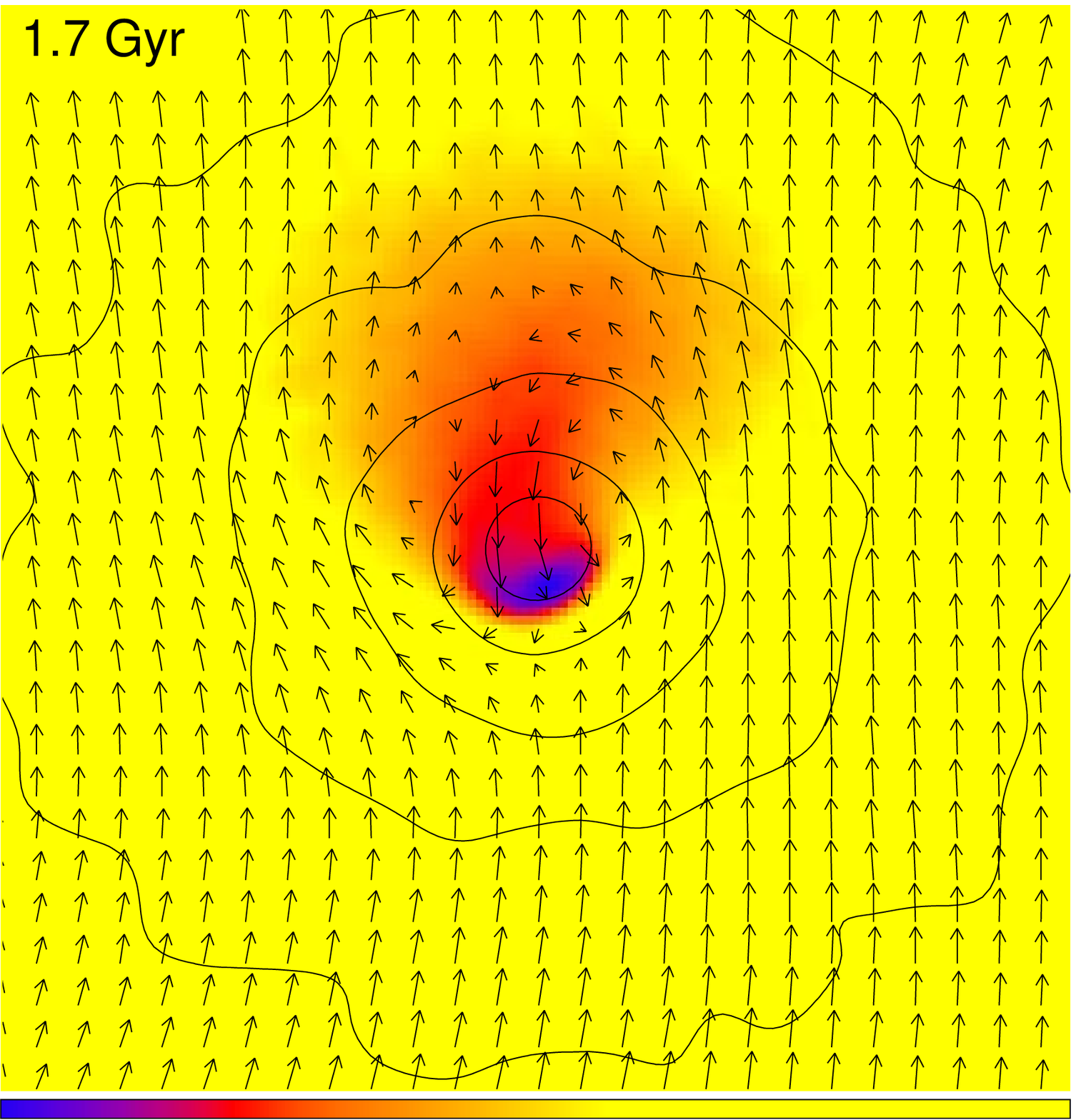}\\[1mm]
\includegraphics[width=6.9cm]{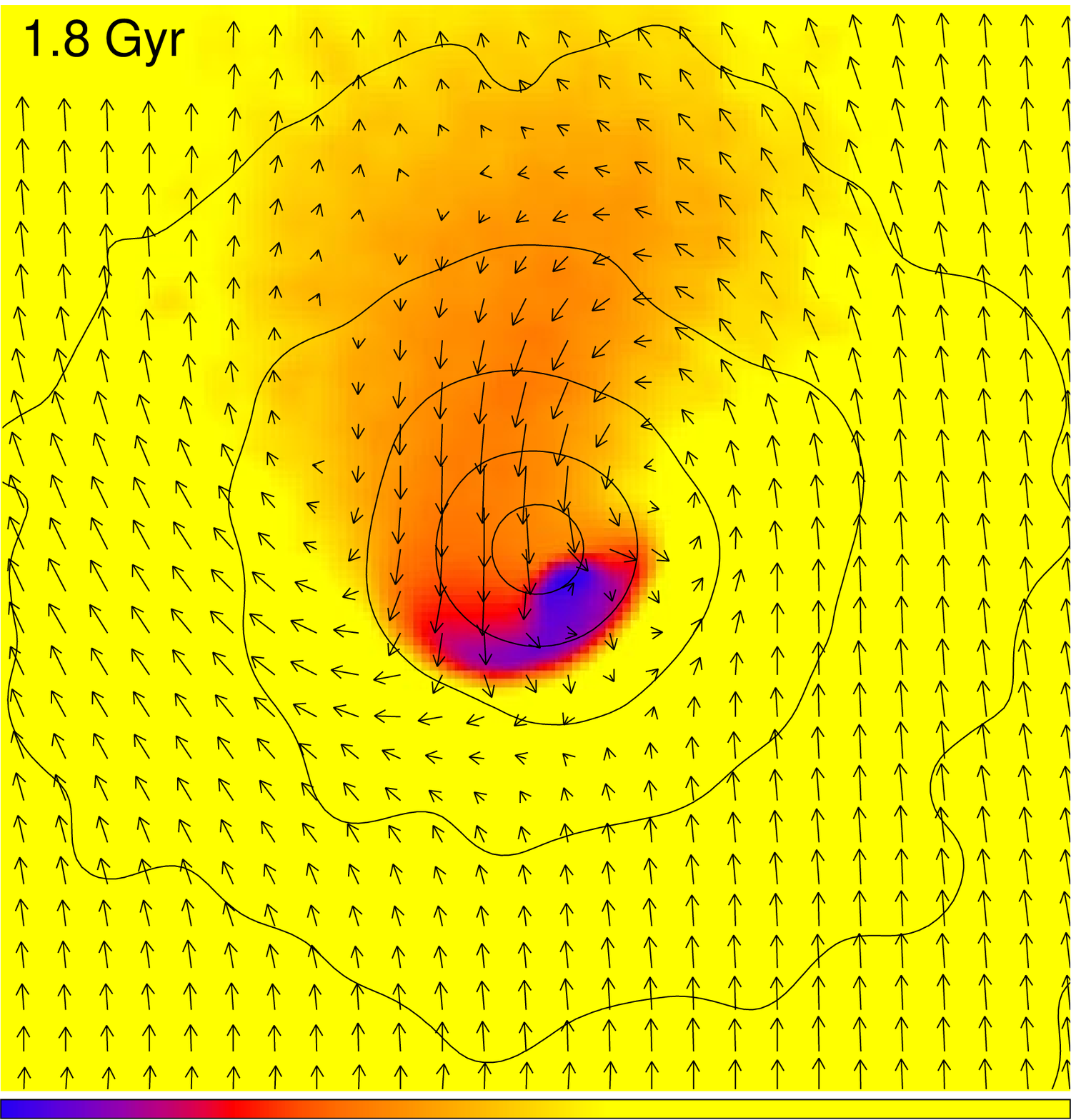}
\includegraphics[width=6.9cm]{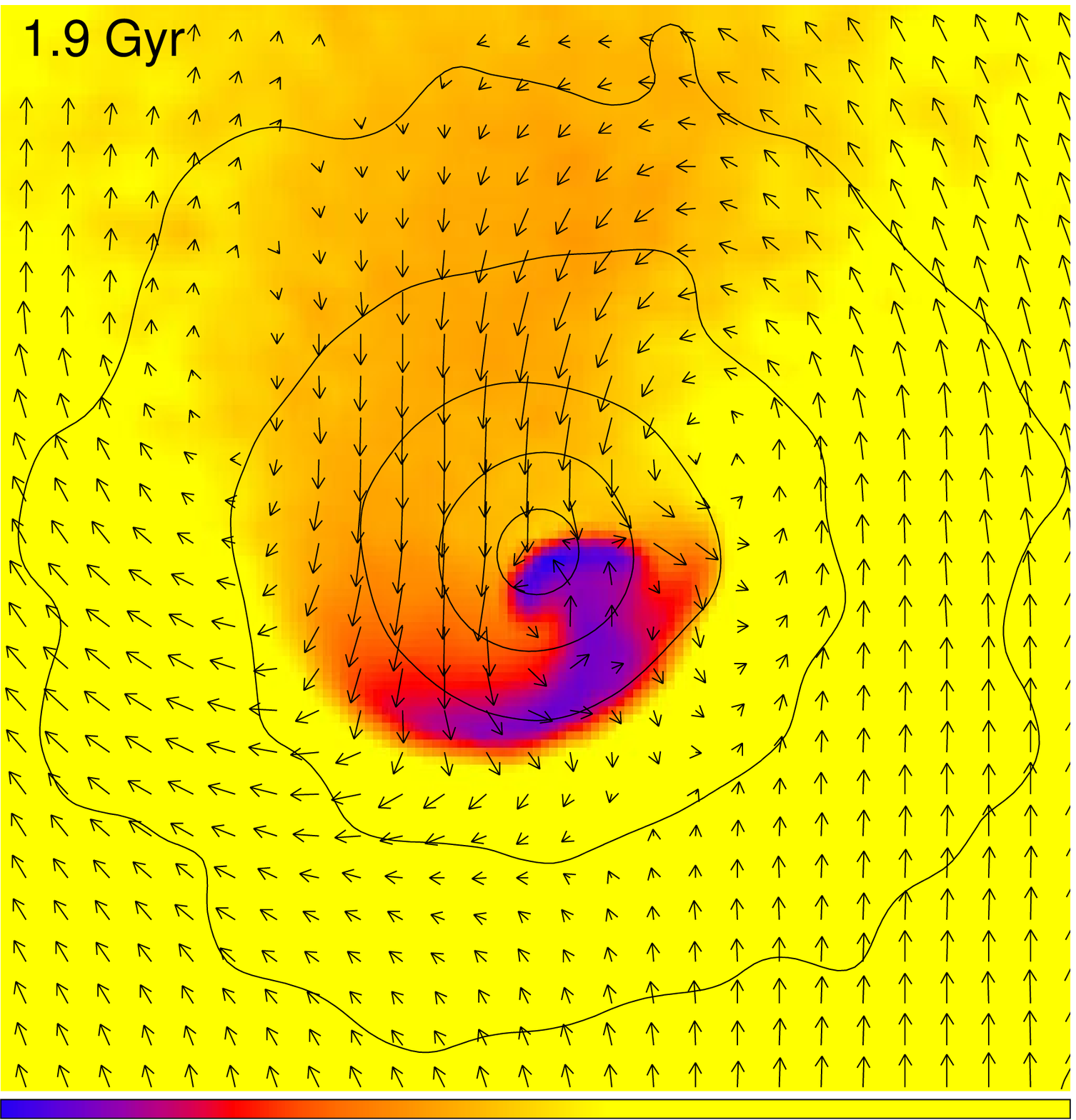}\\[1mm]
\includegraphics[width=6.9cm]{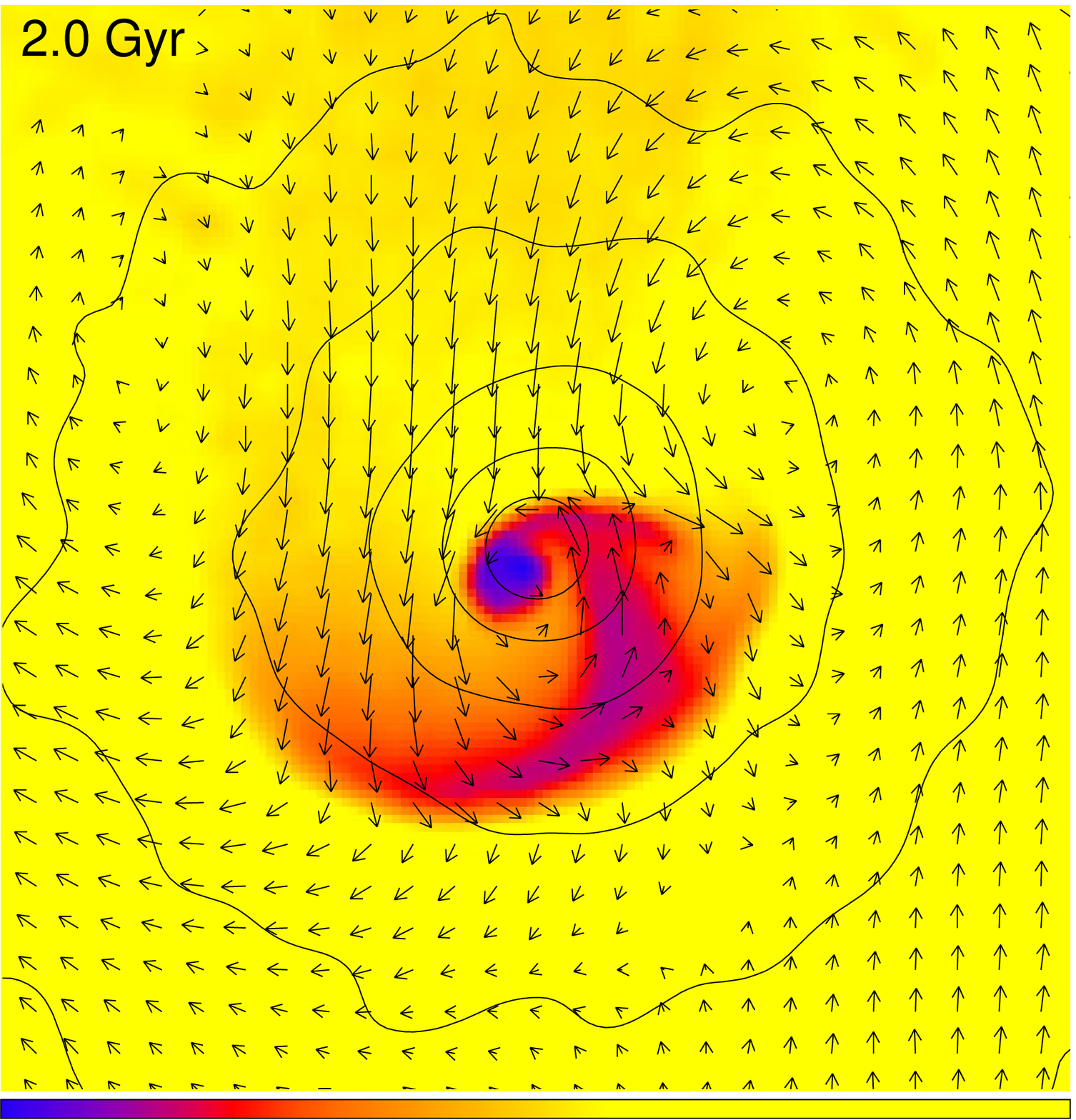}
\includegraphics[width=6.9cm]{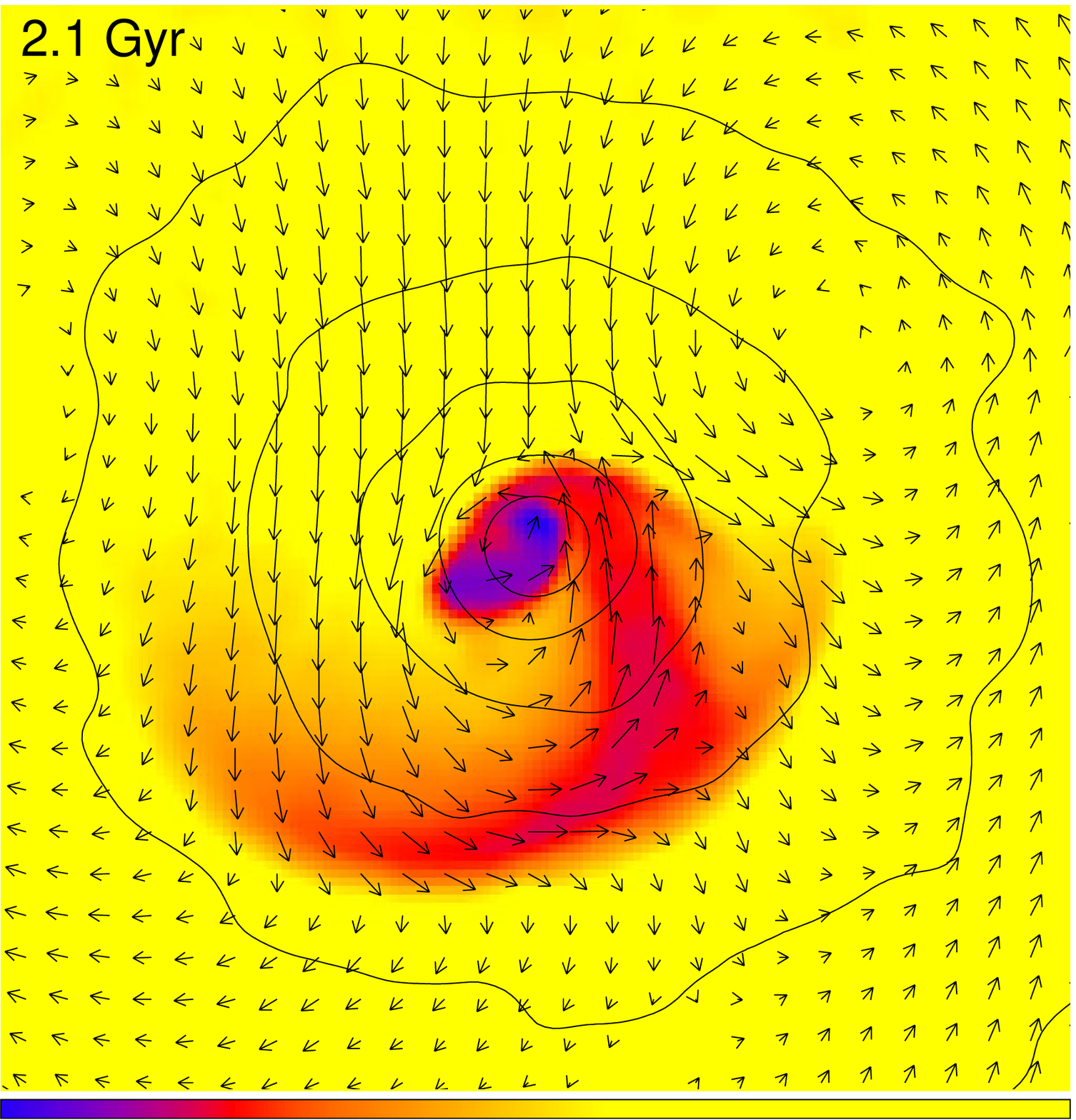}
\caption{
  Zoomed-in gas temperature slices (color; blue is 2 keV, yellow is 7--9
  keV) with DM density contours overlaid (log-spaced by factor of 2), for
  the pure-DM subcluster run shown in Fig.\ \ref{figDM}. The size of the
  panels is 0.25 Mpc, except 3.8 Gyr which is 0.33 Mpc. The gas velocity
  field, relative to the DM peak, is shown by arrows (the longest arrows are
  500 \kms\ and the scale is linear; $v<30$ \kms\ are not shown). For
  comparison, the sound speed in the 7--9 keV gas is 1300--1500 \kms. Time
  labels are the same as in Fig.\ \ref{figDM}.}
\label{figDMzoom}
\end{figure*}

\begin{figure*}
\centerline
{
  \includegraphics[width=7.5cm]{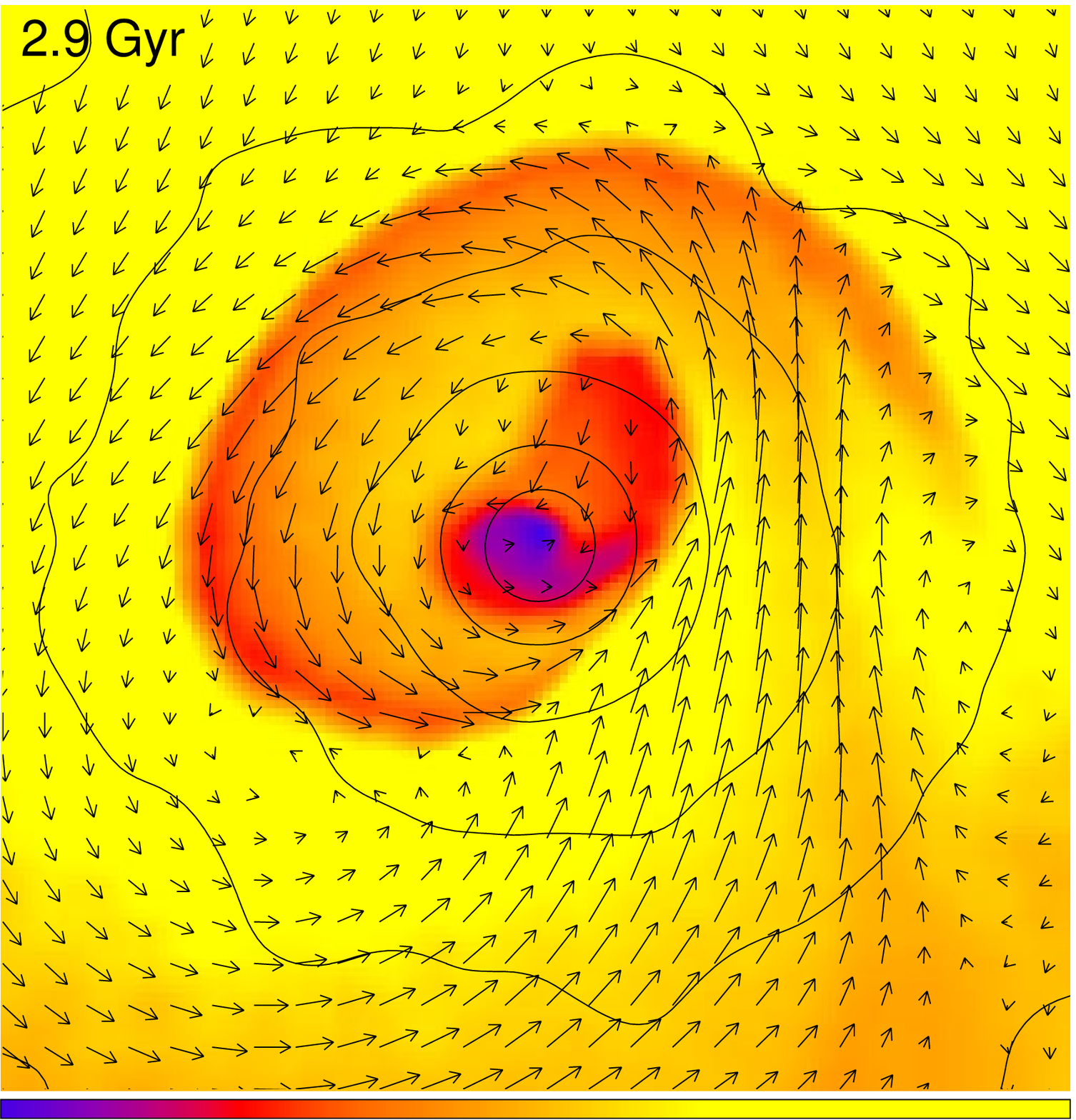}
  \includegraphics[width=7.5cm]{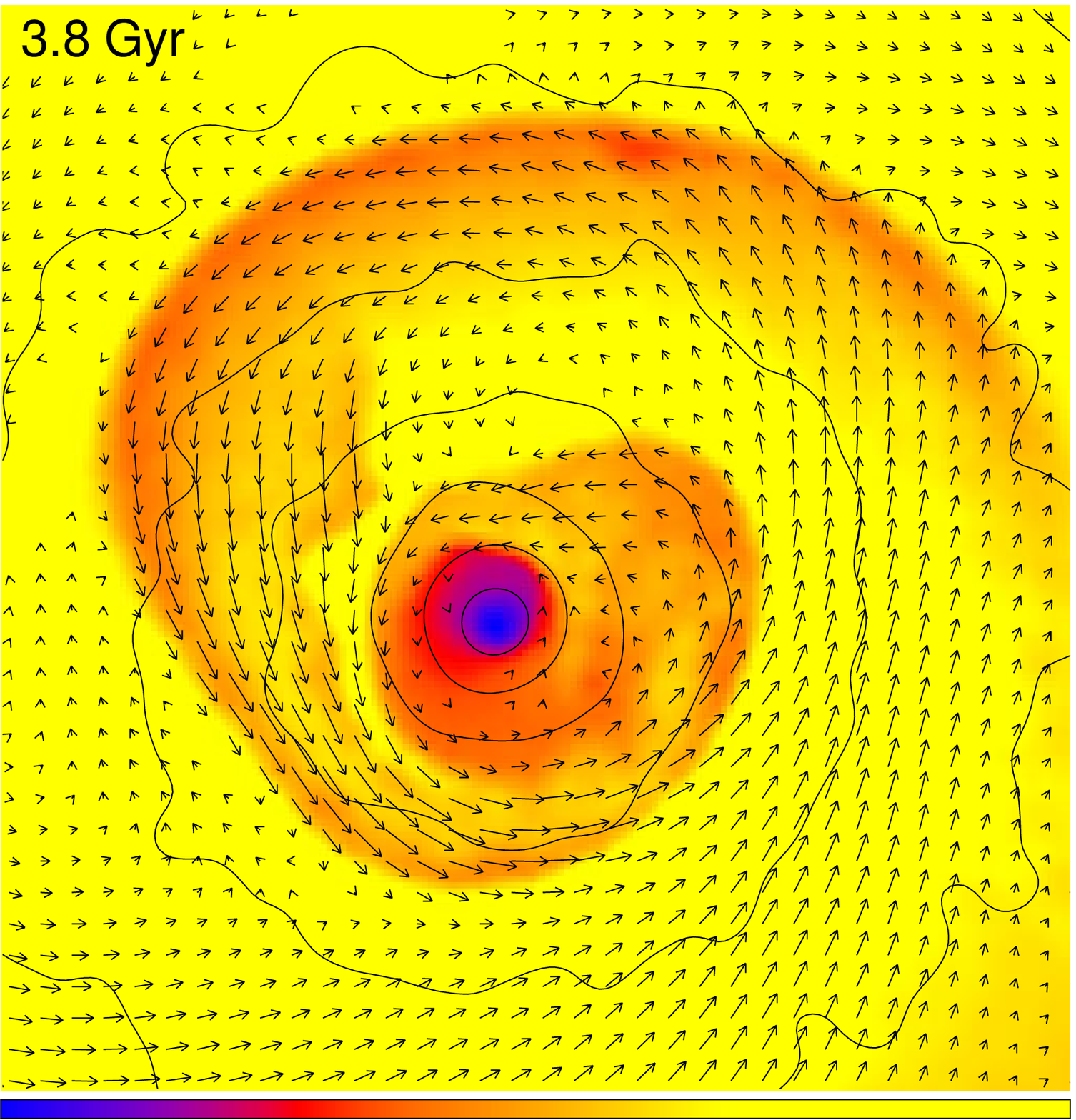}
}

\centerline{Fig. 7. --- continued}
\end{figure*}

\section{Merger with a gasless subcluster}
\label{secDM}

\begin{figure*}
\centering 
\includegraphics[width=5.4cm]{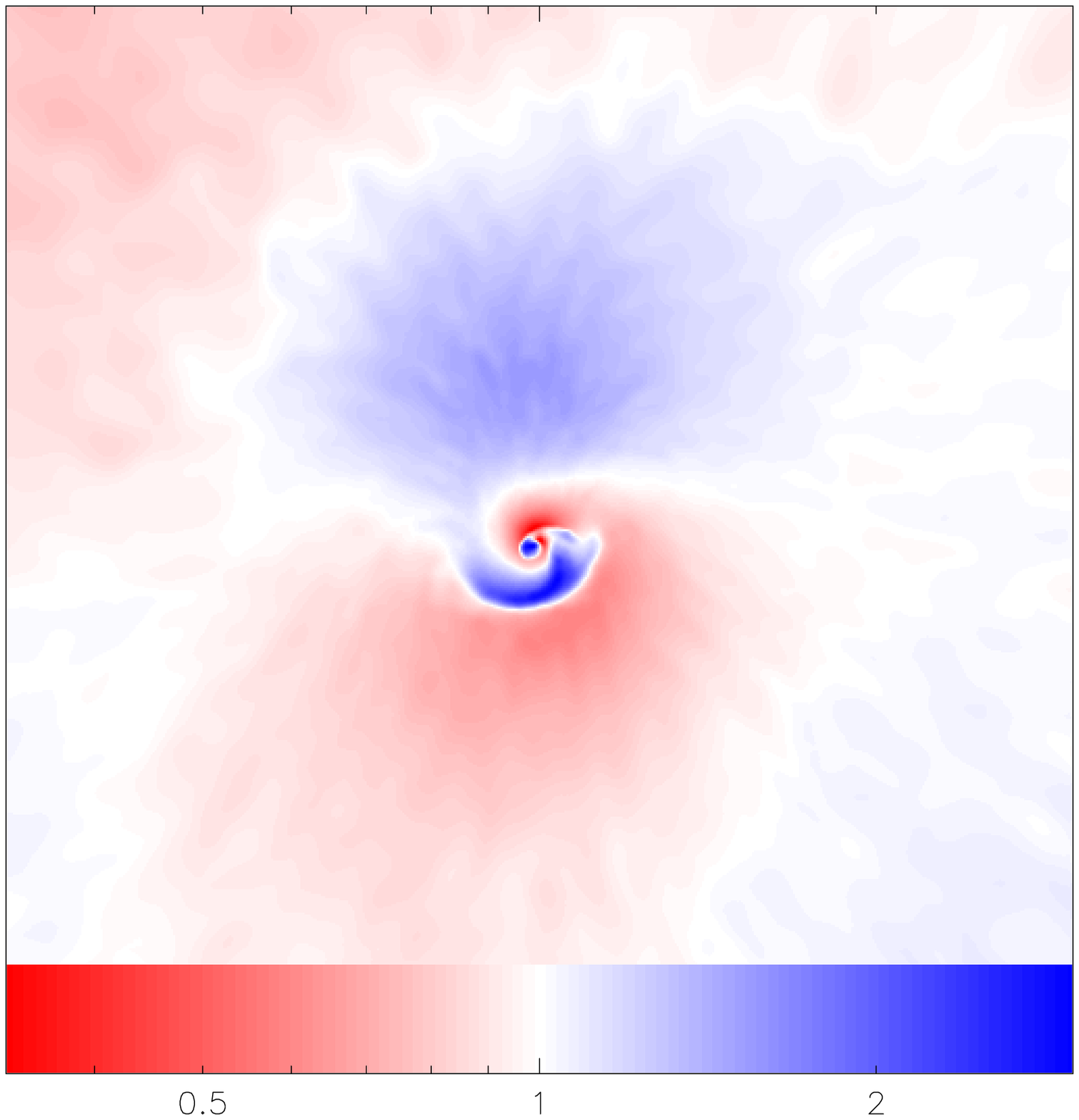}
\includegraphics[width=5.4cm]{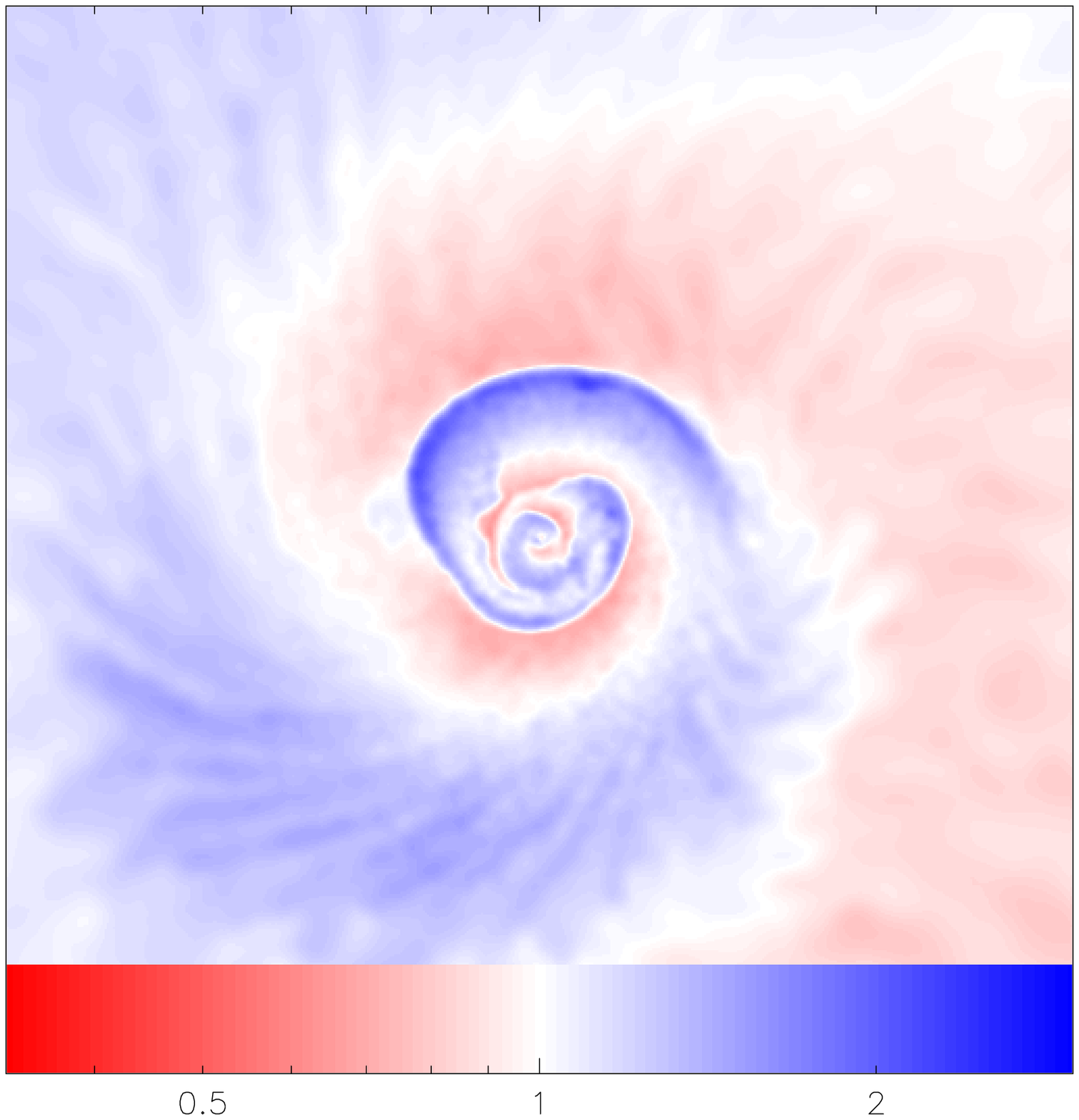}
\includegraphics[width=5.4cm]{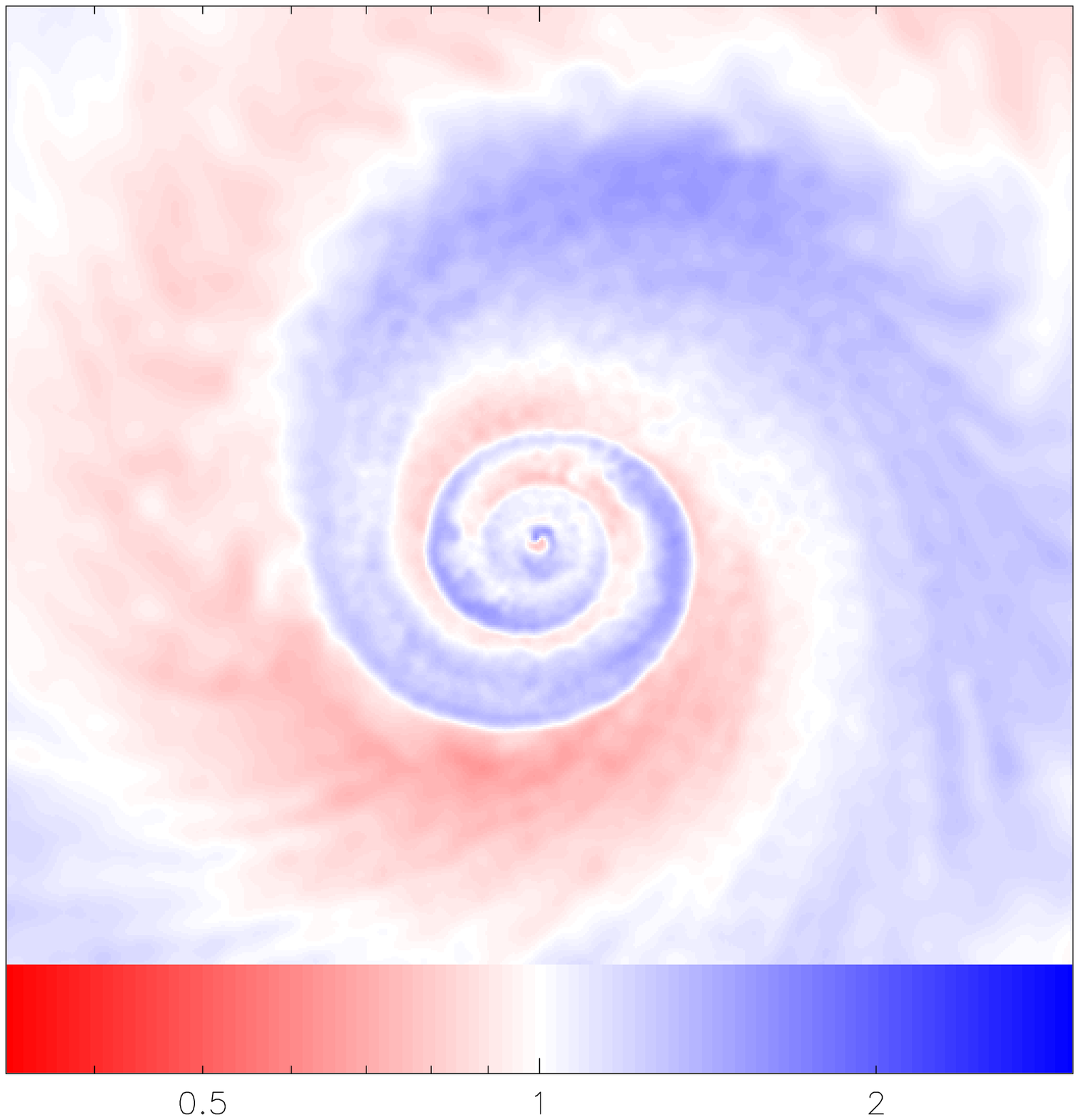}
\caption{The relative radial displacement of the gas particles, $r/r_0$,
  where $r_0$ is the initial radial distance of the particle, for
  $t=2.0,\,3.8,\,6.0$ Gyr, for the DM-only subcluster run (the same slices
  as shown in Fig.\ \ref{figDM}). Both $r$\/ and $r_0$ are relative to the
  DM peak of the main subcluster at the corresponding time. Panels are 1 Mpc
  in size, centered on the DM peak.}
\label{figRR0}
\end{figure*}

\begin{figure*}
\centering
\includegraphics[width=5.1cm]{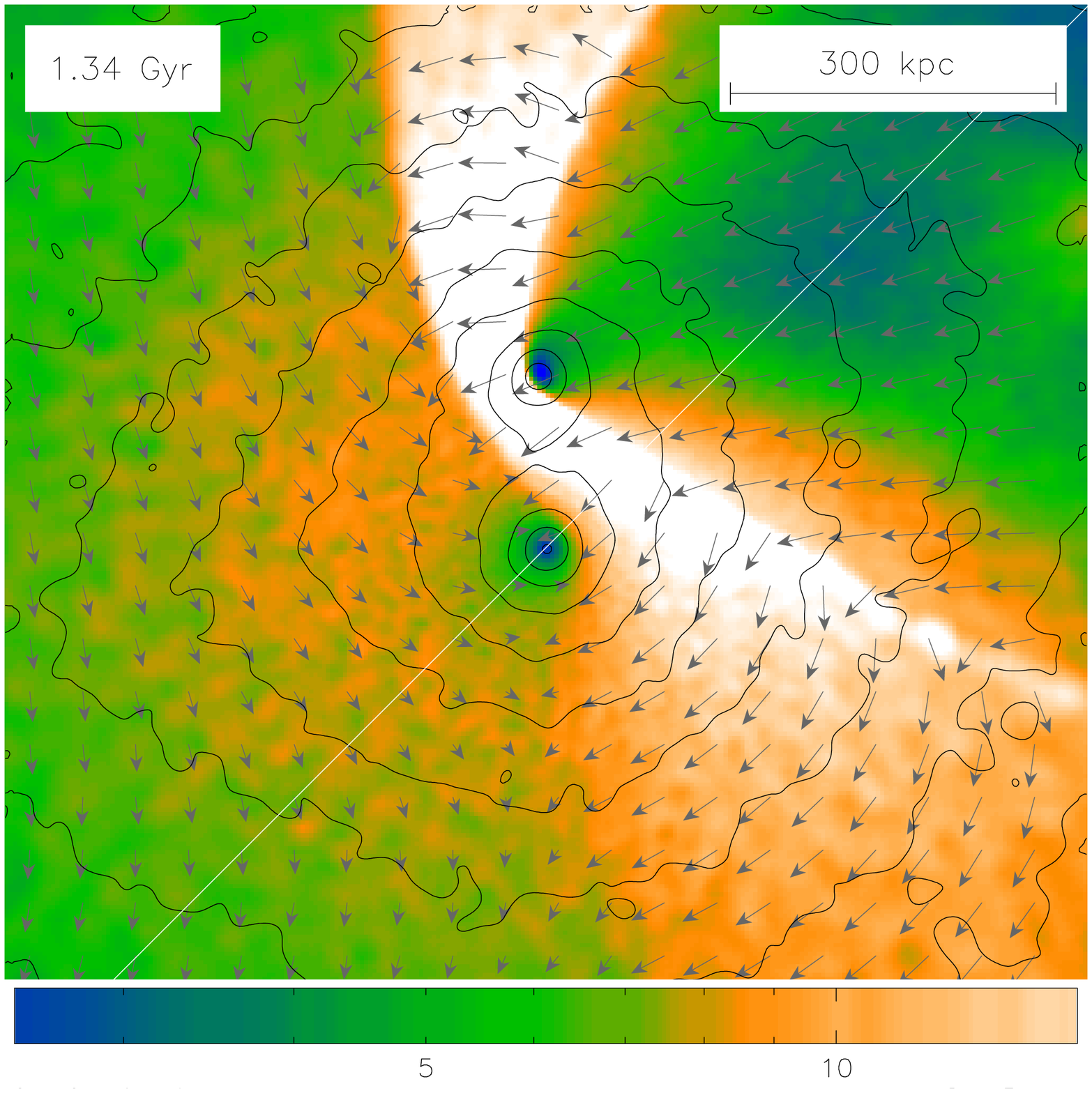}
\includegraphics[width=5.1cm]{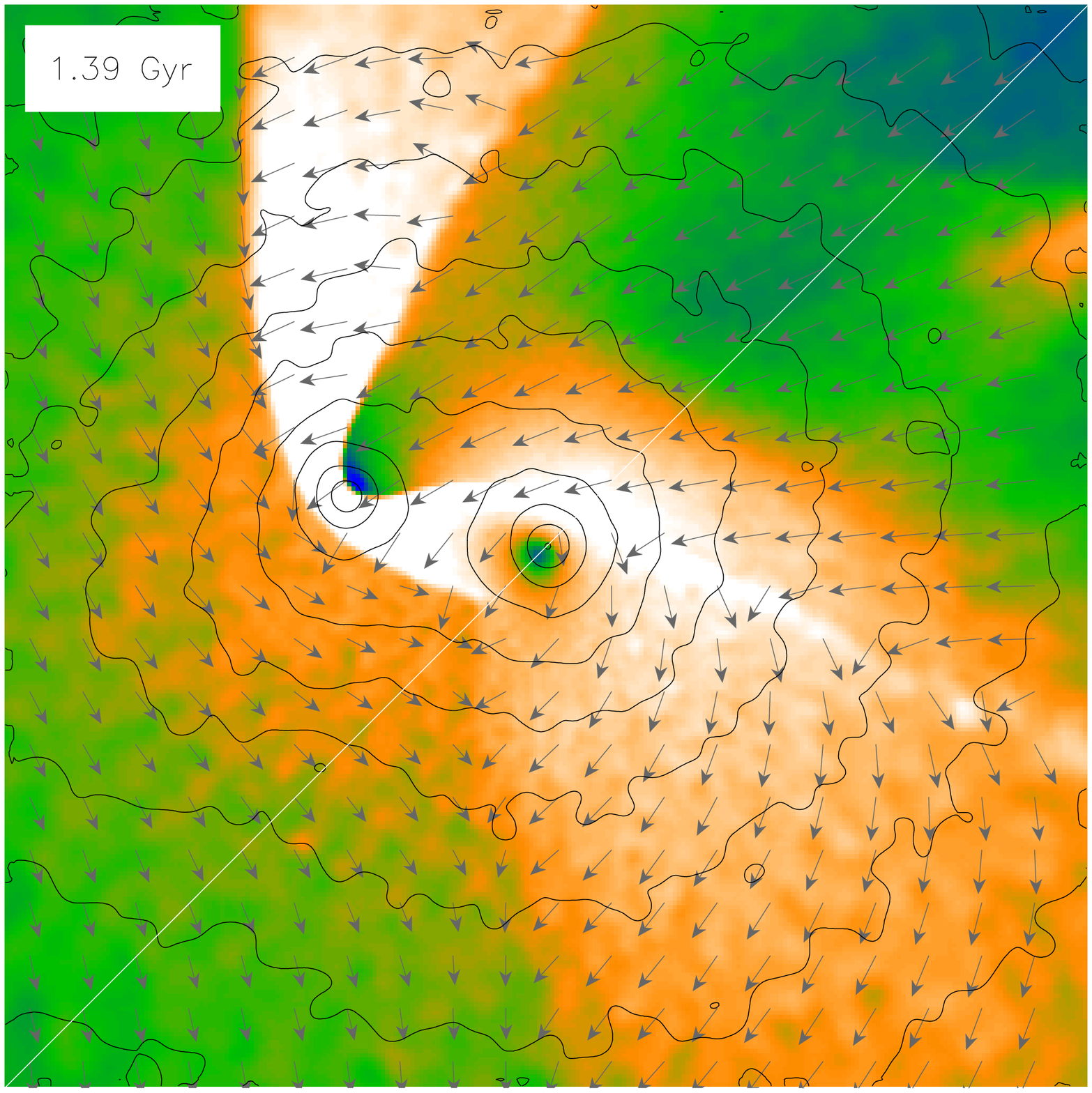}
\includegraphics[width=5.1cm]{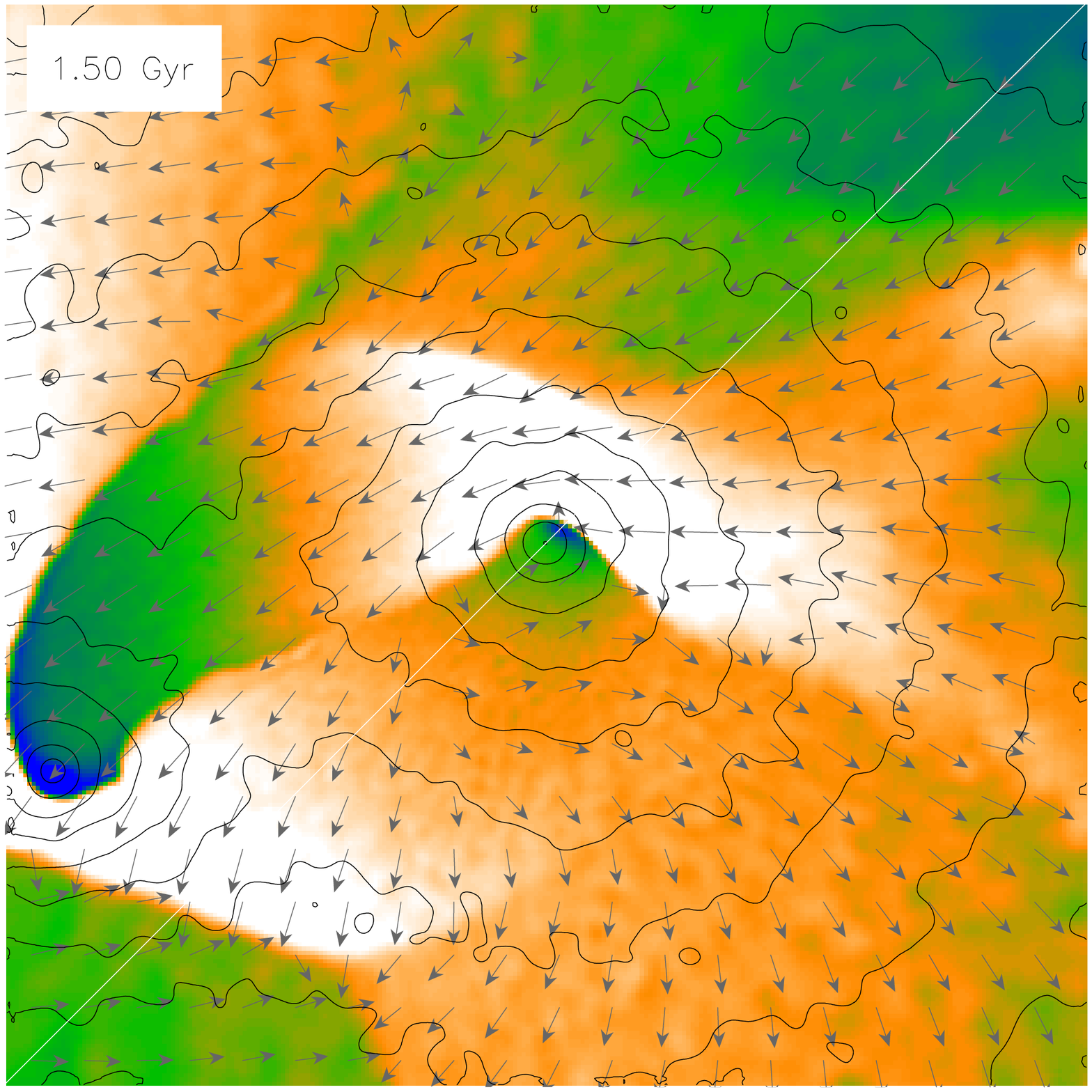}\\
\includegraphics[width=5.1cm]{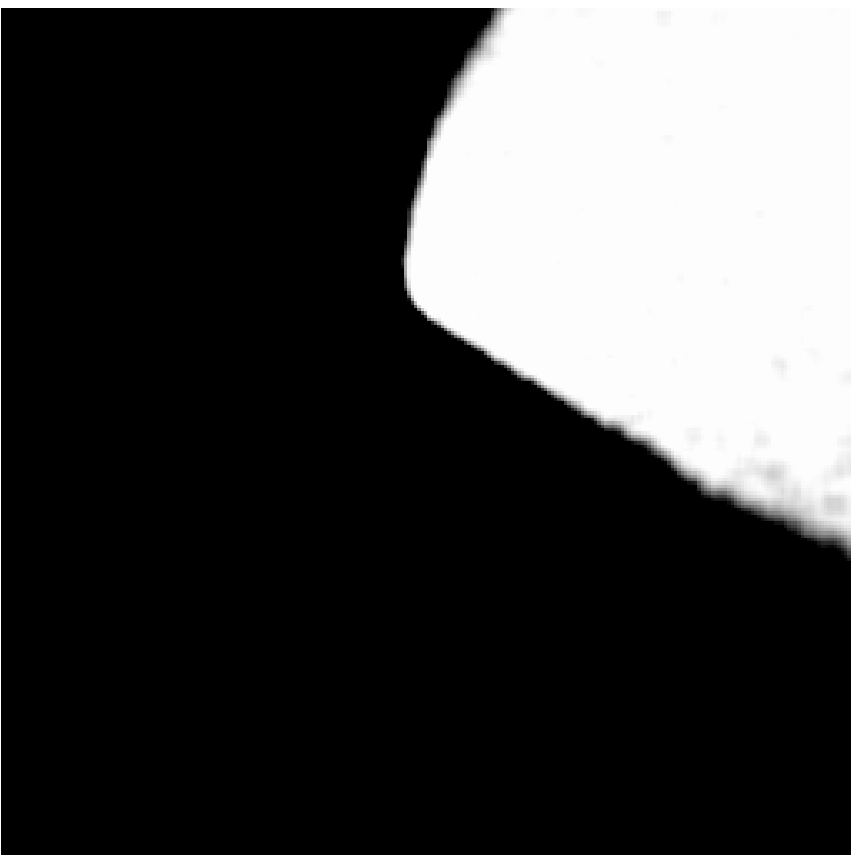}
\includegraphics[width=5.1cm]{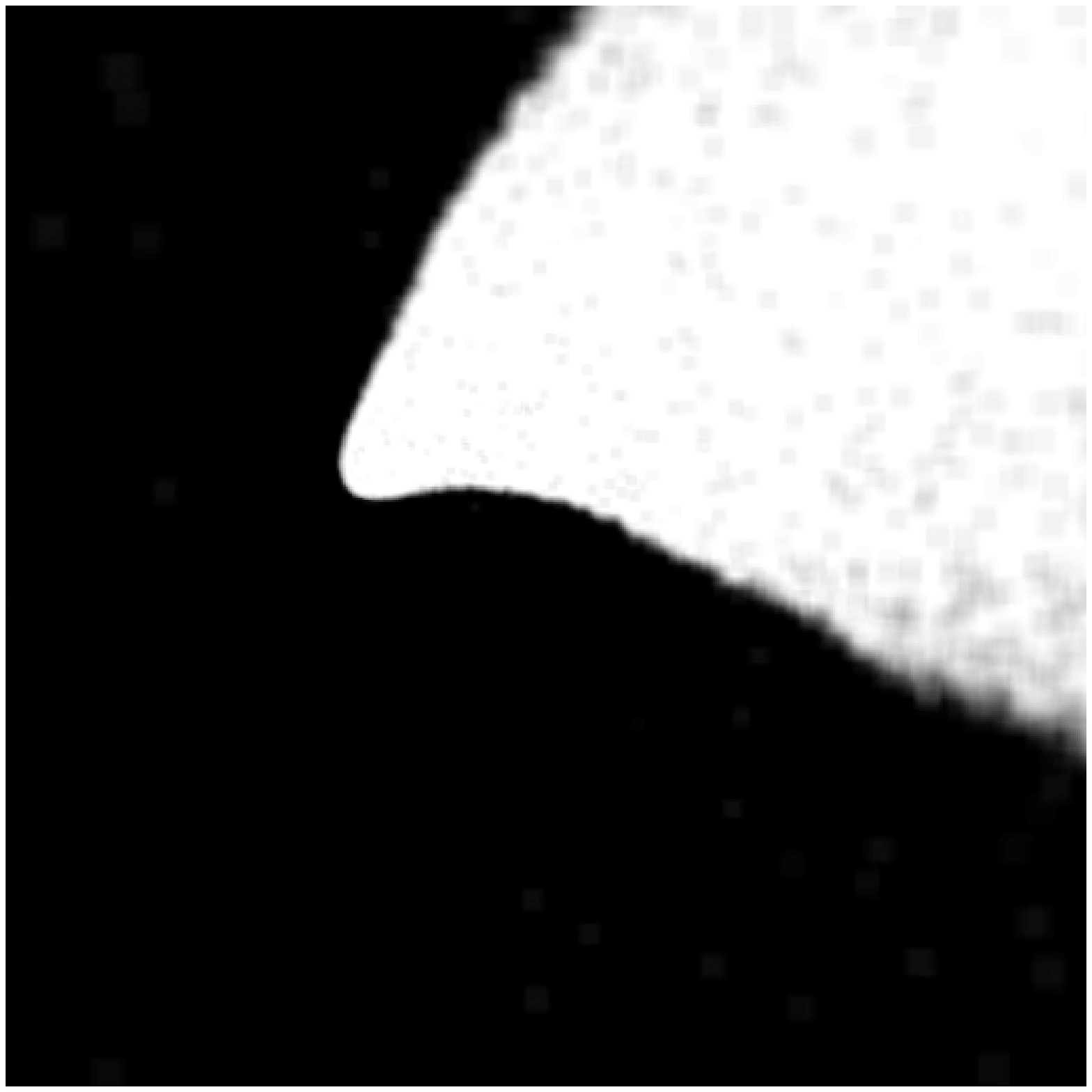}
\includegraphics[width=5.1cm]{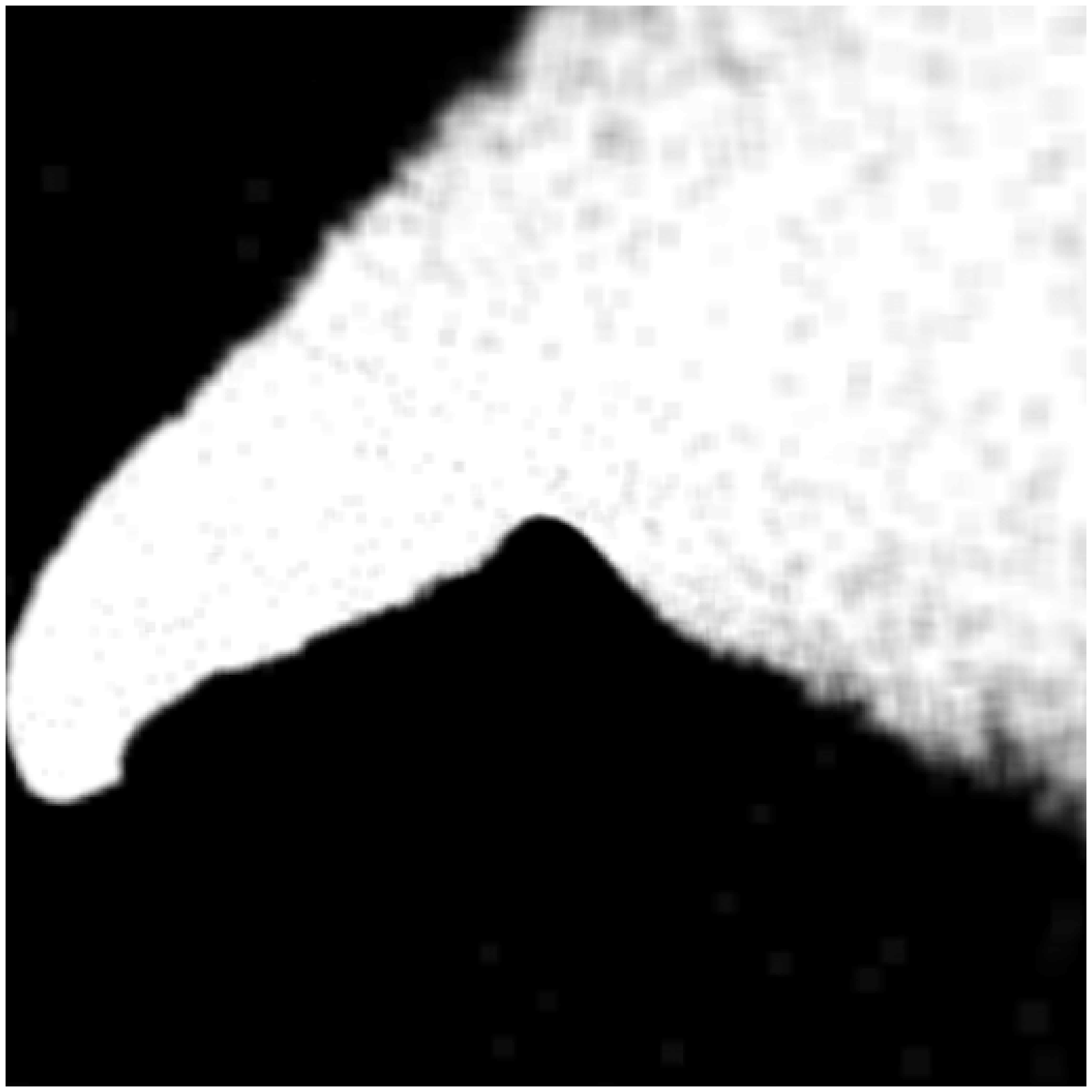}\\[4mm]
\includegraphics[width=5.1cm]{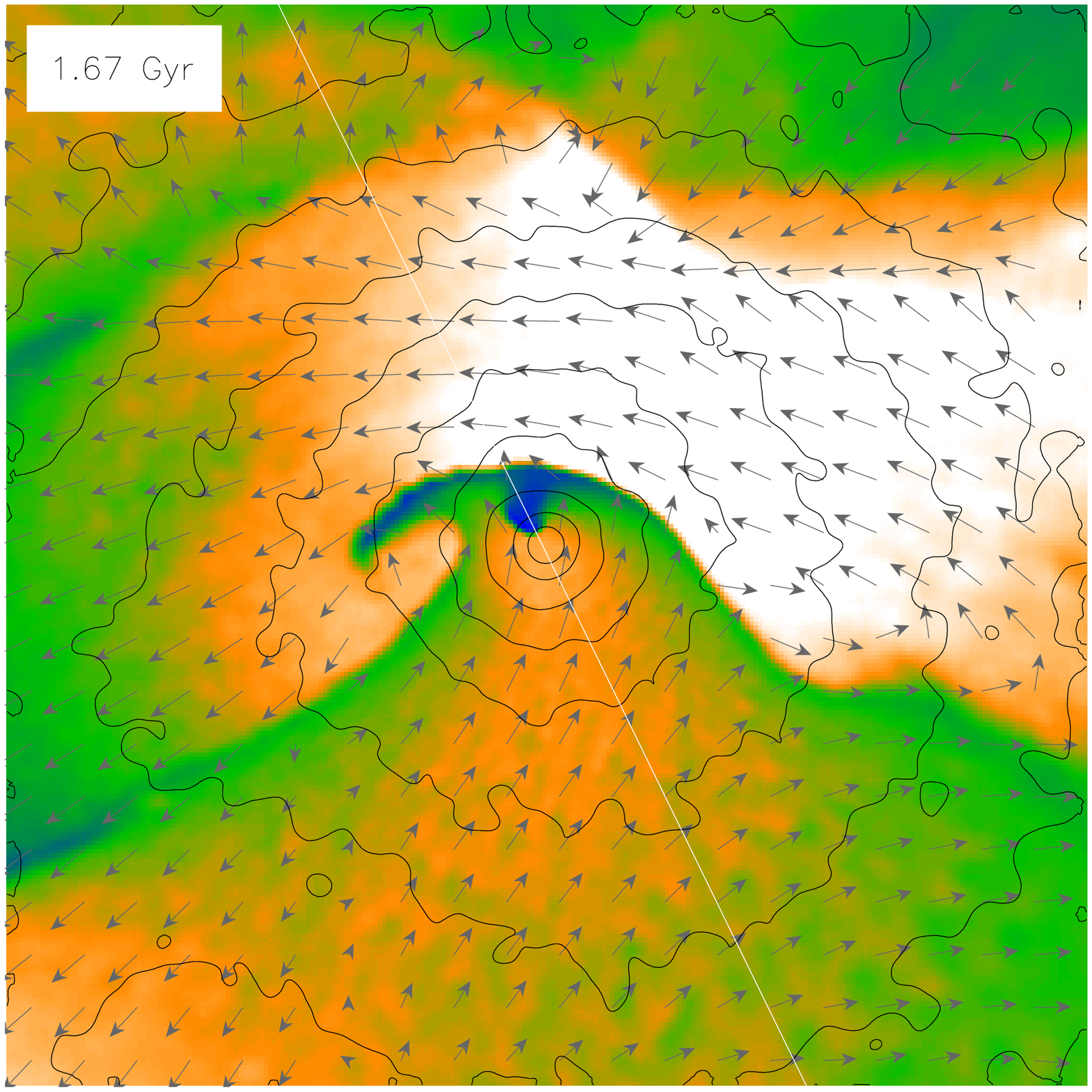}
\includegraphics[width=5.1cm]{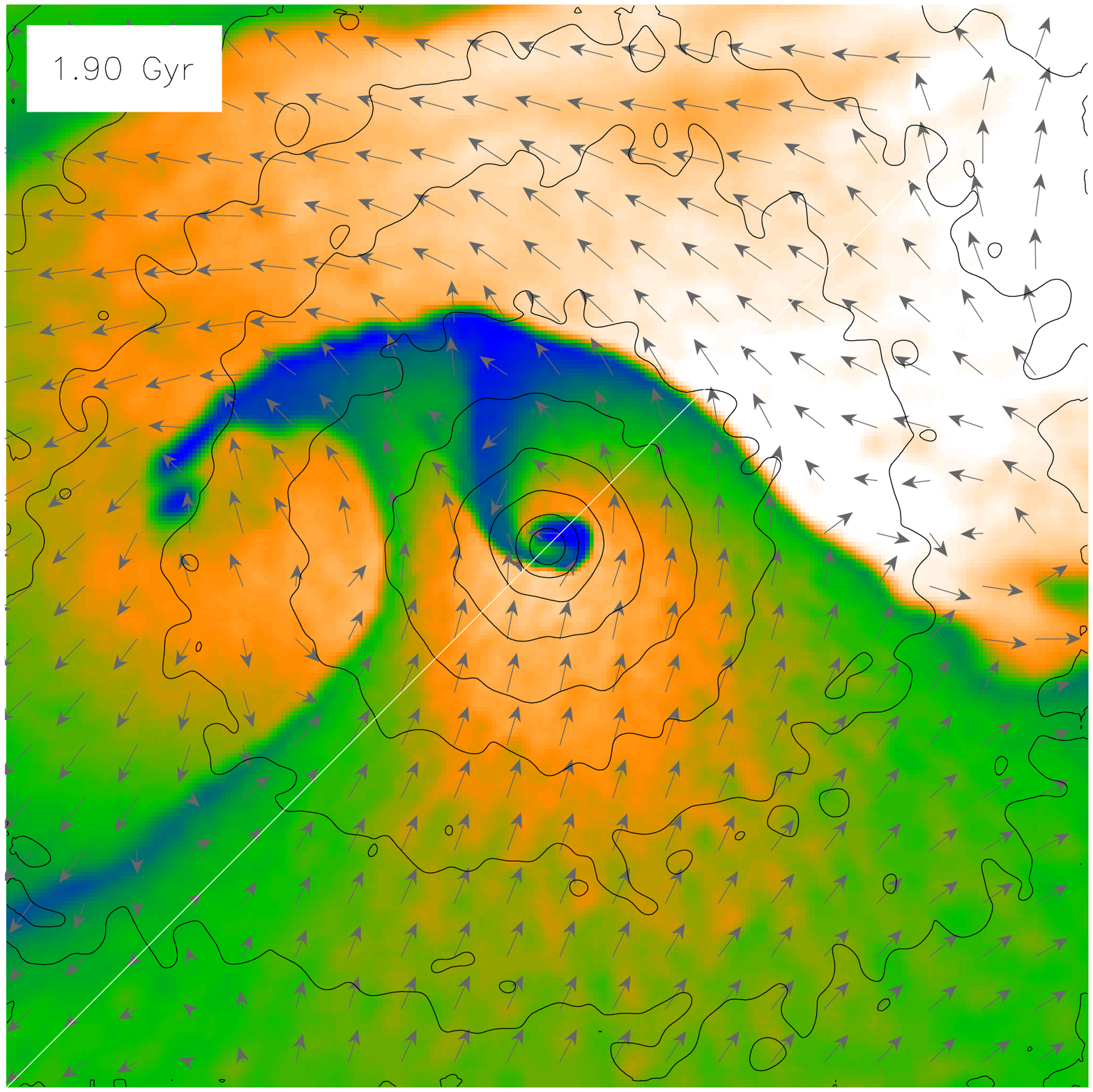}
\includegraphics[width=5.1cm]{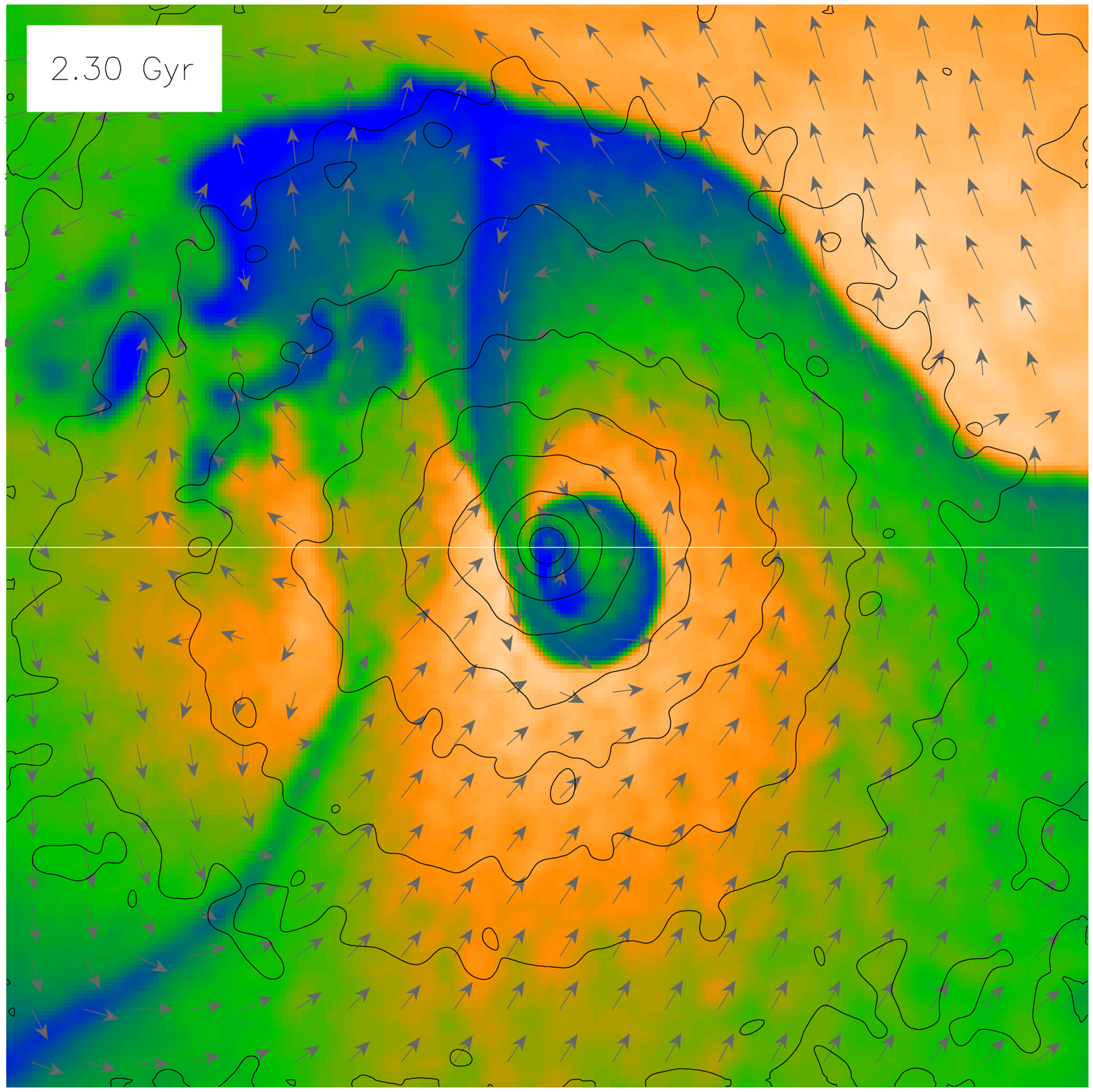}\\
\includegraphics[width=5.1cm]{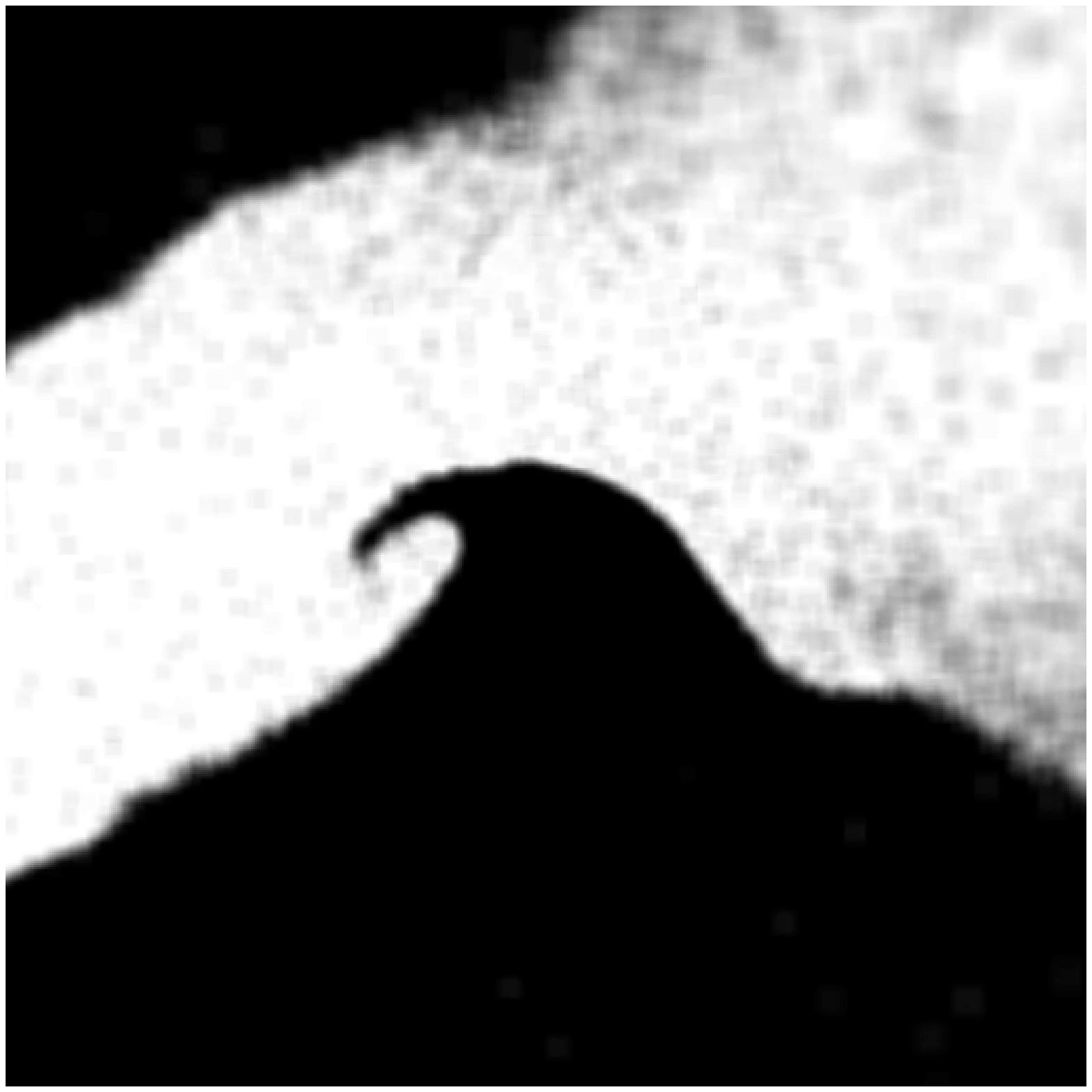}
\includegraphics[width=5.1cm]{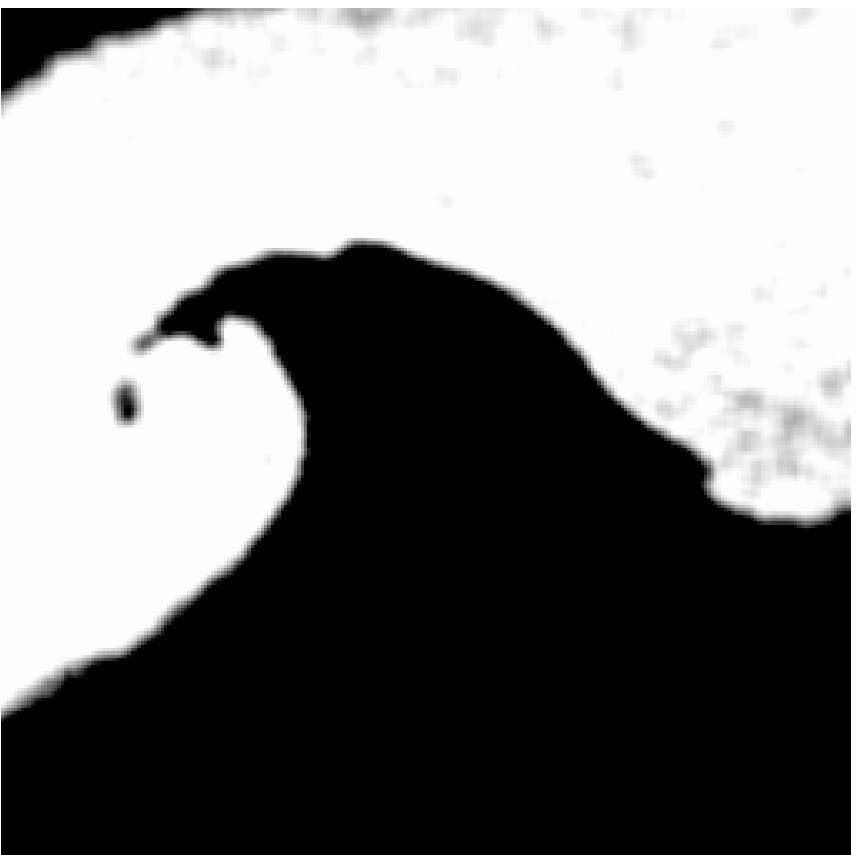}
\includegraphics[width=5.1cm]{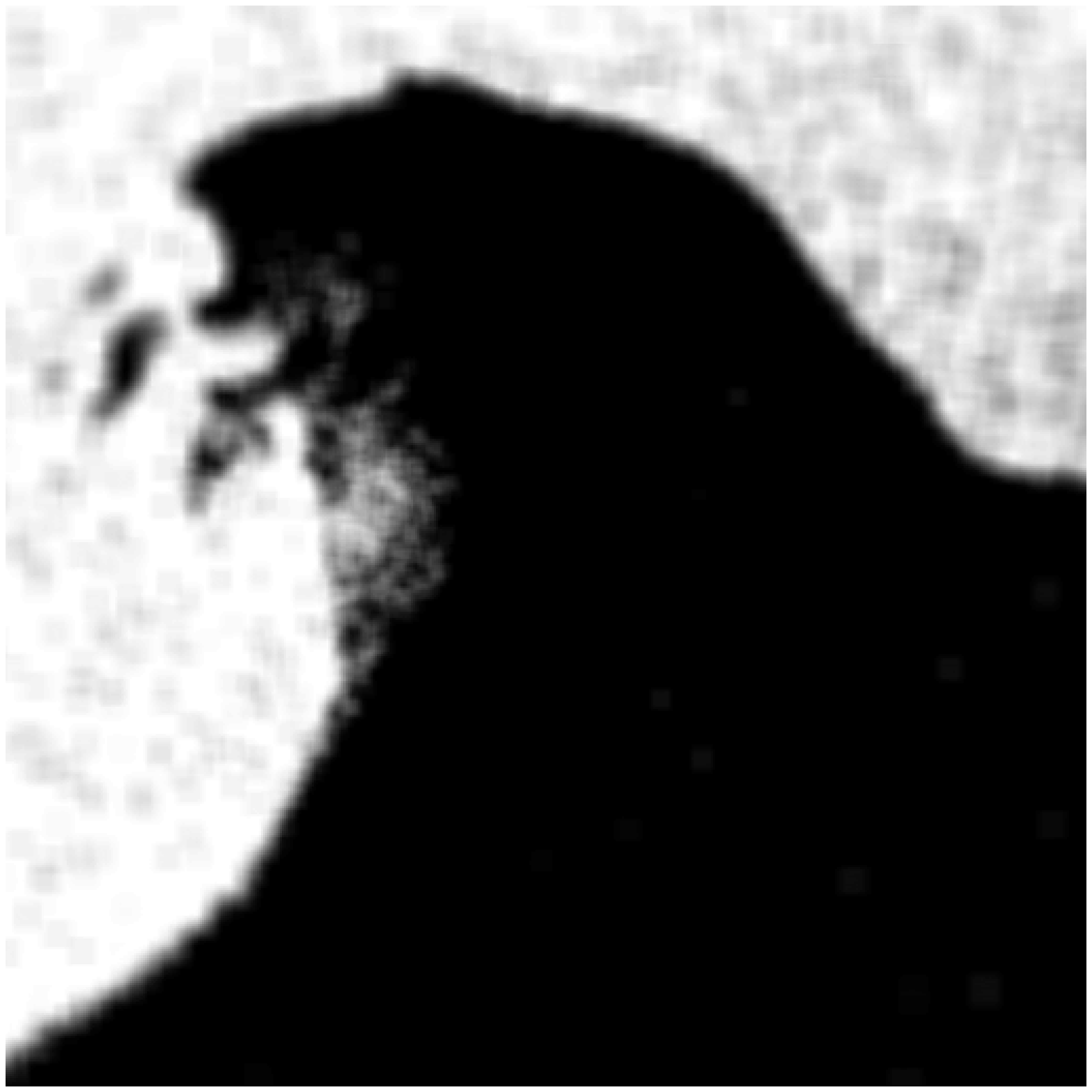}
\caption{
  Temperature slices for the same encounter as in Fig.\ \ref{figDM} ($R=5$,
  $b=500~$kpc), but now the satellite that has its own gas component. A
  black-and-white plot under each image shows the fraction of particles
  initially belonging to the two subclusters (black is main cluster, white
  is subcluster). The panel size is 1 Mpc. The profiles in Fig.\ 
  \ref{figJumpGas} are derived along the white lines in these temperature
  plots.}
\label{figGas}
\end{figure*}

\begin{figure*}
\centering
\includegraphics[width=5.4cm]{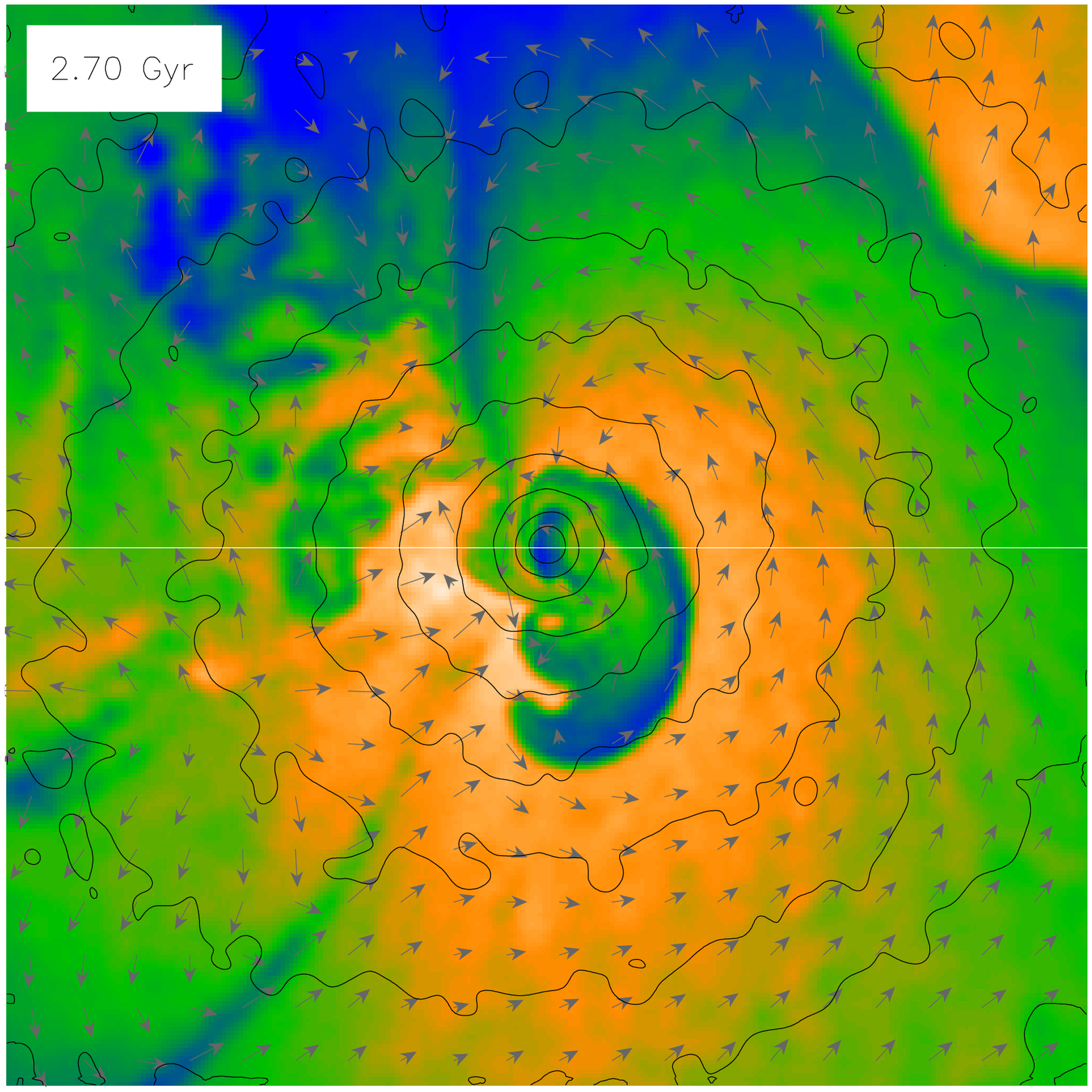}
\includegraphics[width=5.4cm]{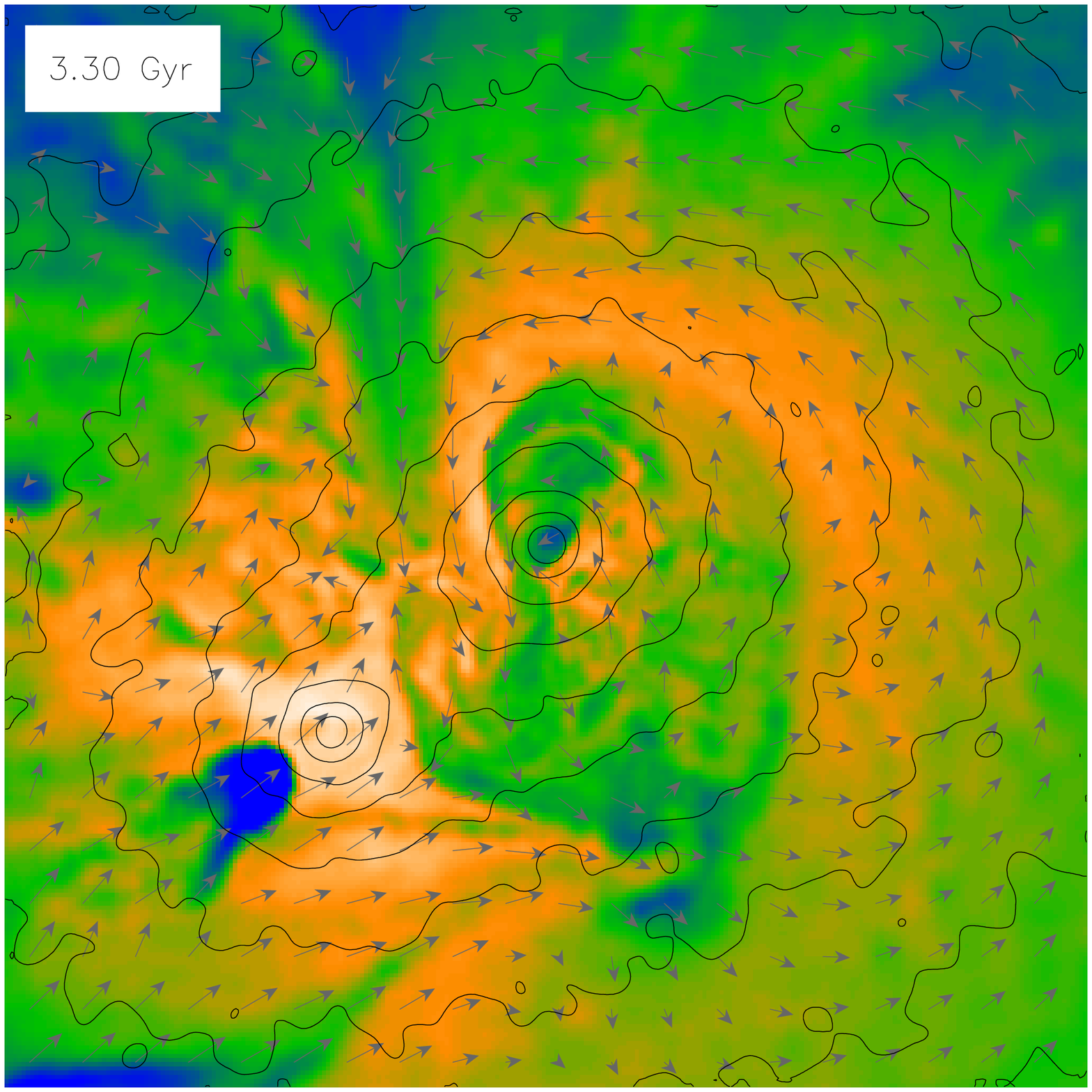}
\includegraphics[width=5.4cm]{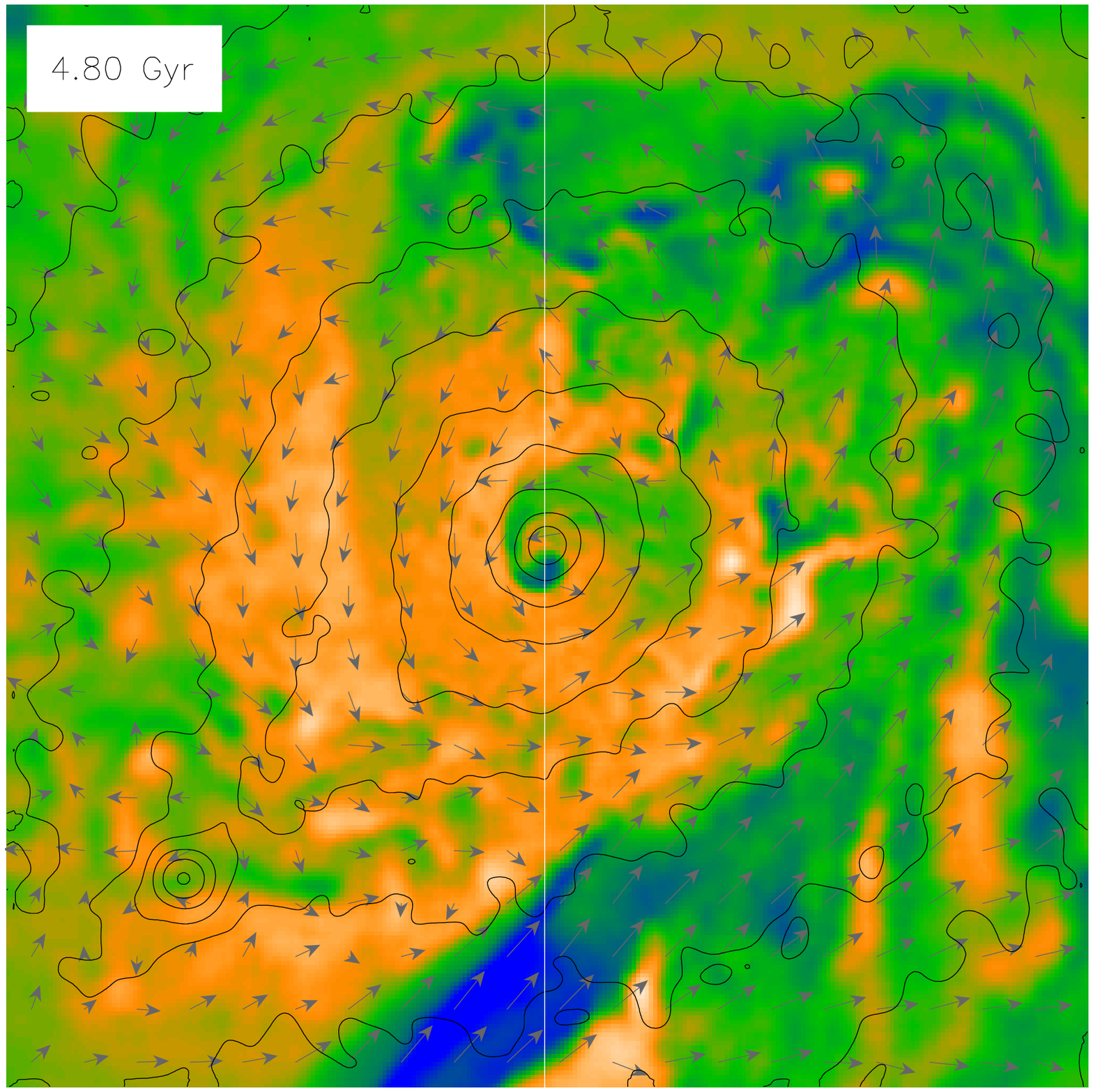}\\
\includegraphics[width=5.4cm]{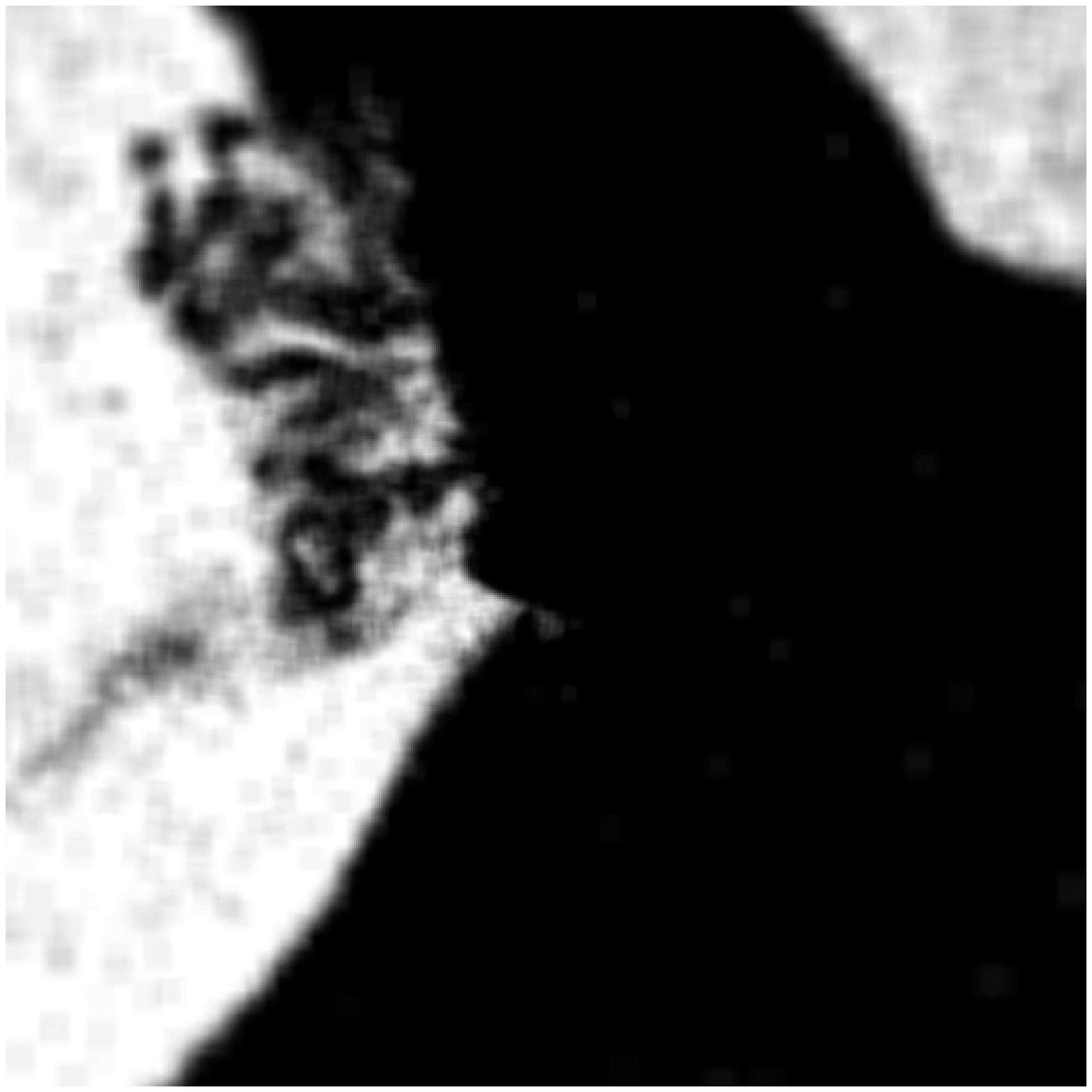}
\includegraphics[width=5.4cm]{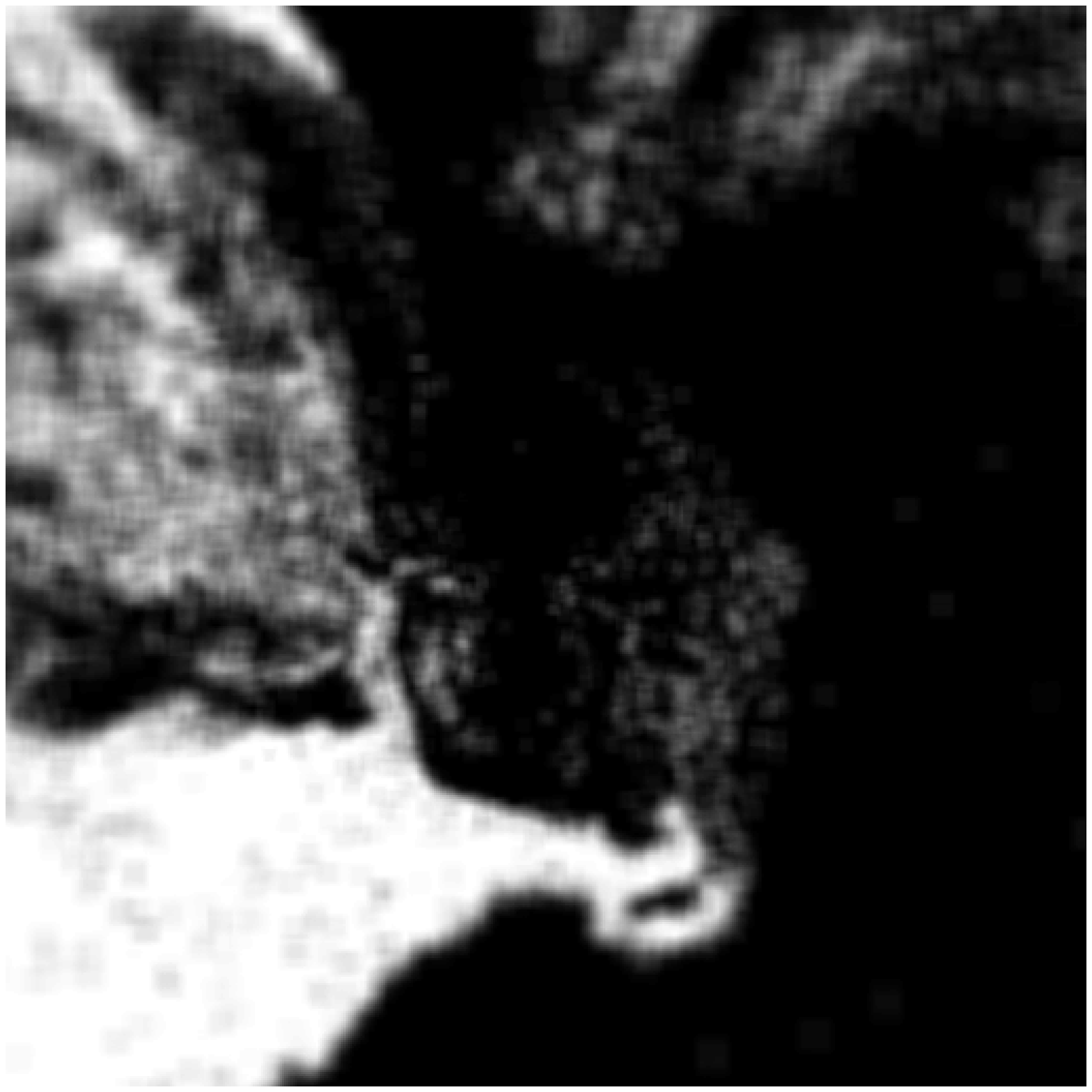}
\includegraphics[width=5.4cm]{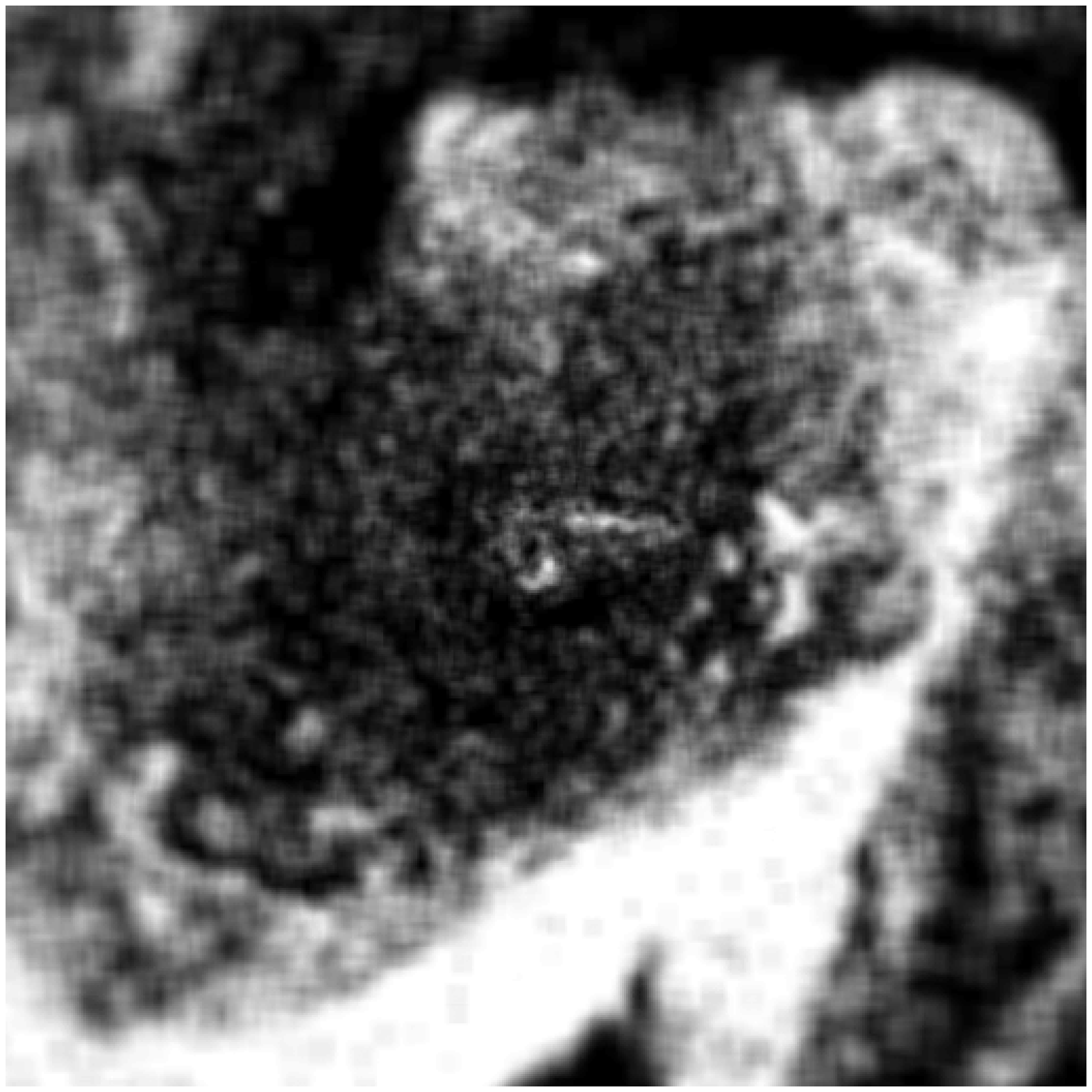}
\centerline{Fig. 9. --- continued}
\end{figure*}

\begin{figure*}
\centering
\plotone{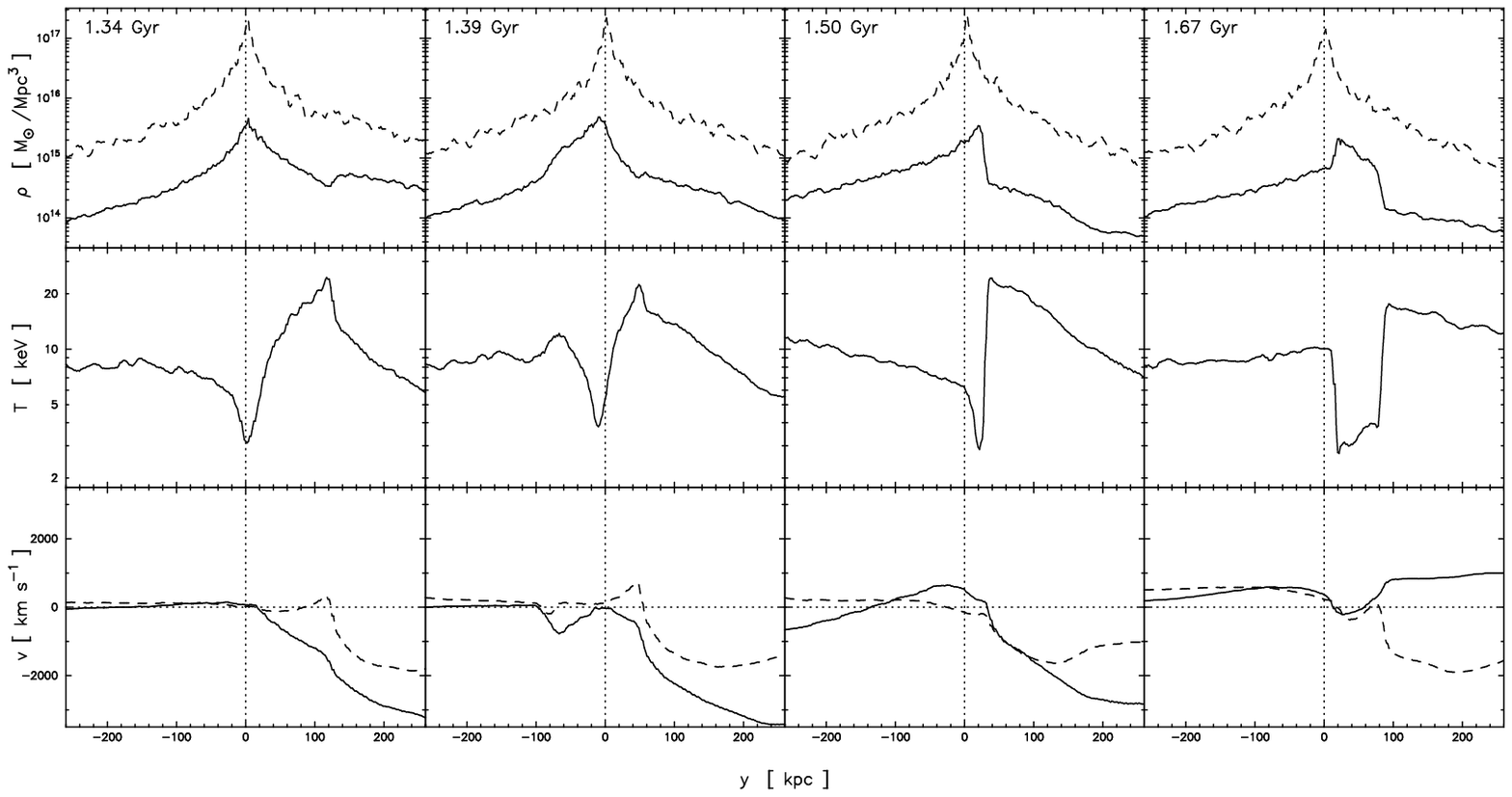}
\plotone{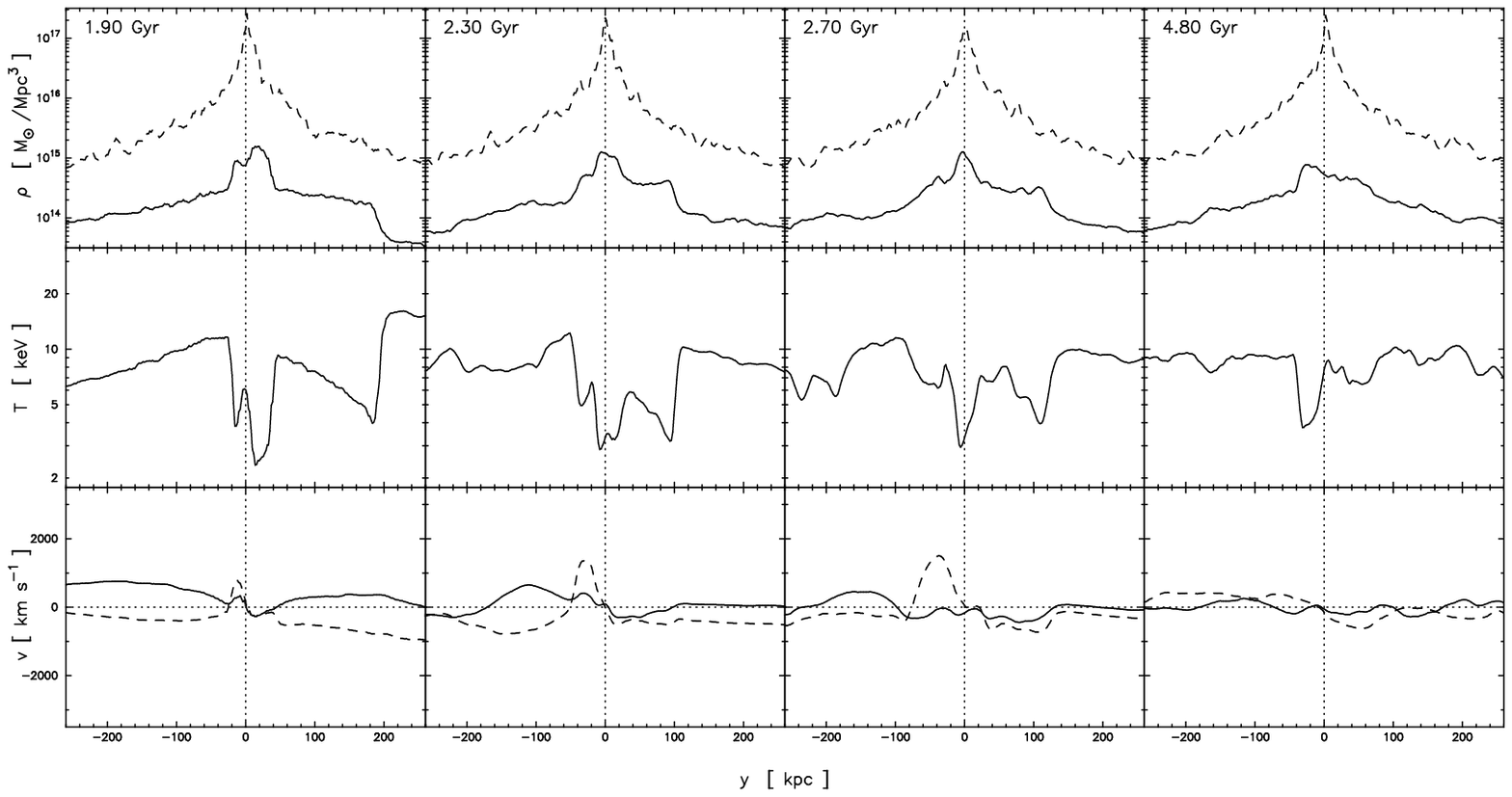}
\caption{Density, temperature and velocity along the white lines in
  Fig.~\ref{figGas}. As in Fig.\ \ref{figJumpDM}, solid and dashed lines in
  the top panels represent gas and DM density, respectively; solid and
  dashed lines in the velocity panels show the components parallel and perpendicular
  to the $y$-axis (white lines in Fig.~\ref{figGas}), relative to the DM peak.
  Note that the velocity scale is much larger than in Fig.~\ref{figJumpDM}.}
\label{figJumpGas}
\end{figure*}

We first consider a simple case (which will also turn out to be the most
relevant), in which the infalling substructure is just a DM halo without any
gas at all. This situation may arise, for instance, if the satellite lost
all its gas due to ram-pressure stripping during an earlier phase of the
merger.  As we will see, such a merger does generate sloshing of the cool
central gas and multiple cold fronts.  Compared to a merger in which both
subclusters have gas (considered in the next section), the hydrodynamics in
this case is relatively simple and the underlying processes can be
identified more easily.

Figure~\ref{figDM} shows the evolution of an encounter with $R=5$ and impact
parameter $b=500$ kpc. While mergers with such massive subclusters may be
relatively rare, this choice allows us to see the effects of the disturbance
more clearly. For these merger parameters, the first core passage of the
satellite takes place at about $t\approx1.37$ Gyr from the start of the
simulation run, at a distance approximately 150 kpc from the minimum of the
gravitational potential.  Different values of $R$ and $b$ lead to different
orbits, with different time and length scales.  The extent and intensity of
the induced sloshing and subsequent cold fronts vary, but the qualitative
behavior is similar.  The cool gas peak is first displaced from the DM peak
(around $t=1.6$ Gyr in Fig.\ \ref{figDM}), then falls back and starts
sloshing in the minimum of the gravitational potential, generating cool
edges. For $b>0$, these edges create a characteristic long-lived spiral
structure. We will now look into each of these steps in detail.

\subsection{Initial gas-DM displacement}
\label{secdecoup}

There are two independent effects acting simultaneously to provide initial
separation of the cool gas from the DM peak during the subcluster flyby.
For the DM-only subcluster considered here, the more important effect is the
gravitational disturbance that the subhalo creates around the moment of the
core passage. It causes the density peak of the main cluster (DM and gas
together) to swing along a trajectory shown in Fig.\ \ref{figtraj} relative
to the center of mass of the main cluster, and consequently, relative to the
matter and gas surrounding the peak (that is, the matter approximately
outside the distance of the closest encounter, 150 kpc, which, for the most
part, does not participate in this swinging motion).  The gas and DM peaks
feel the same gravity force and start moving together toward the subcluster
($t<1.4$ Gyr in Figs.\ \ref{figDM} and \ref{figtraj}).  However, after the
core passage, the direction of this motion quickly changes.  For the gas
peak, this leads to a change of sign of the ram-pressure force --- compare
the gas velocity field outside the cool core in Fig.\ \ref{figDM}, at 1.43
Gyr and 1.6 Gyr, as well as the gas velocity profiles in Fig.\ 
\ref{figJumpDM} (solid lines). Between these snapshots, the gas velocity
around the core changes direction from downward ($v_y\approx -200$ \kms) to
upward ($+200$ \kms; hereafter we will refer to the $y>0$ and $y<0$
directions as ``upward'' and ``downward'', respectively).  As a result, the
cool gas core, previously compressed by ram pressure from above, shoots up
from the potential minimum in a kind of ``ram-pressure slingshot'' (proposed
by Hallman \& Markevitch 2004 for a cold front in A168).  We will see this
effect more clearly in simulations where the subcluster has gas
(\S\ref{secGas}).

There is another mechanism contributing to the initial gas displacement.
Although the subcluster does not have any gas, it drags some of the main
cluster's ICM in a trailing Bondi-Hoyle wake \citep[see
e.g.][]{Sakelliou00}.  The subcluster is supersonic at core passage
($M\simeq 2$), but the ICM disturbance that it creates is sonic and the gas
density is continuous, unlike that in a shock. This wake will become a weak
shock after the core passage, when it starts propagating outwards along the
declining density profile. For the off-center merger considered here, this
wake transfers some angular momentum from the DM satellite to the ICM,
resulting in large-scale rotation of the gas around the main cluster peak.

When this disturbance reaches the central parts of the cluster, it acts to
push the cold gas core out of equilibrium. The thermal pressure profile
along the vertical line passing through the DM peak is shown in Fig.\ 
\ref{figP} for different times around core passage.  Before the wake reaches
the center, one can see the excess pressure from the wake above the cool
core (at $y\approx +200$ kpc for $t=1.39$ Gyr). The pressure difference
changes sign as the wake passes the core, so that at $t=1.5$ Gyr, the
pressure below the core (at $y\approx -100$ kpc) is higher. This is
analogous to the sound wave disturbance proposed by Churazov et al.\ (2003)
and Fujita et al.\ (2004) as a cause of the initial core displacement.  This
change of the gas pressure difference occurs at approximately the same time
as the ram-pressure slingshot described above.  The two effects combine to
displace the cool gas peak upward (in the $y>0$ direction) from the DM peak
by about 15 kpc at $t\simeq 1.6$ Gyr (Fig.\ \ref{figJumpDM}), while the
outlying core gas is displaced to greater distances.  (Note that because
there is also a gas velocity change across the wake, the two effects are not
entirely independent.) Neither the passage of the wake through the core, nor
the motion of the core, create any sharp temperature discontinuities until
$t\approx 1.7$ Gyr (Fig.\ \ref{figJumpDM}), by which time the subcluster has
moved 1 Mpc away from the main peak.

\subsection{Onset of gas sloshing}
\label{secslosh}

After the cool gas peak has been displaced from the potential minimum and
the displacing force has diminished, the gas starts falling back toward the
center. The details of how this happens can be seen in Fig.\ 
\ref{figDMzoom}, which shows a zoomed-in view of the gas temperature and
velocity field in the core at several interesting moments. The outermost
part of the displaced cool gas expands adiabatically as it is carried
further out by the upward flow of the surrounding gas (the orange plume
above the center in the 1.6--1.7 Gyr panels of Fig.\ \ref{figDMzoom}).
However, in a process not unlike the onset of a Rayleigh-Taylor (RT)
instability, the densest, coolest gas quickly starts sinking towards the
minimum of the gravitational potential, against ram pressure from the upward
flow of the surrounding gas.  This is seen most clearly in the 1.6 Gyr and
1.8 Gyr snapshots. At 1.6 Gyr, the cool gas peak starts to flow downward
toward the DM center for the first time. By $t=1.7$ Gyr, the cool gas has
overshot the center and, subjected to ram pressure from the gas on the
opposite (downward) side of the DM center still moving upward, spreads into
a characteristic mushroom structure, behind which the cool gas
continues to flow downward.  At 1.8 Gyr, a new RT tongue develops on the
inner side of the first mushroom head --- the densest, lowest-entropy gas
separates and again starts flowing back toward the potential minimum.
Meanwhile, the rest of the gas in the mushroom head, and the gas with still
higher entropy that by now has started to flow in from above (downward
arrows in the orange area in the 1.8 Gyr panel), continues to move outwards,
expanding adiabatically as it moves into the lower-pressure regions of the
cluster.  A snapshot later (1.9 Gyr panel), the coolest gas on its way in
(up) encounters ram pressure from the hotter gas that by now has a
significant downward velocity, and again develops a classic mushroom
structure seen in many hydrodynamic simulations (e.g., Heinz et al.\ 2003;
Takizawa 2005).  The stem of the mushroom is the forward flow of the cool
gas and the head is where it is slowed by ambient ram pressure and spread
sideways, creating characteristic eddies in the velocity field.

Every time the velocity of the densest, lowest-entropy gas is reversed
w.r.t.\ that of the outer, higher-entropy gas in the course of the above
RT-like process, the gas parcels with different entropies are quickly
brought into contact, creating a cold front.  The emergence of these
discontinuities (at least at the resolution of our simulations) is clearly
seen in the gas density and temperature cross-sections shown in Fig.\ 
\ref{figJumpDM}. We will discuss the precise mechanism by which the
discontinuities arise in \S\ref{secOrigin}.

As we see, immediately after the disturbance from the subcluster flyby, the
cool central gas starts sloshing in the minimum of the gravitational
potential, creating edge-like discontinuities on progressively smaller
linear scales at each new passage through the center.  Sloshing of the
densest gas that is closest to the center occurs with a smaller period and
amplitude than that of the gas initially at greater radii. Note that the
oscillation of the DM peak caused by the subcluster flyby has a much longer
period (Fig.\ \ref{figtraj}), of order 1 Gyr, than the 0.1 Gyr timescale of
gas sloshing.  Indeed, as seen in Fig.\ \ref{figDMzoom}, the DM distribution
in the core stays centrally symmetric, while the gas sloshes back and forth
in its potential well. This is contrary to the picture proposed by Tittley
\& Henriksen (2005) (based on lower-resolution simulations) where the DM
peak oscillates back and forth and drags the gas peak along, exposing it to
ram pressure.  The DM peak motion is indeed one of the two causes of the
initial DM-gas decoupling (as discussed above in \S\ref{secdecoup}), and in
the long term, the DM oscillations should continue to feed kinetic energy to
the sloshing gas. However, sloshing itself in our simulations is a
hydrodynamic effect in a quasi-static central gravitational potential.

\subsection{The spiral structure}
\label{secspiral}

As mentioned above, the off-center flyby of the subcluster transfers angular
momentum to the gas near the core via the gas wake.  When the cool gas is
displaced from the center for the first time, it acquires angular momentum
from the gas in the wake and does not fall back radially.  As a result, the
subsequent cold fronts are not exactly concentric (e.g., the 1.9--2.1 Gyr
panels in Fig.\ \ref{figDMzoom}) but combine into a spiral pattern.
Initially, this spiral pattern does not represent any coherent spiraling
motion --- each edge is an independent structure. In fact, the gas on the
opposite ends of each edge flows in the opposite directions.  However, as
the time goes by and the linear scale of the structure grows, most clockwise
motion (i.e., that against the average angular momentum of the outer gas)
subsides and the ``mushrooms'' become more and more lopsided (compare the
1.9 Gyr and 2.9 Gyr panels in Fig.\ \ref{figDMzoom}).  On large scales, the
spiral does indeed become a largely coherent spiraling-in of cool gas ---
the mushroom stems, through which the low-entropy gas flows toward the
mushroom cap, shift more and more to the edge of the cap.  In the inner
regions of the ``spiral'', there are still parts flowing in the opposite
direction (e.g., at 2.9 Gyr).

Meanwhile, the satellite proceeds on its orbit, reaching apocenter at
$t\approx2.7~$Gyr and $r\approx1.7~$Mpc, and falling back to the core at
$t\approx4~$Gyr (Fig.\ \ref{figDM}).  During all this time, cold fronts in
the core are the only visible trace of the encounter.  The density and
temperature contrasts across the cold fronts decrease from a factor of 3 at
$t=1.7~$Gyr to about a factor of 2 at $t=2~$Gyr and $\sim1.5$ at
$t=3.8~$Gyr. The cold fronts generated by the first flyby of the dark matter
satellite survive its second passage, which generates new sloshing in the
inner part (although less significant, because the satellite has lost some
of its mass).  Temperature and density jumps at a 20\% level remaining from
the first encounter are still seen even at $t=6$ Gyr, almost 5 Gyr after
that encounter (this is, of course, academic, since any real cluster will
probably experience a few more mergers during this period).

One interesting question that may arise from looking at the evolution of the
spiral temperature structure (Figs.\ \ref{figDM} and \ref{figDMzoom}) is how
the small initial displacement of the central cool gas apparently succeeds
in spreading this low-entropy gas out to large radii against convective
stability in the radially increasing entropy profile.  In fact, this is not
the case. Figure \ref{figRR0} shows the ratio of the radial distances of the
gas particles at present to their initial distances from the center of the
undisturbed cluster. All cool structures are indeed comprised of gas that
has moved away from the center.  However, the amplitude of the radial change
is diminishing with time. At 2 Gyr, sloshing displaces the gas along the
radial coordinate by up to a factor of 3, while at 6 Gyr it is not more than
a factor of 1.5.  That is, as expected, the gas that initially splashes out
to large distances eventually settles back in the center.  The cool gas that
forms the large-scale cold fronts at later stages has not come all the way
from the center, but from only slightly smaller radii (although possibly
from the opposite side of the cluster). It expanded adiabatically, which
enhanced the temperature contrast.  While Figs.\ \ref{figDM} and
\ref{figDMzoom} show that each cold front indeed starts near the center and
expands continuously, the gas at the front is not the same at every moment.
As seen most clearly in the 1.9--2.1 Gyr panels of Fig.\ \ref{figDMzoom},
there is circulation inside the cold front, in which the lowest-entropy gas
continuously flows back from the front toward the center.  It is replaced at
the front by higher-entropy gas that arrives later and whose origin traces
back to greater radial distances.  Finally, the oscillation of the DM peak
on a Gyr timescale relative to the bulk of the cluster gas also contributes
to widening the spiral pattern.

We note here that our conclusion about the longevity of the spiral structure
may be affected by two numeric effects. As will be discussed in
\S\ref{secnum}, the finite resolution and the artificial viscosity employed
in the SPH code suppress small-scale turbulence that might disrupt the sharp
features and mix the gas with different entropies (although this is not
necessarily a problem, because viscosity in real clusters is not known and
may be comparable). It also injects entropy into the gas wherever there is
velocity shear, including at cold fronts. This entropy increase may make it
slightly more difficult for that gas to sink back to the center.  However,
the origin and the near-term evolution of the cold fronts should be
reproduced qualitatively correctly.

\begin{figure}
\centering
\plotone{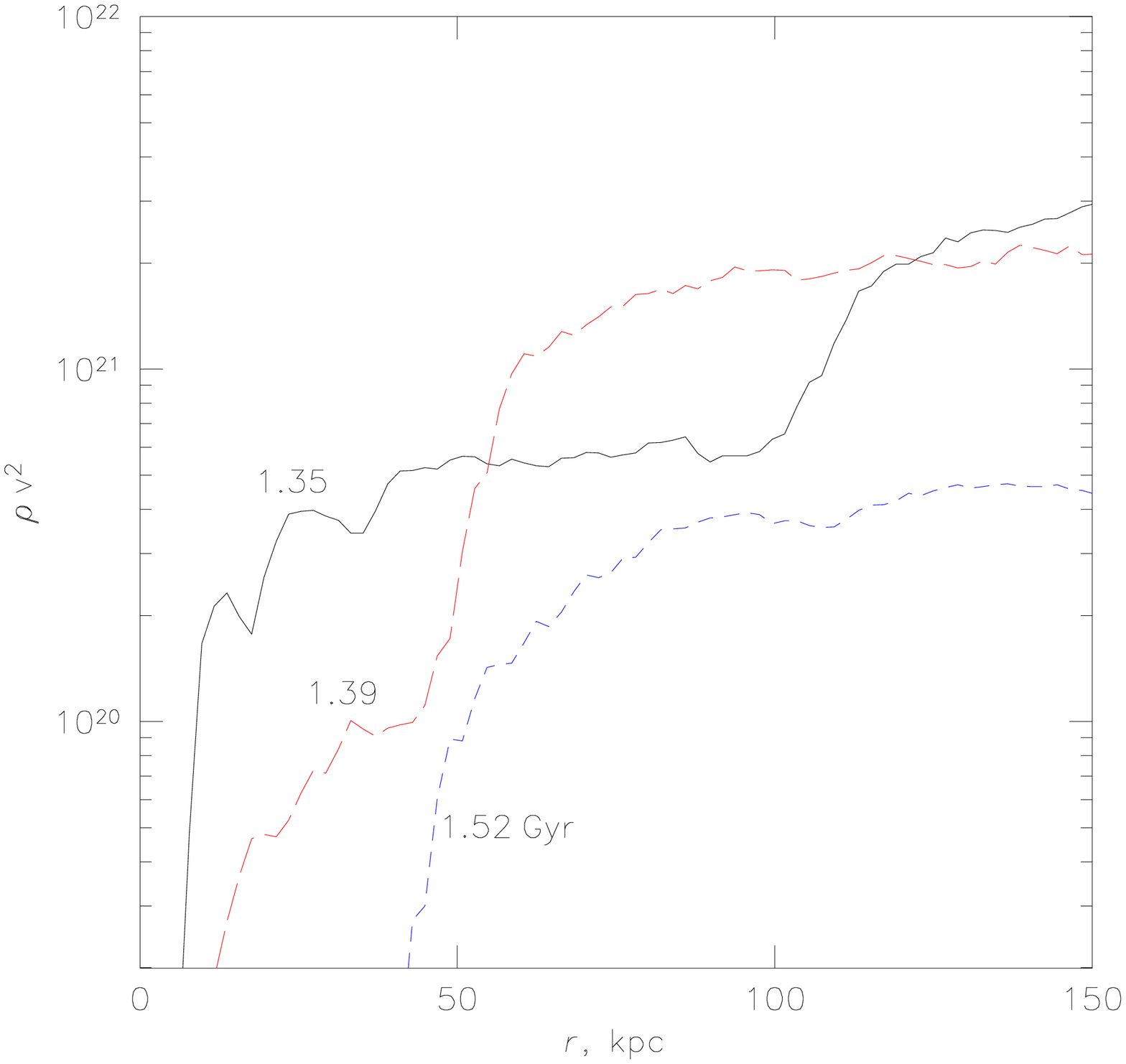}
\caption{Ram pressure exerted on the upper-right side of the
  main cool gas core during the subcluster fly-by, for the gas run shown in
  Fig.\ \ref{figGas}.  The quantity shown is the product of the gas density
  and the square of the radial component of the inflow velocity (relative to
  the DM peak) along the diagonal line from the DM peak to the upper-right.
  Units are arbitrary.  The distance is from the DM peak, the labels show
  moments of time. At $t=1.35$ Gyr, the pressure on the core is exerted by
  the shocked main cluster gas; the second bump around $r=100$ kpc is the
  approaching gas stripped from the subcluster (at 1.39 Gyr, it moves to
  $r=50$ kpc). After reaching a maximum around 1.35 Gyr, the ram pressure
  generally declines rapidly with time, resulting in a slingshot-like
  rebound of the main cluster gas.}
\label{figramp}
\end{figure}

\section{The effect of subcluster gas}
\label{secGas}

If the infalling subcluster has retained its gas, the picture is very
different from that described in \S\ref{secDM}. Figure \ref{figGas} shows a
simulated encounter with the same parameters as the one described in the
previous sections ($R=5$ and $b=500~$kpc), but now the infalling satellite
has its own gas component.  The most important difference is that the gas of
the subcluster now displaces the gas of the main cluster, creating a shock
front --- a region of high-density, hot gas moving with the same velocity
($M\approx 2$ at core passage in our fiducial setup), which should be
compared to a more subtle sonic wake that accompanied the DM-only
subcluster. The shock and the gas stripped from the subcluster are now the
dominant agents disturbing the gas at the main core, much more significant
than the slow orbital motion of the main DM peak. If the merger is head-on,
they can sweep the cool gas from the main cluster completely.  As long as
the merger is off-center and the central gas survives the shock passage, the
gas starts sloshing in a process qualitatively similar to that seen in the
DM-only subcluster run.

The 1.3--1.5 Gyr panels in Fig.\ \ref{figGas}, as well as the gas density
and velocity profiles through the center plotted in Fig.\ \ref{figJumpGas},
show again the effect of a ram-pressure slingshot. To illustrate it more
clearly, Fig.\ \ref{figramp} shows profiles of the quantity which
approximately represents ram pressure on the upper-right side of the main
core, at several interesting moments of time. At core passage ($t=1.35$
Gyr), the ram pressure on the cool density peak is at its maximum; it is
strong enough to compress and displace much of the main core gas to the
lower-left from the DM peak (this displacement can be seen in the 1.39 Gyr
panel in Fig.\ \ref{figGas}).  As soon as the subcluster moves away, ram
pressure drops very quickly (although not monotonically, because initially it
is exerted by the shocked gas and then by the stripped subcluster gas, as
can be understood from Fig.\ \ref{figramp}).  This causes the gas in the
main core to rebound and overshoot the DM peak.  Around $t=1.5-1.6$ Gyr,
there is an interesting additional effect --- the rebounding dense core
sticks out into the flow of the subcluster gas (see the 1.5 Gyr gas fraction
panel in Fig.\ \ref{figGas}), making the subcluster ICM flow around it.
This creates an ``airplane wing'' effect, additionally lifting this gas up.

The first sharp cold front (Fig.\ \ref{figJumpGas}) forms at $t\approx 1.5$
Gyr as a result of this rebound motion.  Compared to the DM-only subcluster
run, the cold front forms earlier and on the opposite side from the core ---
because the initial gas-DM displacement is now caused by ram pressure from
the subcluster, not by the change in the direction of motion of the main DM
core as it was in \S\ref{secdecoup}. This front separates the main cluster
and the subcluster ICM.  The tangential flow of the subcluster gas along its
surface creates an eddy seen in the 1.67--1.9 Gyr panels of Fig.\ 
\ref{figGas}.  Note that not just the cool gas from the core, but the bulk
of the main cluster gas, initially pushed to the south by the subcluster
flyby, rebounds and takes part in the upward flow.  However, the densest,
coolest gas quickly develops a RT-like tongue (the 1.67 Gyr panel in Fig.\ 
\ref{figGas}) and starts flowing back into the potential minimum (such cool
filaments behind cold fronts were also seen in the simulations by Mathis et
al.\ 2005). From this moment on, the picture is qualitatively similar to our
DM-only subcluster run (\S\ref{secDM}). This cool gas reaches the potential
minimum and starts sloshing in it, generating mushroom-head cold fronts and
RT tongues on smaller and smaller scales, similar to Fig.\ \ref{figDMzoom}.
Inside each of these mushroom heads, as well as inside the first
``slingshot'' cold front, the lowest-entropy gas is constantly flowing
toward the DM peak along the RT-like tongues and is replaced at the cold
front with the newly arrived, higher-entropy gas, as described in
\S\ref{secspiral}. This allows the cold fronts to expand outward.  Note that
the ``slingshot'' cold front separates gases from the two merging
subclusters, while the ``sloshing'' cold fronts in the center arise entirely
within the ICM of the main cluster, as in \S\ref{secDM}.

The ``slingshot'' cold front moves beyond $r=500$ kpc approximately 1 Gyr
after the core passage (where it still exhibits a factor of $\sim2$ density
jump).  Compared to the DM-only subcluster run, the initial disturbance was
much stronger and the sloshing cold fronts evolve much faster.  In addition,
the velocity field is more chaotic due to the eventual infall of the gas
stripped from the subcluster, which destroys most of the central pattern of
``sloshing'' cold fronts in about 2 Gyr (the 3.3 Gyr panels in Fig.\ 
\ref{figGas}).

\subsection{Stripping of the subcluster gas}
\label{secsubcl}

The subcluster gas in the simulation shown in Fig.\ \ref{figGas}
illustrates a classic ram pressure stripping scenario. As the subcluster
falls in for the first time, its outer gas is stripped and forms a
comet-shaped tail (1.34--1.5 Gyr gas fraction panels in Fig.\ \ref{figGas}).
At $t=1.5$ Gyr, the cool dense gas in the main cluster core presents an
obstacle to the smooth supersonic flow of this stripped gas.  This gives
rise to a second bow shock clearly seen in the 1.5 Gyr panel ``in front of''
the main core, making the picture symmetric.

The front tip of the gas still remaining in the subcluster core is a sharp
cold front caused by ram pressure stripping early in the merger.  At core
passage, where the infall velocity and the ambient gas density are at their
maximum, ram pressure on this remaining gas is so high that the gas is
pushed back from the subcluster's DM peak.  At $t\simeq 1.4$ Gyr, our
subcluster looks much like the ``bullet'' in 1E\,0657--56, which exhibits a
shock front, a cold front, and cool gas lagging the DM peak (Markevitch et
al.\ 2002).

As the subcluster moves out, the ram pressure rapidly diminishes, and we
again see the effect of the ram-pressure/gravity slingshot, which at $t=1.5$
Gyr returns the cool gas to the DM peak and later makes it overtake the DM
peak, as seen, e.g., in the A168 cluster (Hallman \& Markevitch 2004) and in
the Mathis et al.\ (2005) simulations.

Around $t=3.4$ Gyr, the subcluster falls in for the second time.  By this
time it has lost a large fraction of its DM mass and is not able to retain
any of its gas after the second core passage. In the 3.3 Gyr panel of Fig.\ 
\ref{figGas}, we see how the cool gas peak is completely detached from the
DM peak as soon as they enter the dense region of the main cluster.
Eventually, this most persistent parcel of cool gas falls to the center of
the main cluster, where it starts sloshing in the potential minimum (4.8 Gyr
panels in Figs.\ \ref{figGas} and \ref{figJumpGas}). By that time, the main
cluster's own cool central gas has escaped from the core.  At this stage,
the cluster looks relatively relaxed on a large scale (\S\ref{secObs}); the
most notable feature is this sloshing cool gas in the center which
originally belonged to the infalling subcluster. This curious replacement of
the gas in the center does not occur for all combinations of the merger mass
ratios and impact parameters.  We should also note that artificial viscosity
and the finite resolution of our simulations inhibit small-scale turbulence,
which may act to mix the ICM from the two subclusters earlier.

\section{Exploring the parameter space}
\label{secexpl}

\begin{figure*}
\centering
\includegraphics[width=5.4cm]{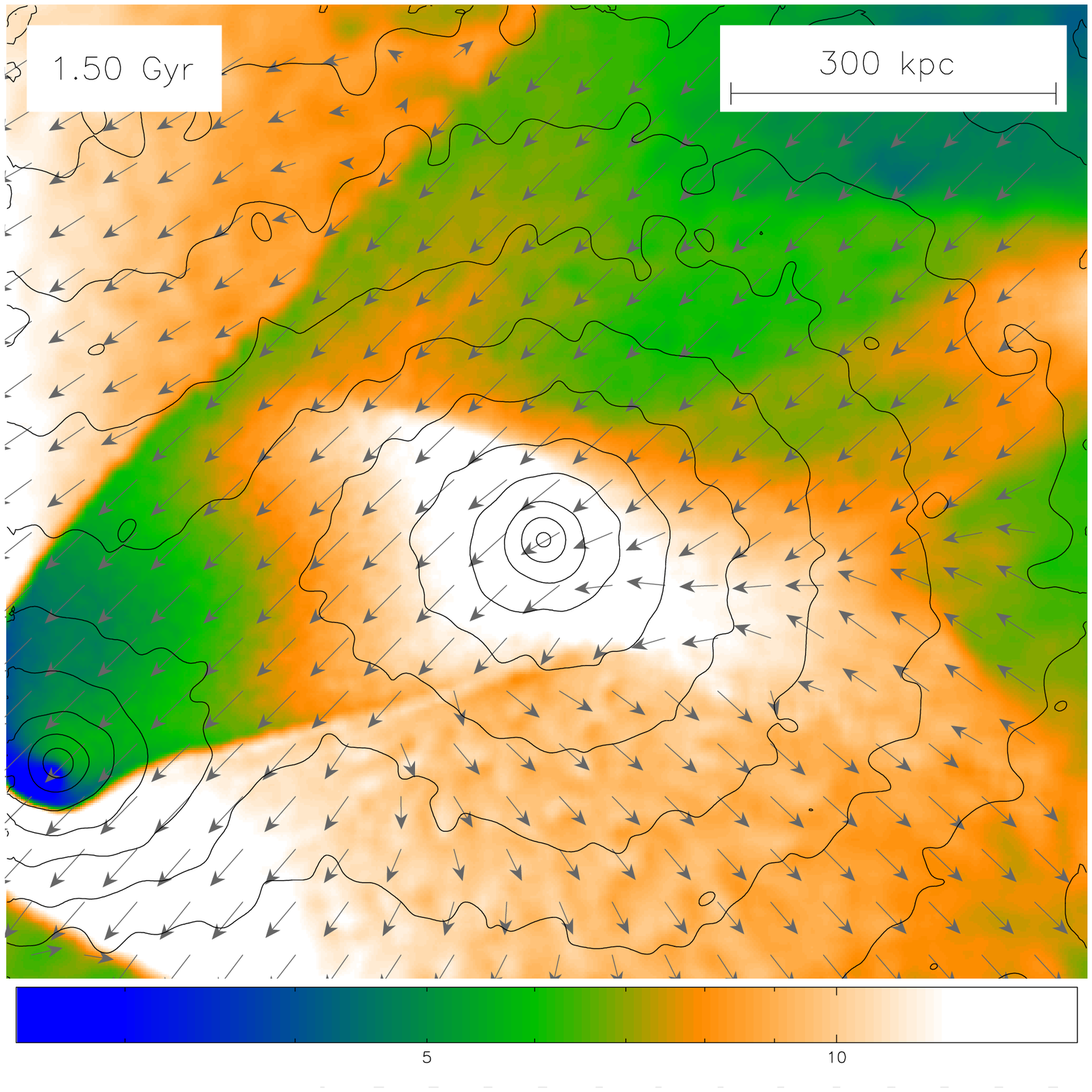}
\includegraphics[width=5.4cm]{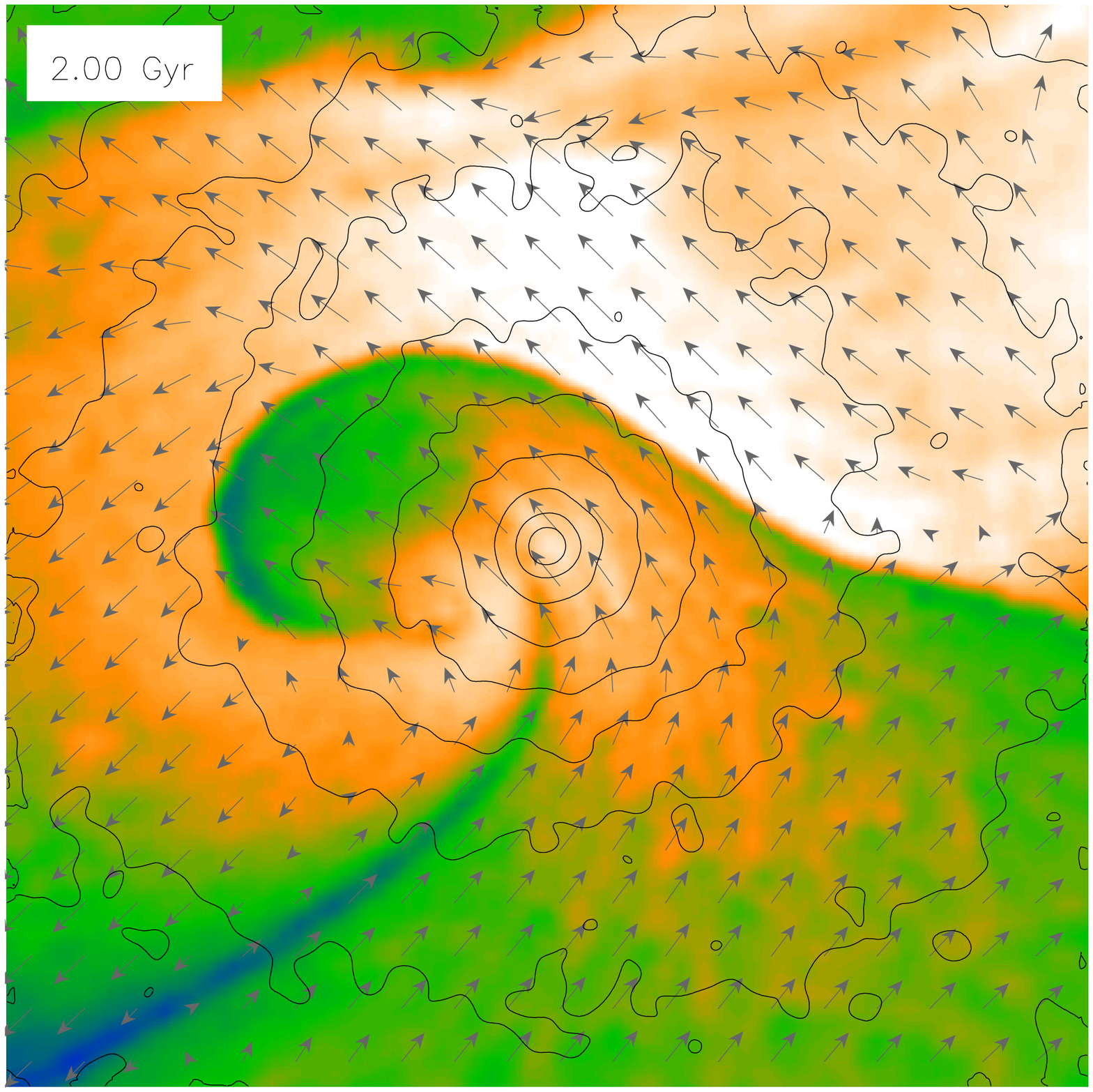}
\includegraphics[width=5.4cm]{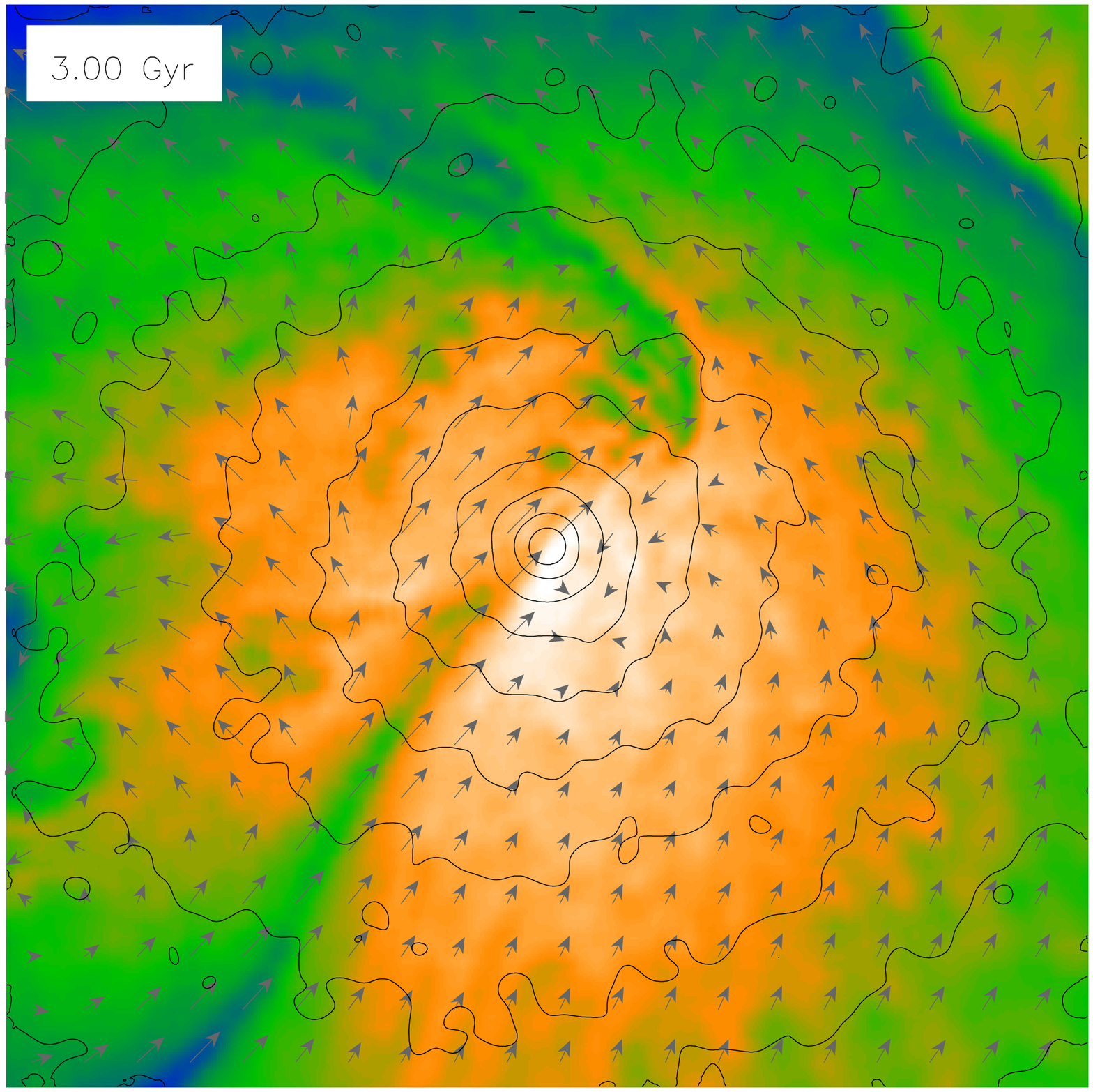}\\
\includegraphics[width=5.4cm]{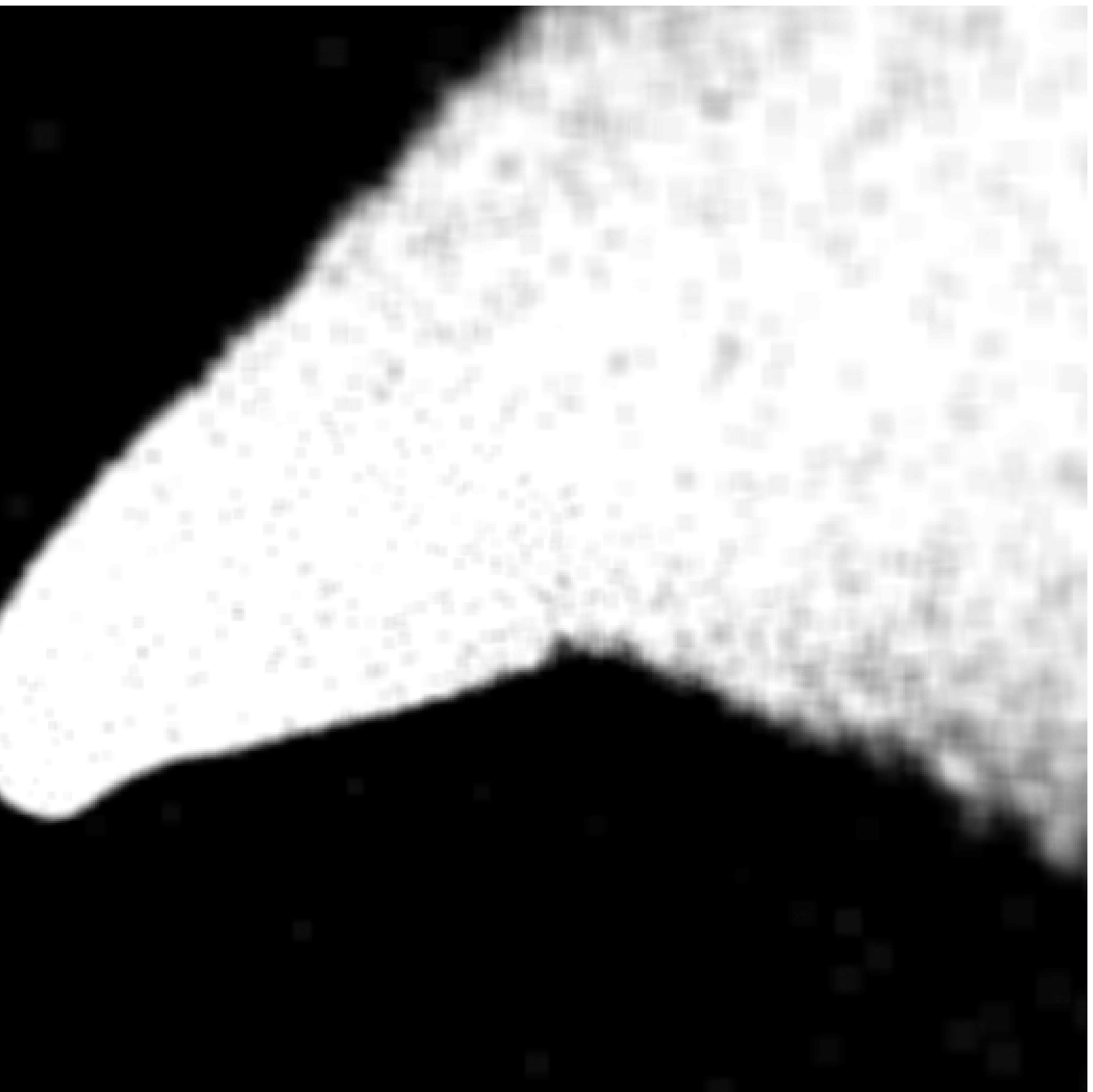}
\includegraphics[width=5.4cm]{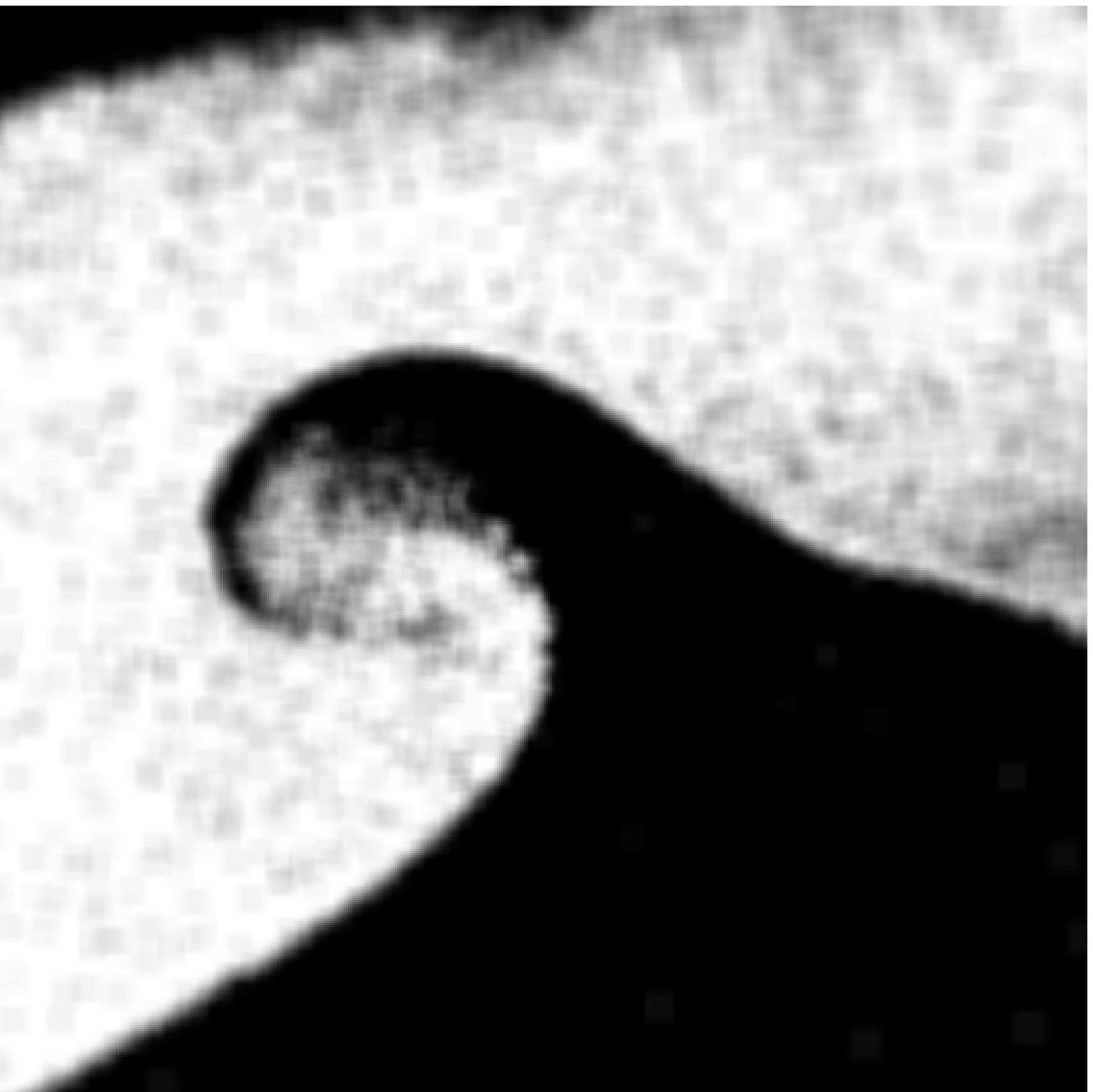}
\includegraphics[width=5.4cm]{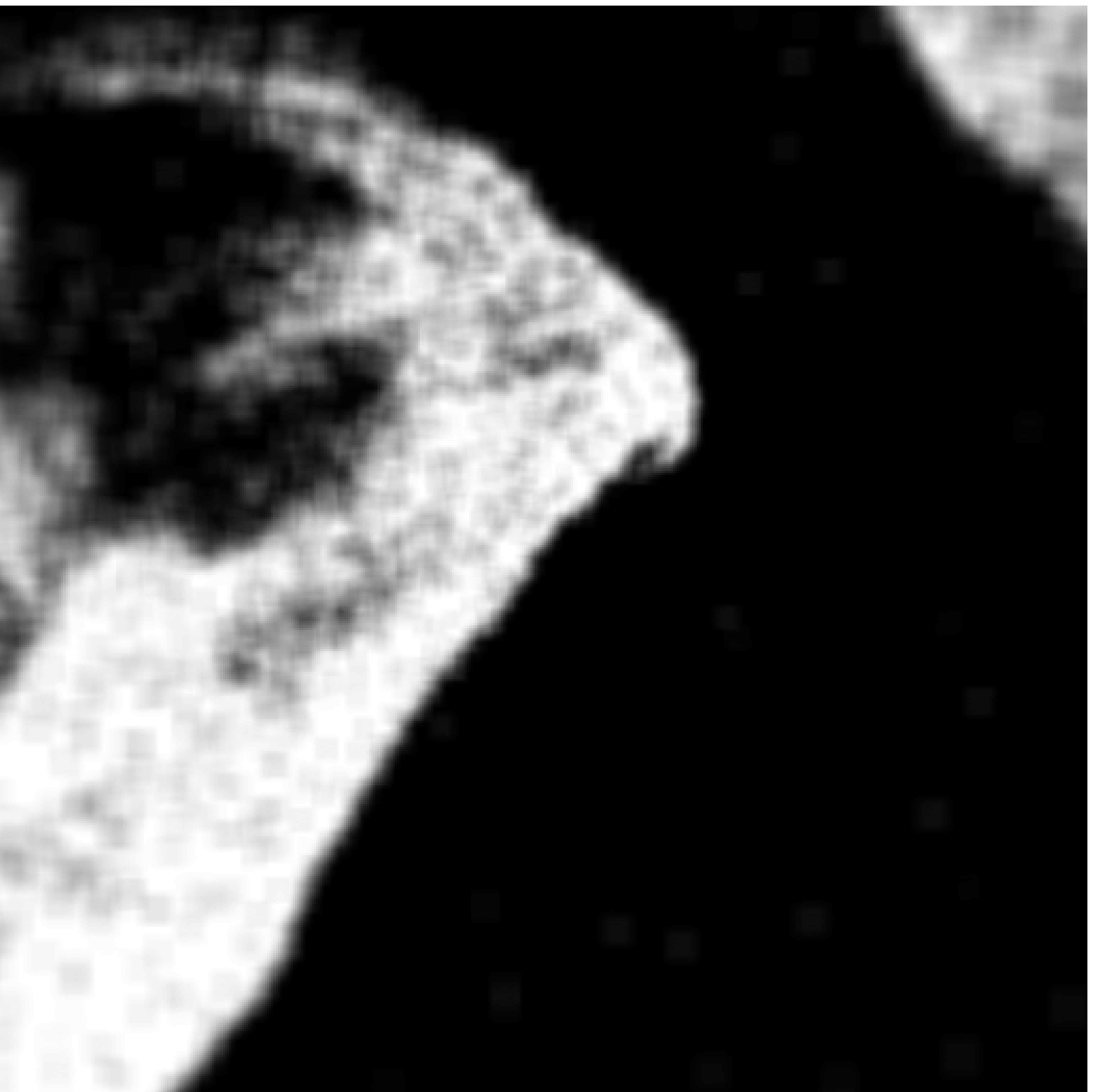}
\caption{
  Encounter with $R=5$, $b=500~$kpc (as in Fig.\ \ref{figGas}), but
  the main cluster has a flat gas density profile without a central
  temperature drop (see Fig.\ \ref{figIC}). The satellite, however, still
  possesses a cool gas core.  A plot under each image shows the fraction of
  particles belonging to the two subclusters. Even though there is no cool
  gas in the main cluster, its entropy still declines toward the center, and
  the low-entropy gas still creates a cold front (middle panel), which
  quickly moves outwards.}
\label{figCore}
\end{figure*}

\begin{figure*}
\centering
\includegraphics[width=5.4cm]{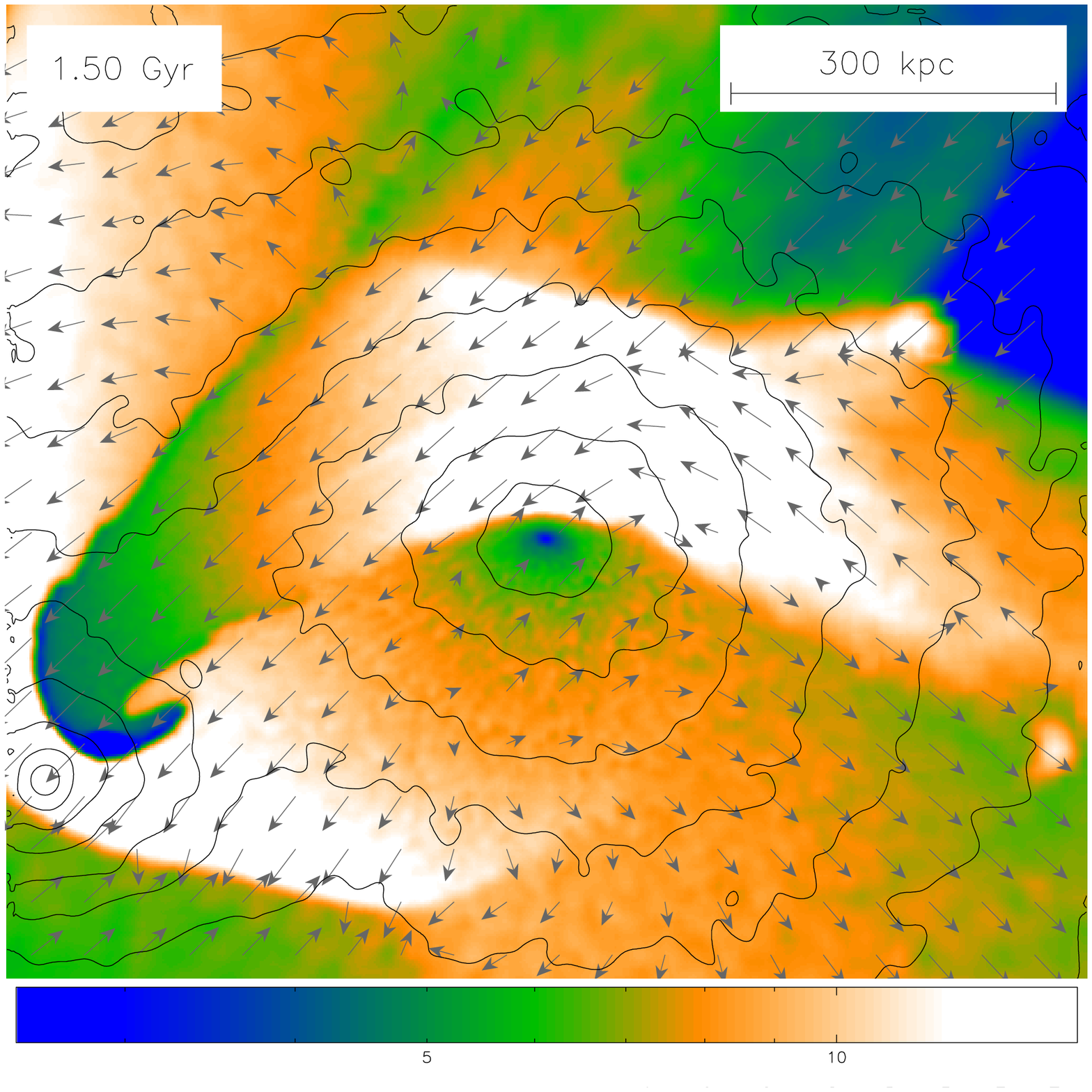}
\includegraphics[width=5.4cm]{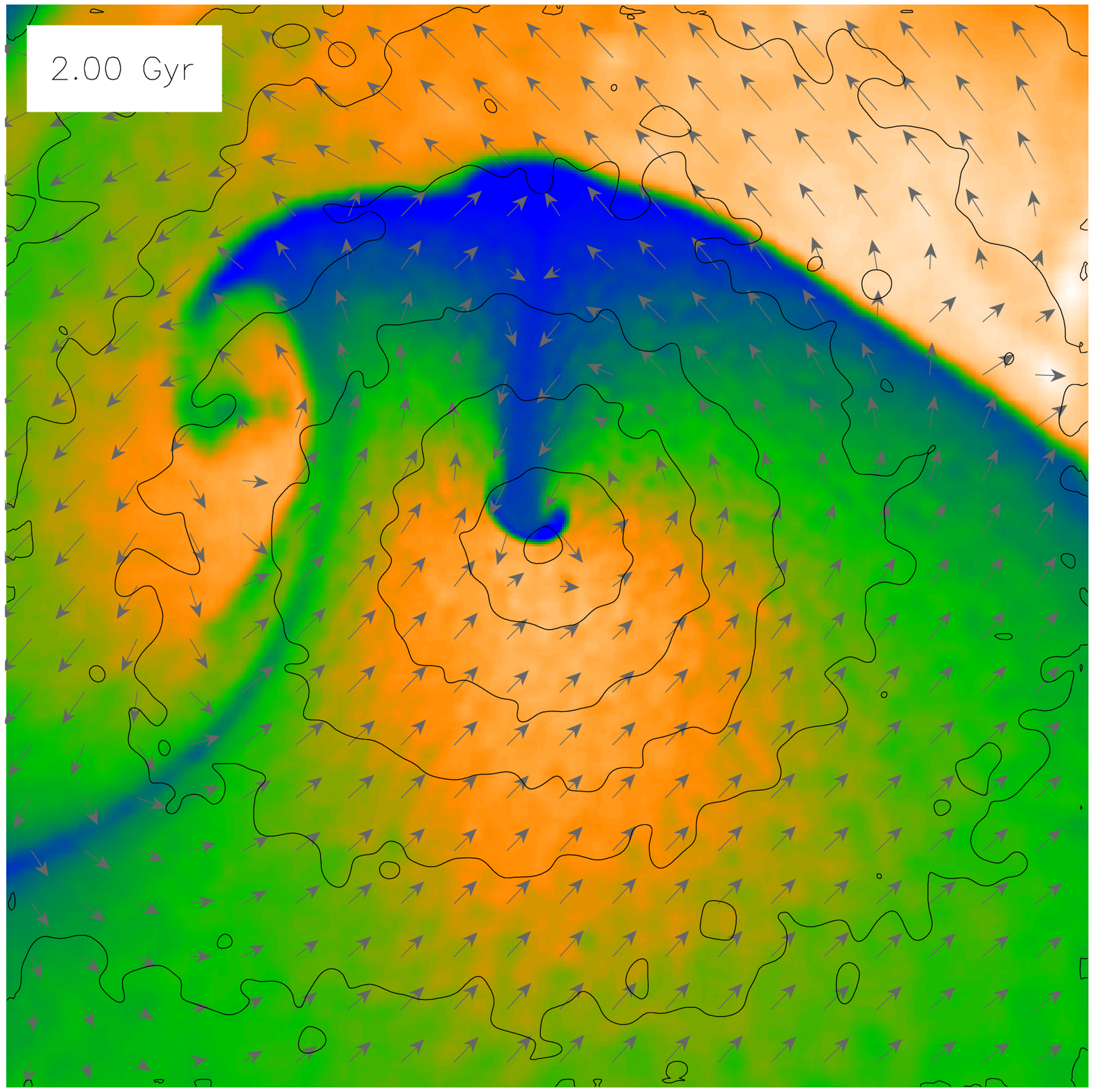}
\includegraphics[width=5.4cm]{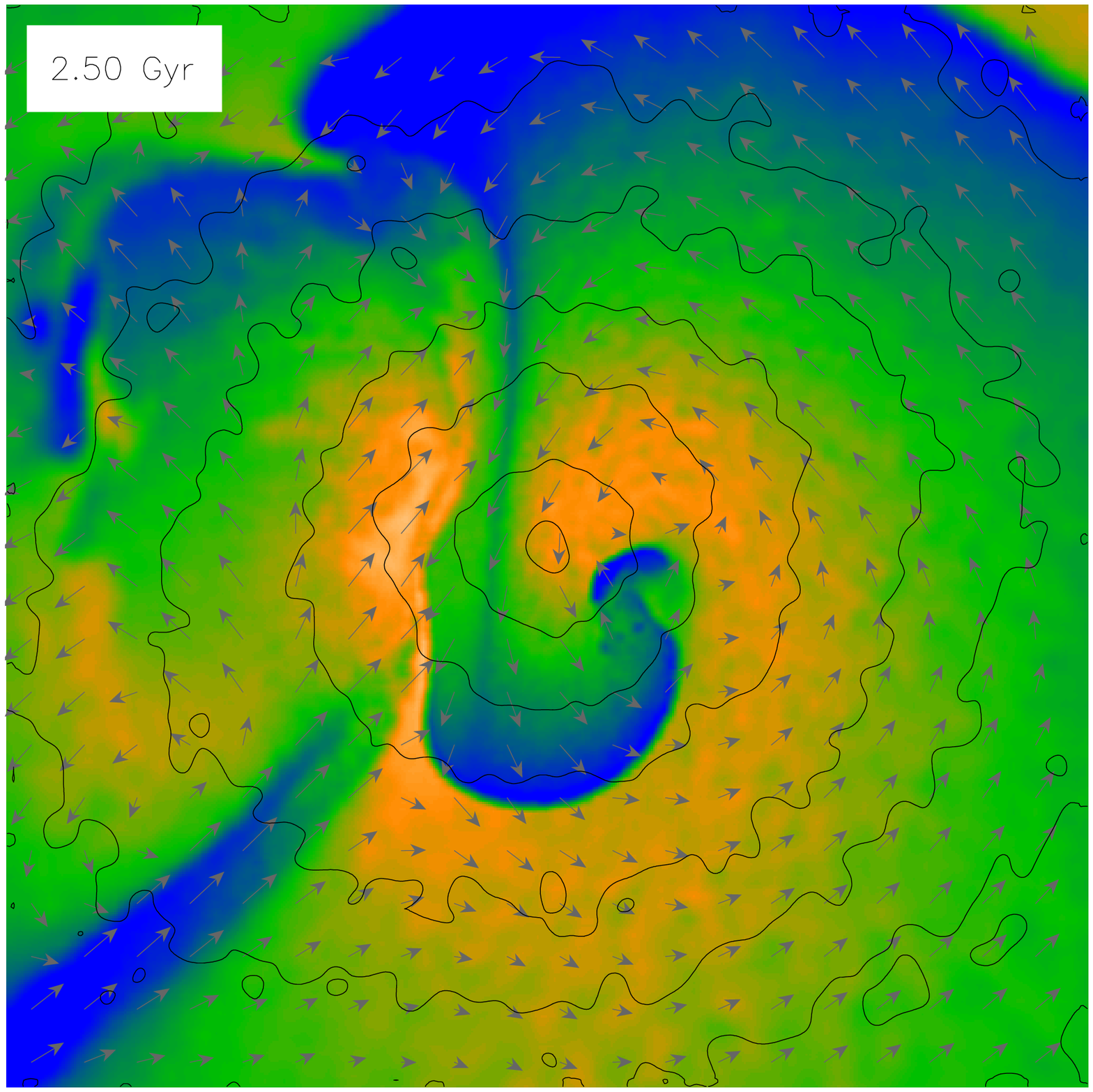}
\caption{
  Encounter with $R=5$, $b=500~$kpc (as in Fig.\ \ref{figGas}), but
  now the DM density profile of the main cluster has a flat central core
  instead of a cusp (eq.\ \ref{eqFlat}), while the gas in both subclusters
  has cool peaks. The behavior is qualitatively similar to the run with a
  cuspy DM profile, but the amplitude of sloshing is much larger.}
\label{figFlat}
\end{figure*}

\begin{figure*}
\centering
\includegraphics[width=5.4cm]{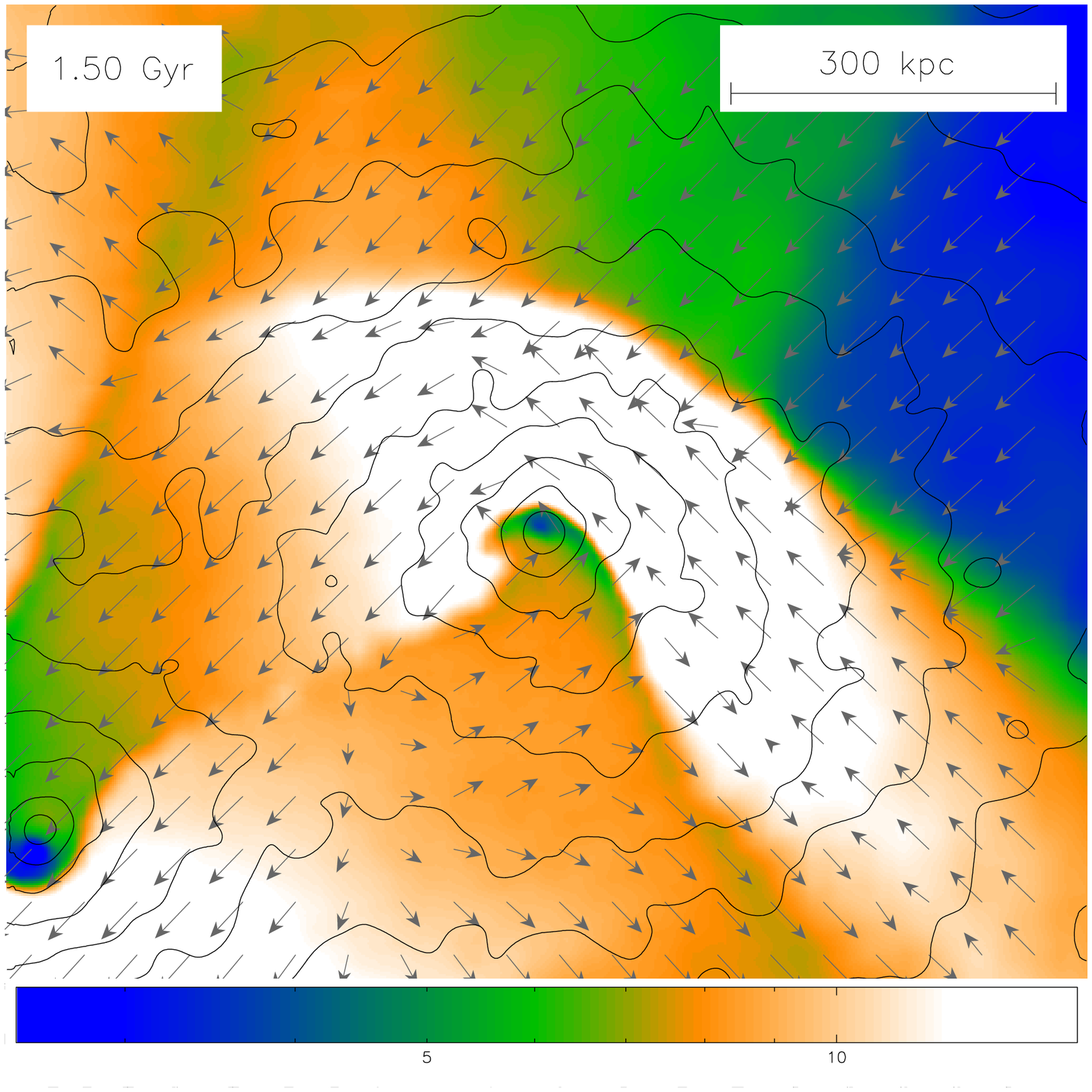}
\includegraphics[width=5.4cm]{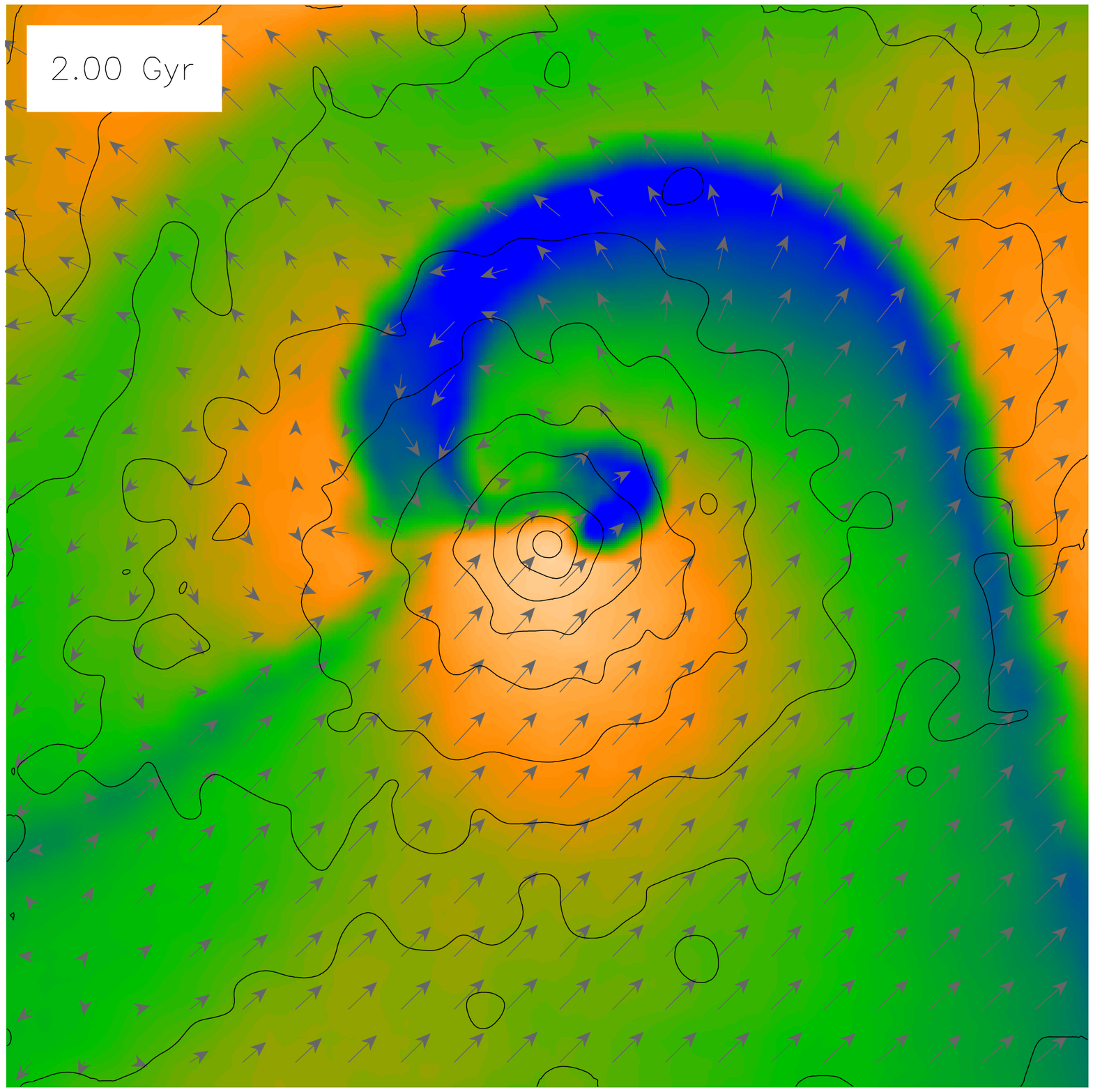}
\includegraphics[width=5.4cm]{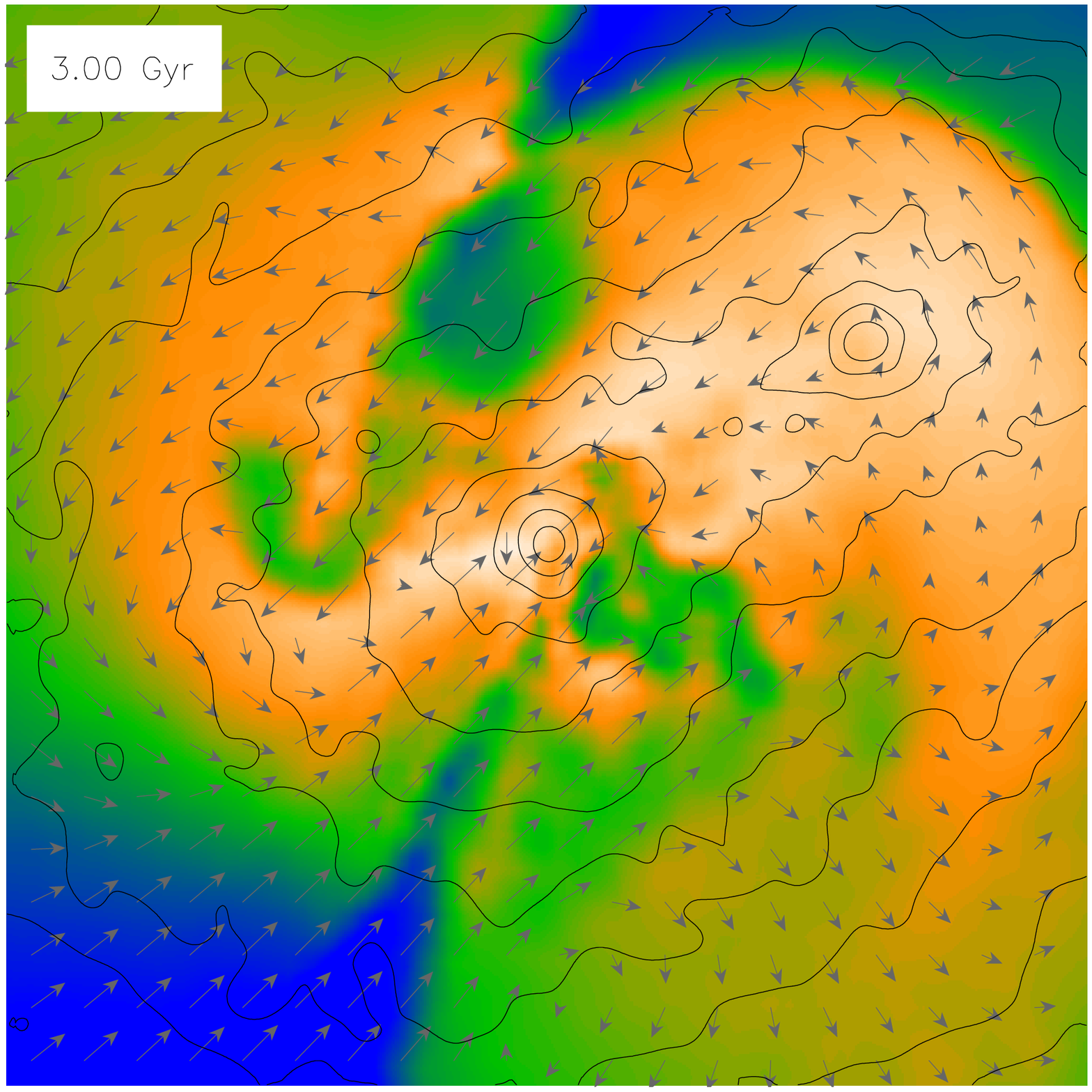}\\[4mm]
\includegraphics[width=5.4cm]{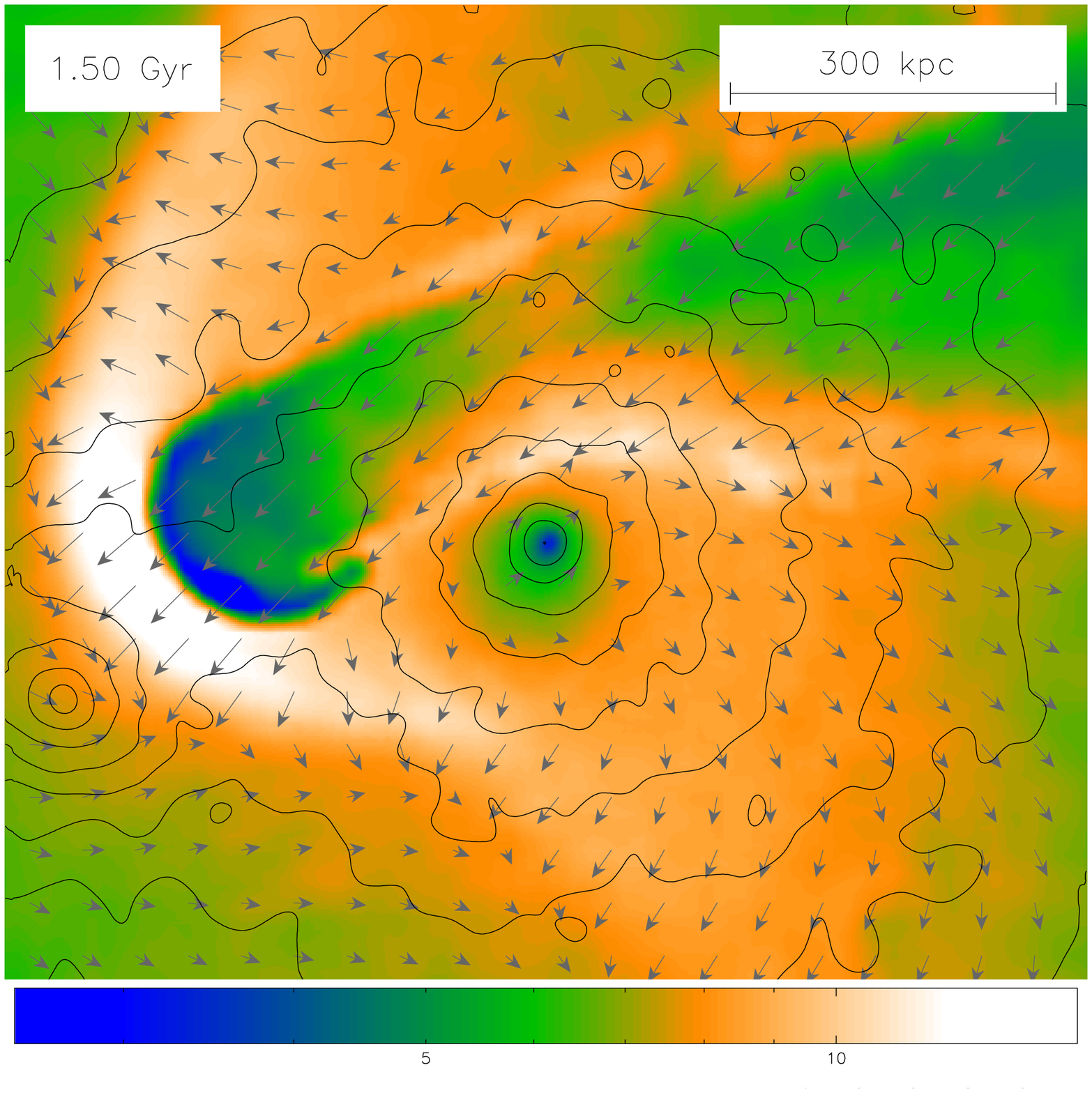}
\includegraphics[width=5.4cm]{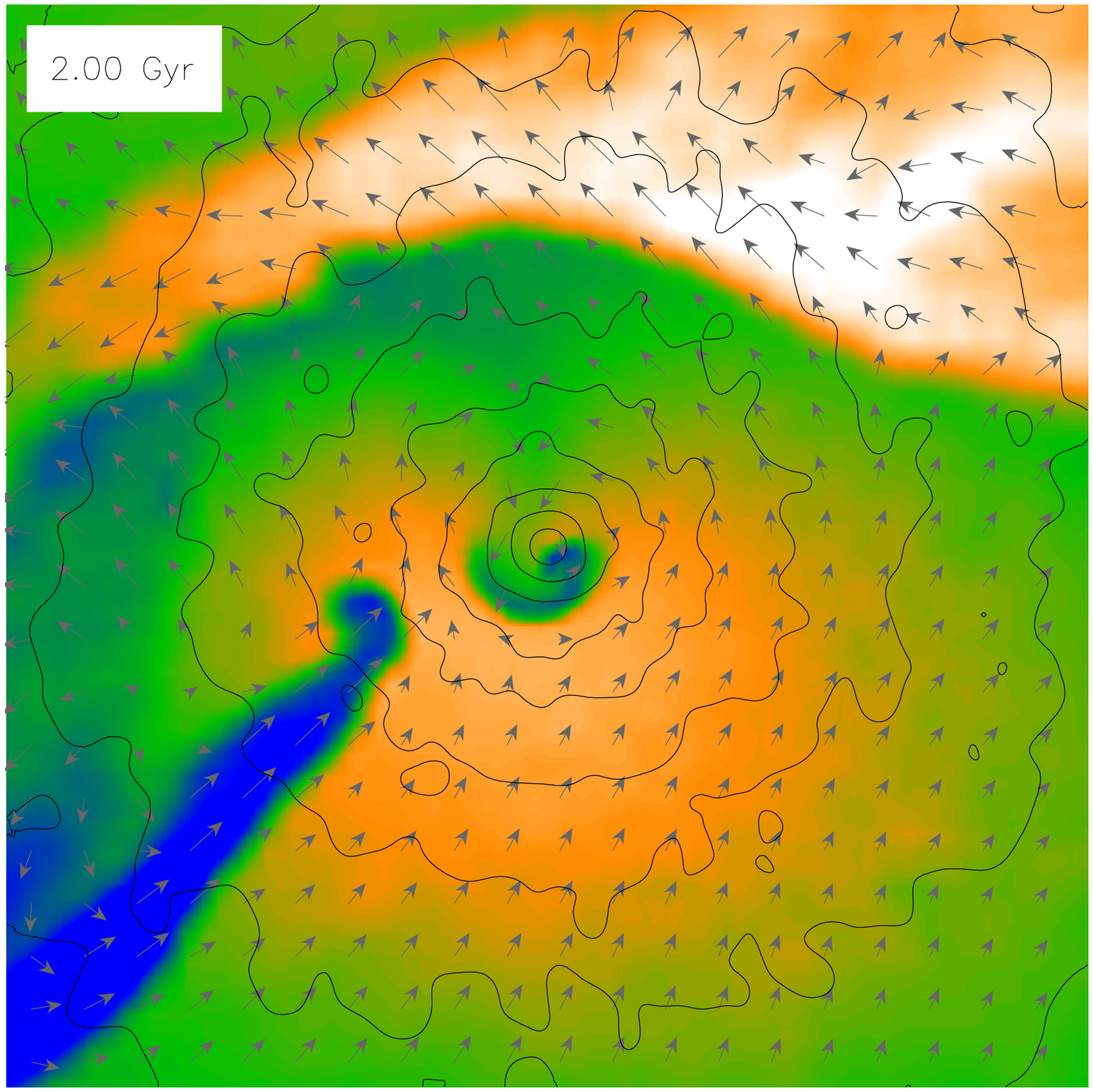}
\includegraphics[width=5.4cm]{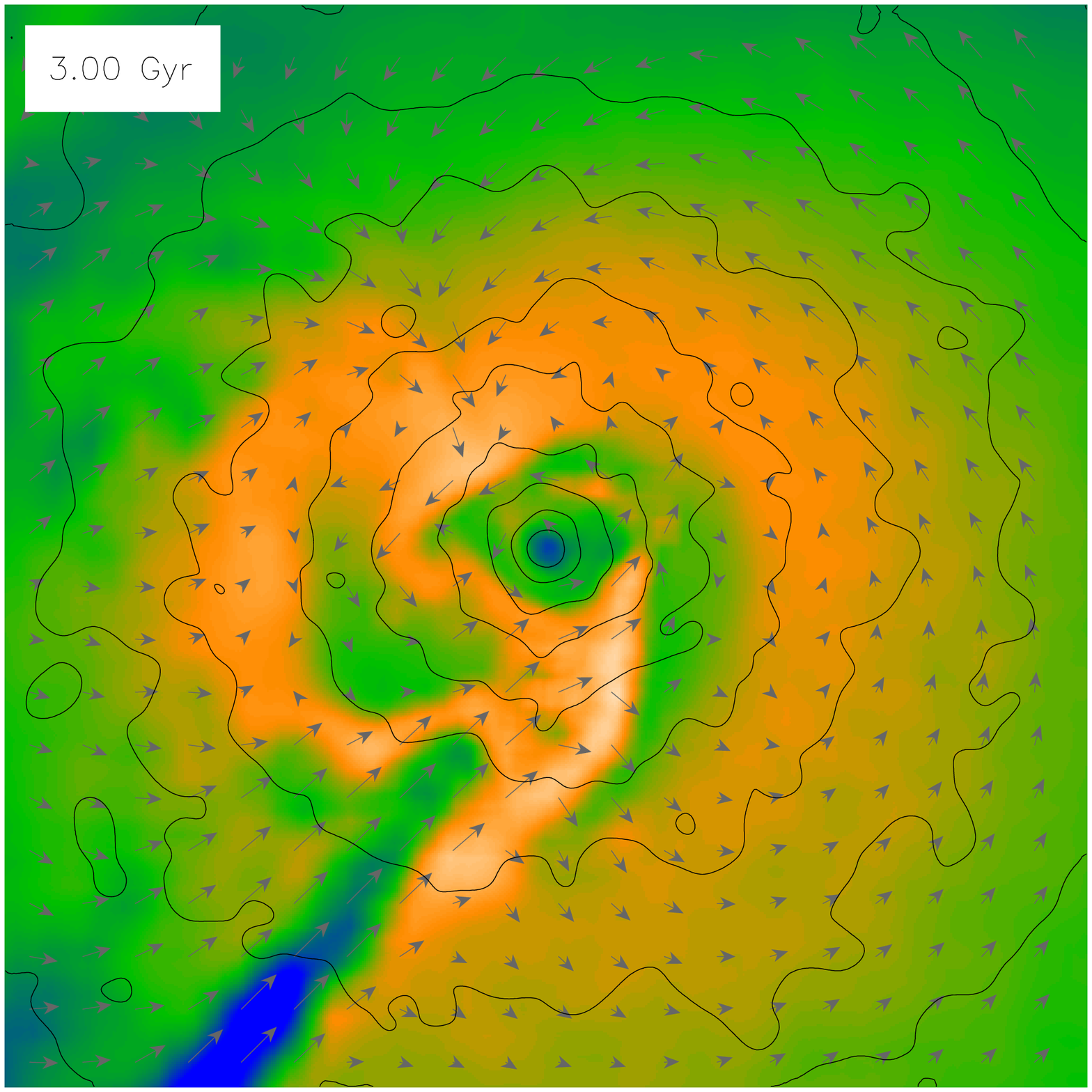}\\[4mm]
\includegraphics[width=5.4cm]{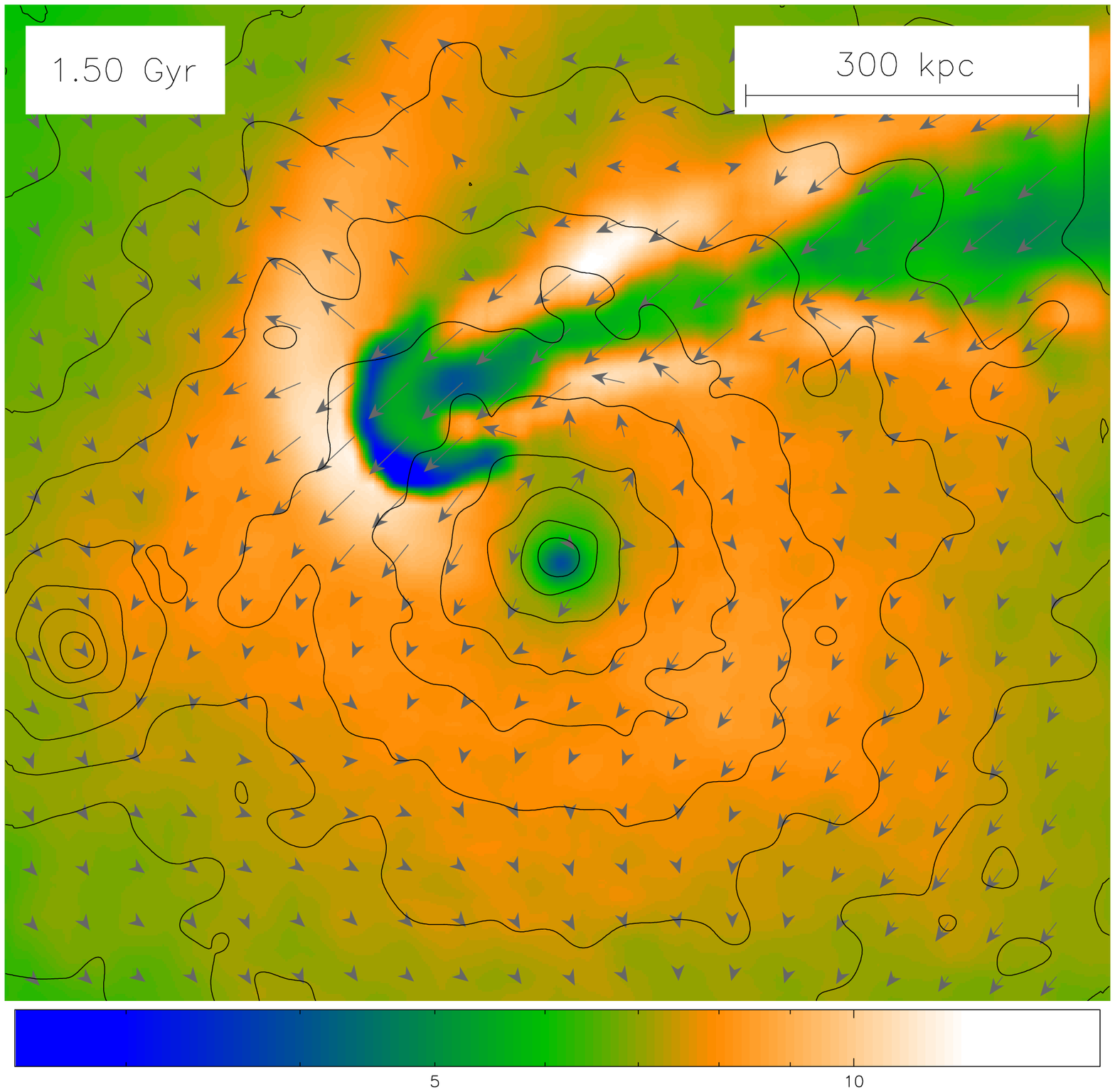}
\includegraphics[width=5.4cm]{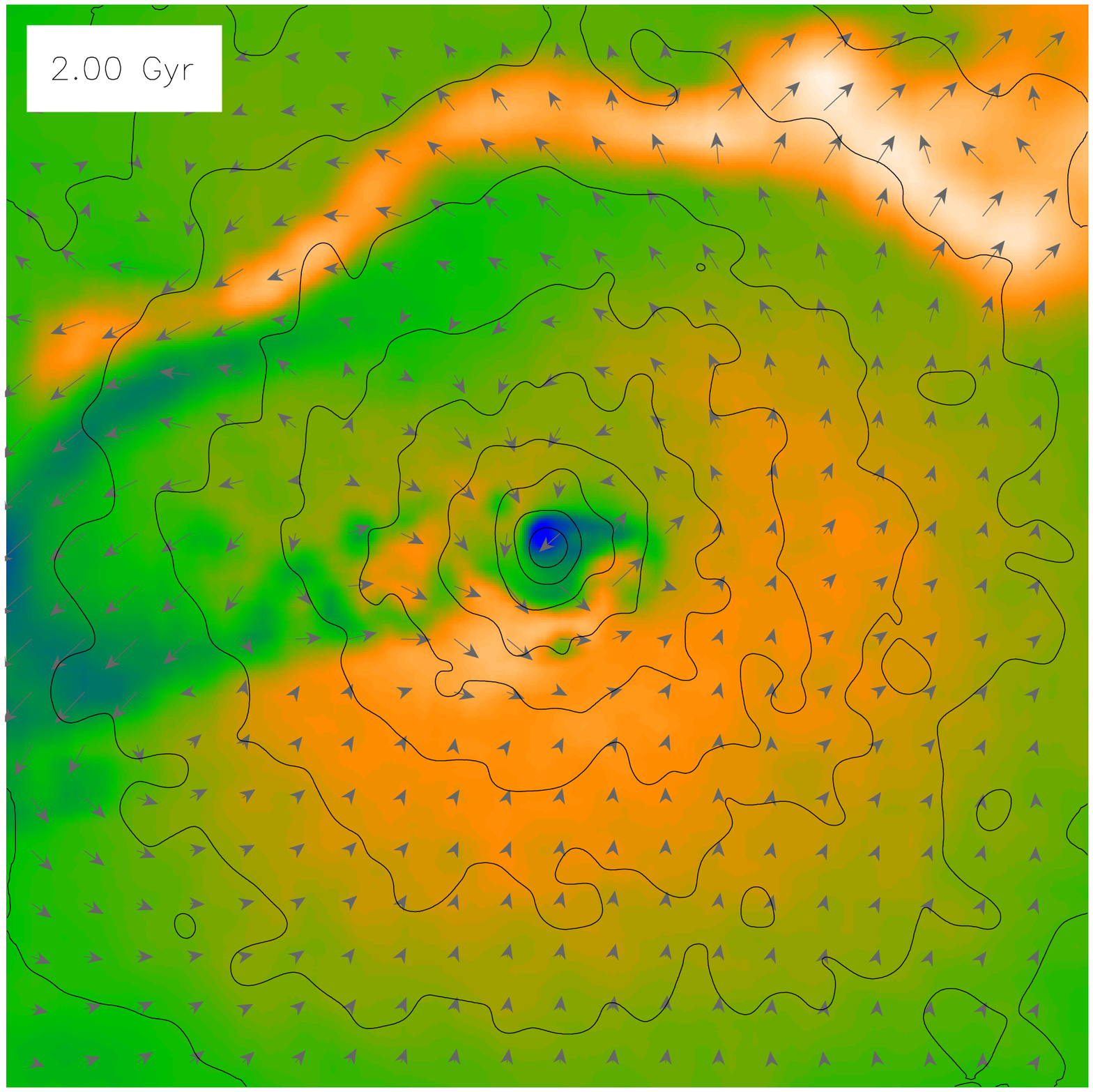}
\includegraphics[width=5.4cm]{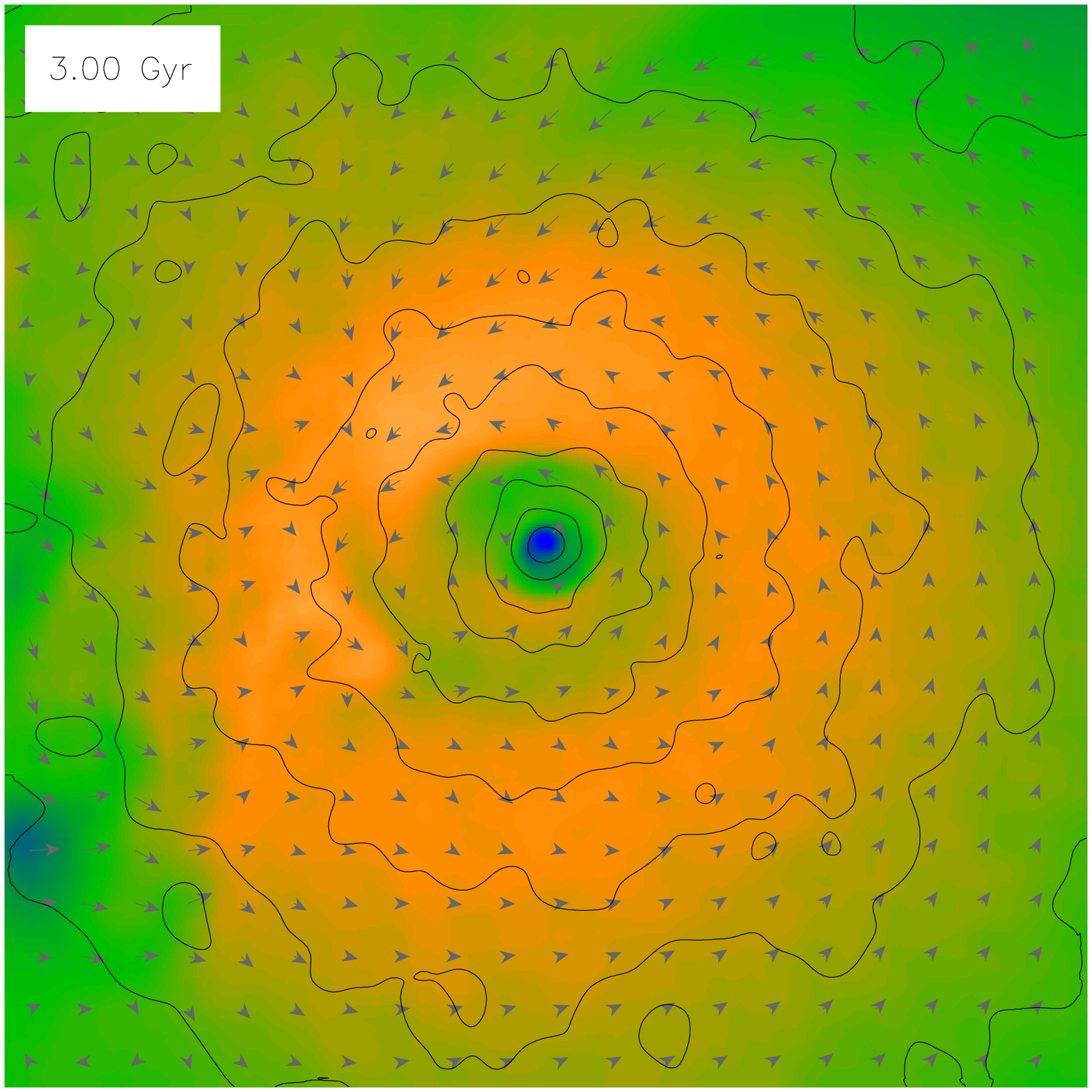}
\caption{
  Encounters with different mass ratios: $R=2$ (top), 20 (middle) and 100
  (bottom); $b=500~$kpc in all cases.  Both subclusters have central DM
  cusps and cool gas peaks.  These simulations have been run at a lower
  resolution, with $2\times10^6$ (gas + DM) particles.}
\label{figR}
\end{figure*}

\begin{figure*}
\centering
\includegraphics[width=5.4cm]{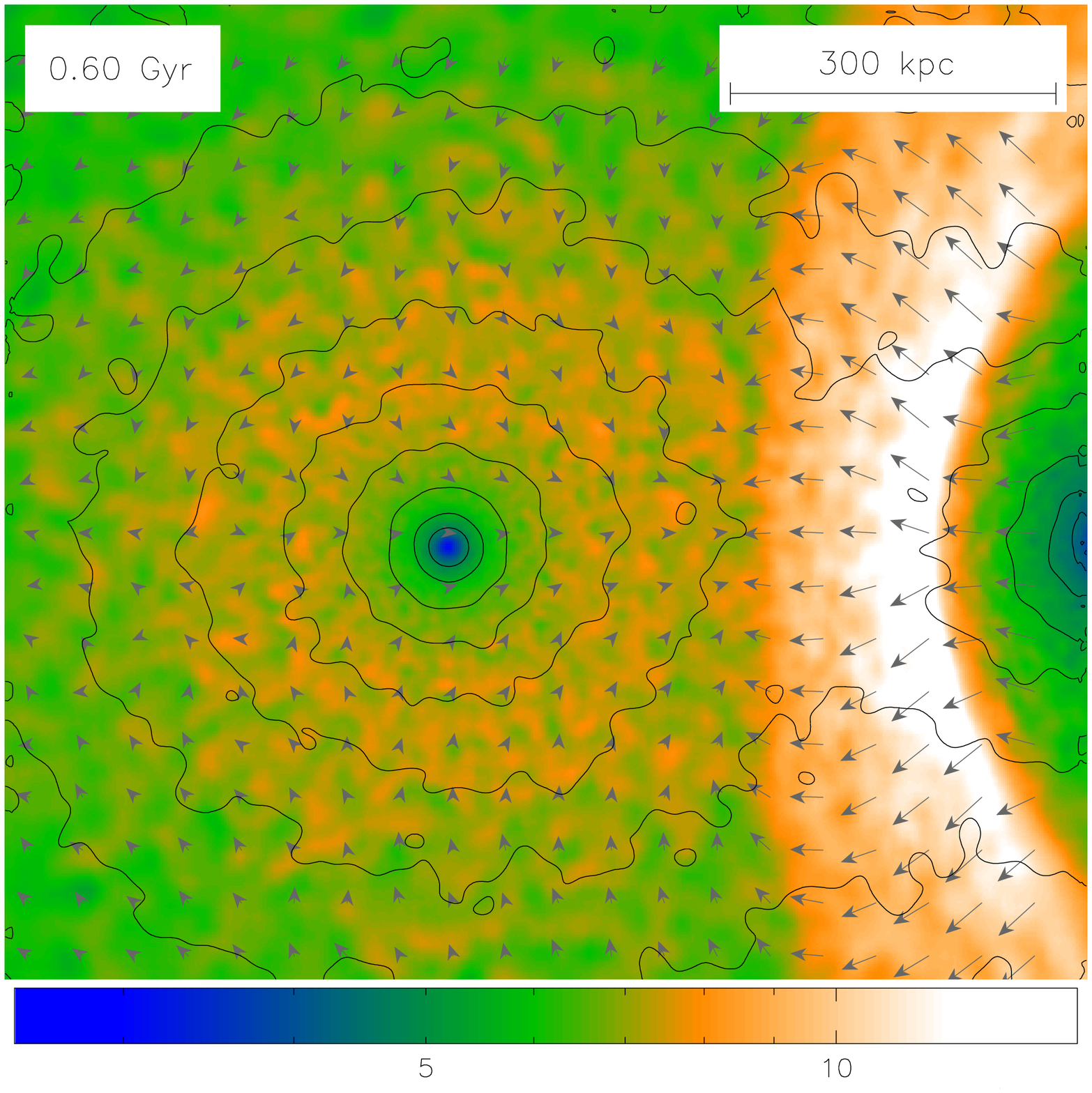}
\includegraphics[width=5.4cm]{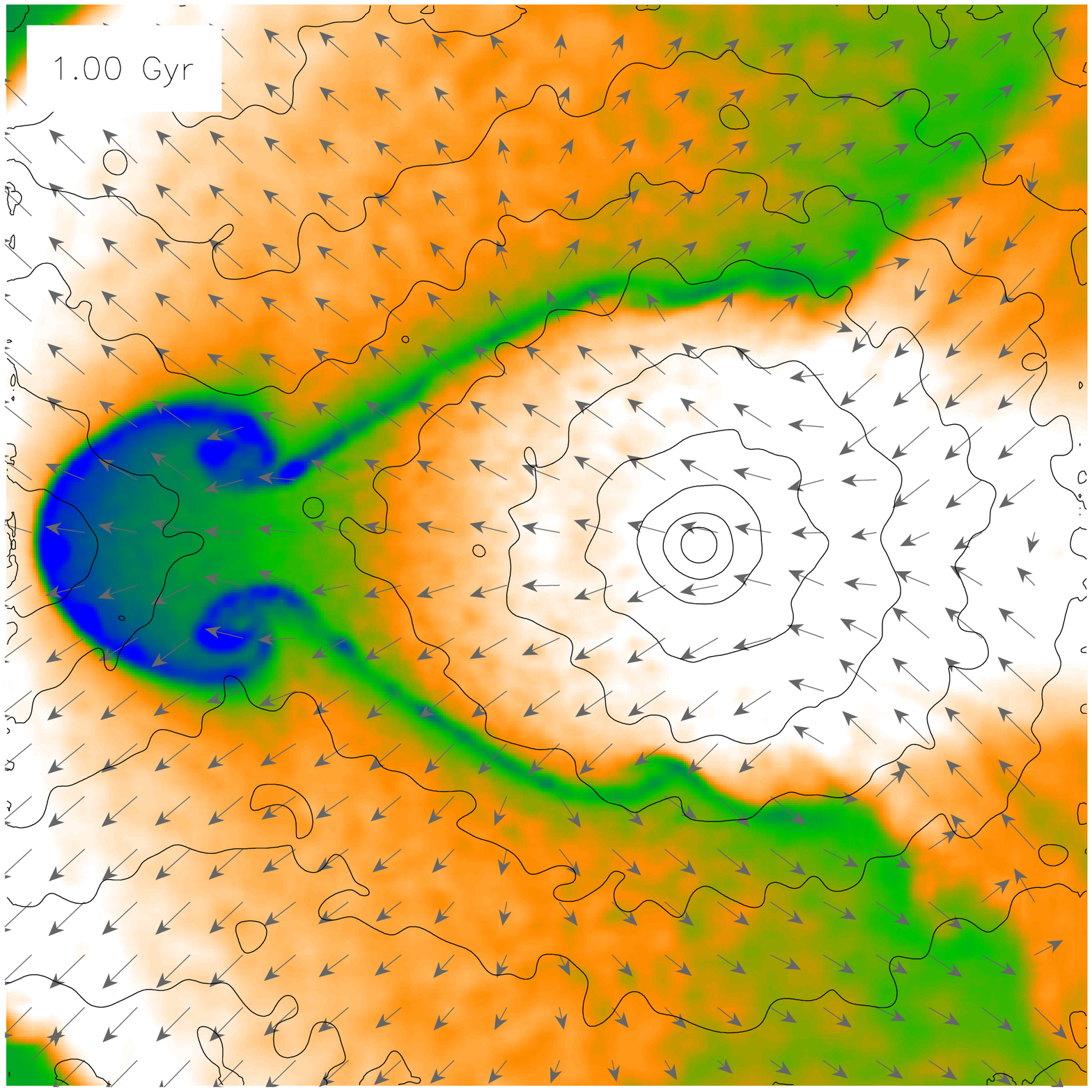}
\includegraphics[width=5.4cm]{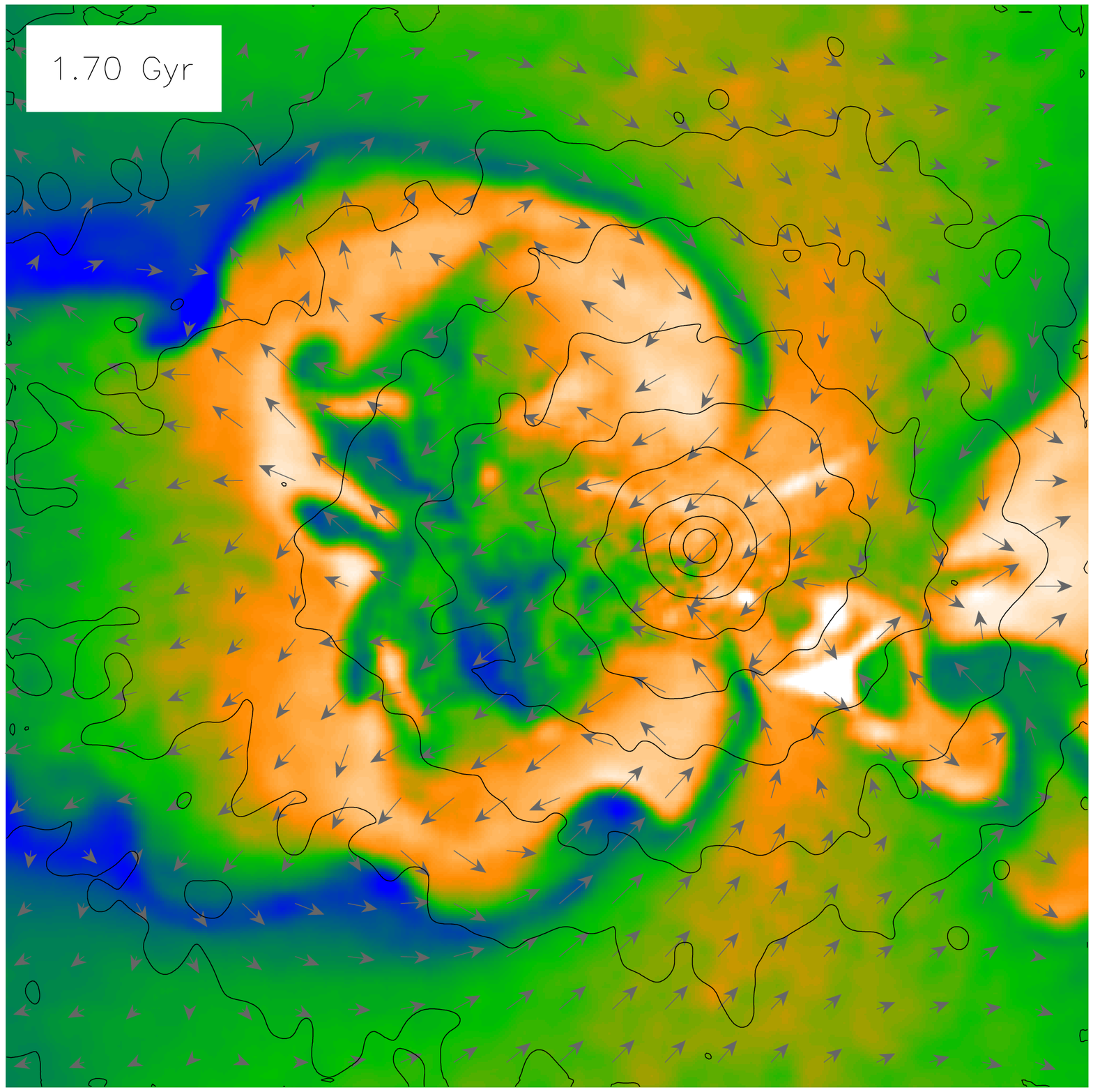}\\[4mm]
\includegraphics[width=5.4cm]{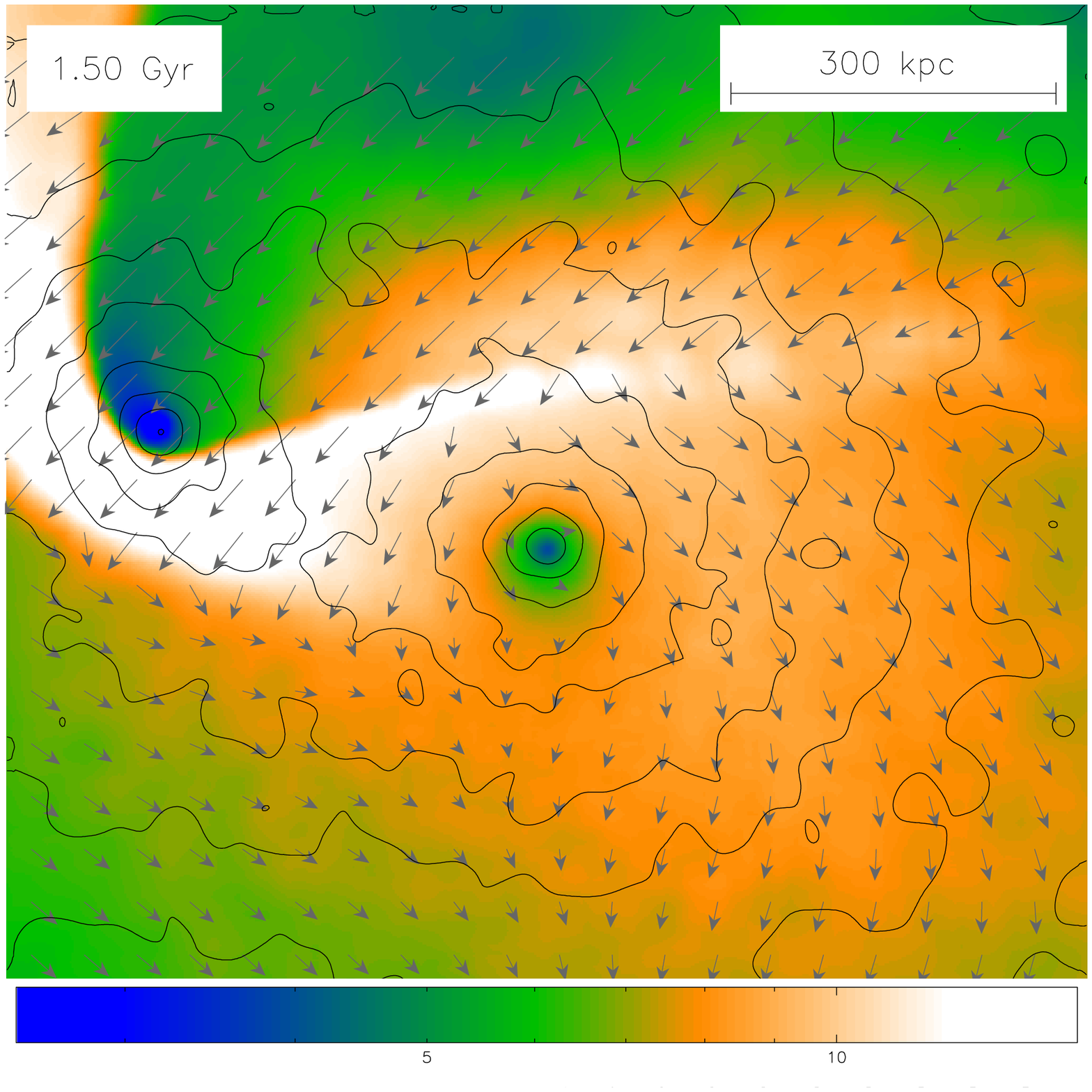}
\includegraphics[width=5.4cm]{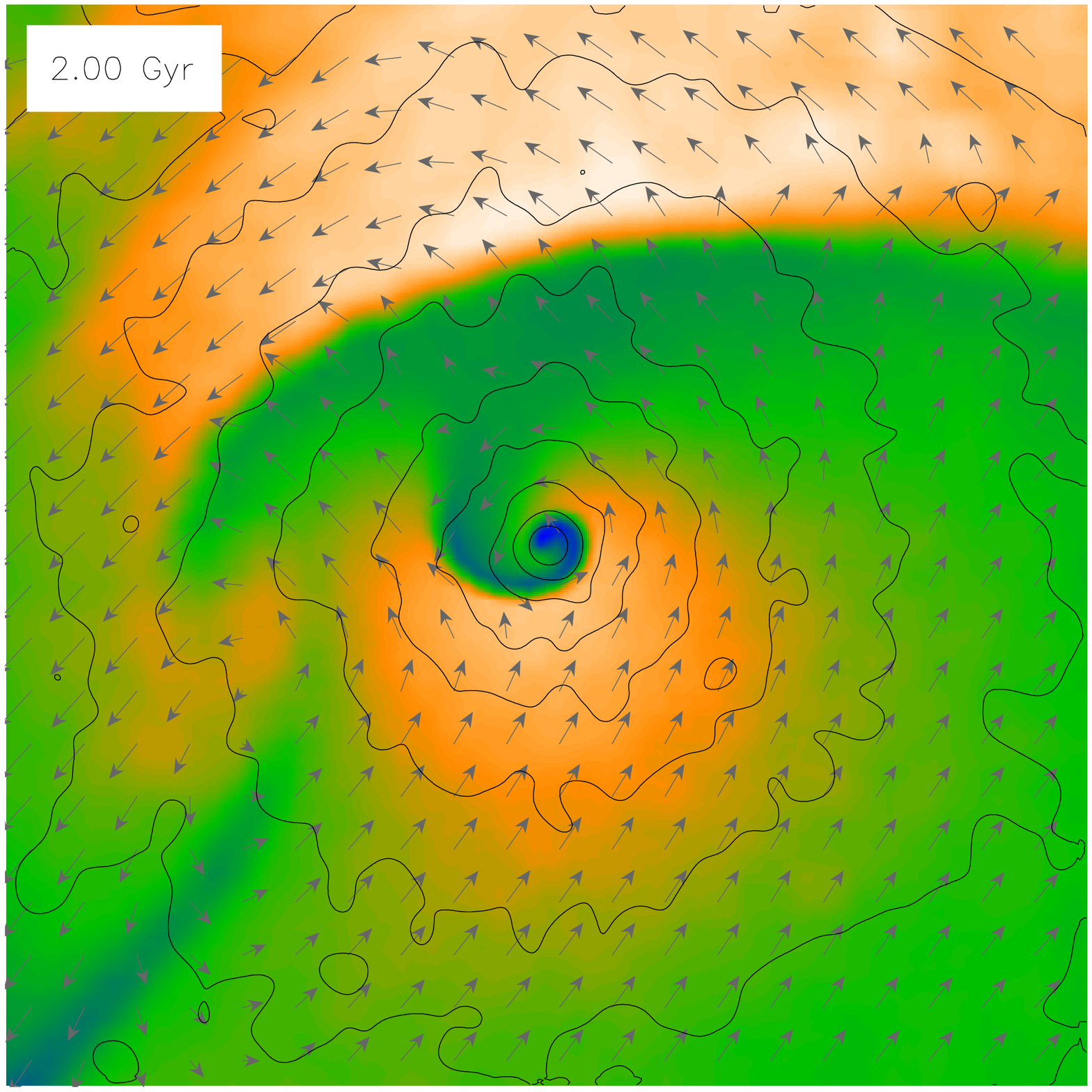}
\includegraphics[width=5.4cm]{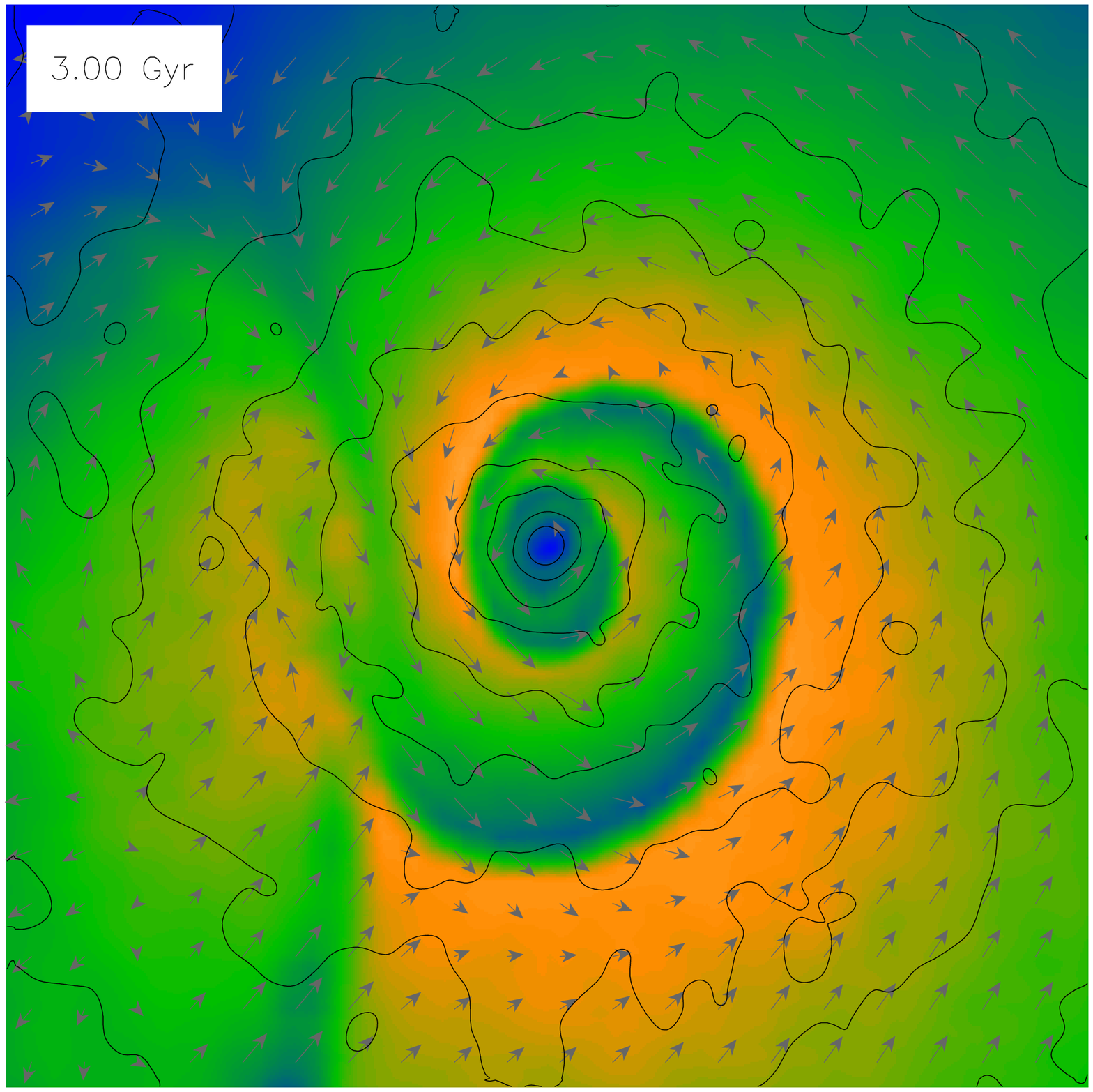}
\caption{
  Collisions with different initial impact parameters, for $R=5$ and DM+gas
  subcluster run. Head-on collison (top) and encounter with $b=1~$Mpc
  (bottom).  The top run has $N=10^7$ gas particles, the bottom run has
  $N=10^6$ particles.}
\label{figB}
\end{figure*}

\subsection{Main cluster without cooling flow}
\label{secCore}

For a contact discontinuity at constant pressure, a density and temperature
jump by a factor of two corresponds to an entropy ratio of $2^{5/3}\sim3$.
The obvious condition for the formation of a cold front is that gases with
such different entropies should exist in the cluster in order for a merger
to bring them into contact.  Clusters with ``cooling flows'' do have steep
radial entropy gradients in the centers which, as we saw above, are
sufficient to produce cold fronts when disturbed. However, about 1/3 of the
nearby clusters do not have cooling flows, instead exhibiting roughly
isothermal flat gas cores. In such clusters, the gas is fairly well fitted
by a simple $\beta$-model, $\rho_{\rm gas} \propto
(1+r^2/r_c^2)^{-3\beta/2}$ with $r_c\sim100$ kpc. The gas entropy is still
increasing with radius in the cores of such clusters, but much slower than
in the cooling flow clusters.  It is interesting to see if such low entropy
contrast is sufficient to generate sloshing and cold fronts.

We have performed an experiment in which the initial density and temperature
profiles for the main object are derived using $c=1$ in equation
(\ref{eqT}), which corresponds to no cool central component. Both DM
profiles are kept the same, and the subcluster still possesses a cool core
with $c=0.17$. All other merger parameters are also the same.  For $c=1$ and
our fiducial DM profile for the main cluster, the gas temperature and
density profiles given by expressions (\ref{eqT}) and (\ref{eqRhog}) are
similar to a $\beta$-model with $r_c=115$ kpc.  For this profile, the
entropy increase by a factor of 3 from its central value is reached at
$r=560$ kpc, while between $r=0$ and $r_c$, it increases only by a factor of
1.6.  For comparison, for our default cooling flow profile with $c=0.17$,
the entropy rises by a factor of 3 already at $r\sim70$ kpc.

The results are shown in Fig.\ \ref{figCore}. The gas core of the main
cluster is now easily pushed downward from the DM peak by the ram pressure
of the stripped subcluster gas. As this ram pressure diminishes, we see the
development of a ``slingshot'' cold front (2.0 Gyr panel in Fig.\ 
\ref{figCore}), similar to that in our simulations with a cool gas peak
(compare to the 1.9 Gyr panel in Fig.\ \ref{figGas}).  As before, this cold
front is a contact discontinuity between the main cluster and the stripped
subcluster gases.  With time, it travels beyond $r=1$ Mpc, maintaining a
density contrast around a factor of 2.

However, no RT instabilities form inside the region delineated by this cold
front --- there is just not enough entropy contrast for them to develop.  So
no cool gas is flowing back to the DM peak and there is no sloshing seen in
the main cluster's core, until the stripped dense, cool central gas from the
subcluster falls there several gigayears later.  This experiment shows that
although the cluster without a steep central entropy decline can produce a
large, prominent cold front of the ``slingshot'' origin in the course of a
merger, ``sloshing'' cold fronts do require a cooling-flow type gas profile.

A cooling flow-like temperature drop is not required, as long as there is an 
entropy gradient. \citet{RickerSarazin01} simulated a merger in an 
isothermal cluster with a cuspy potential, which shows the formation of 
similar multiple mushroom heads. The specific entropy declines toward the 
center less steeply than in a cooling flow cluster, hence the linear 
scale of the resulting sloshing is bigger (note also that this merger 
involves a gas-containing subcluster, so the initial disturbance was 
greater).

\subsection{Main cluster without DM cusp}
\label{secFlat}

It is also interesting to see how the shape of the central DM profile
affects the generation of cold fronts --- in particular, if the DM
density cusp, $\rho_{\rm dm}\propto r^{-1}$, in the center of a Navarro, Frenk
\& White (1995) halo or the \citet{Hernquist90} profile used here,
is a necessary condition.  For this, we consider a limiting case where the
main cluster DM profile is given by
\begin{equation}
\rho_{\rm dm}(r)=\frac{3M_0}{4\pi a^3}\frac{1}{(1+r/a)^4},
\label{eqFlat}
\end{equation}
which features a constant-density core at $r\ll a$.

The gas still has a cooling-flow like density peak at the center. While this
situation does not seem likely in nature, it is a useful test case.  The gas
temperature profile is given by equation (\ref{eqT}), and the density
profile is computed numerically from the hydrostatic equilibrium equation.
We have set the free parameters of the model to the values
$M_2=1.1\times10^{15}\Msun$, $a=\ac=300~$kpc and $c=0.3$, so that the
initial temperature and density profiles of the ICM are as close as possible
to those of the previous experiments in \S\S\ \ref{secDM}-\ref{secGas} (and
to those observed in real cooling flow clusters). The subcluster has the
same gas and DM profiles as before, and the mass ratio and the impact
parameter of the merger are the same as in \S\S\ref{secDM}-\ref{secGas}.

Results of this simulation are plotted in Fig.\ \ref{figFlat}.  All the
salient features observed in a cluster with the cuspy DM profile
(\S\ref{secGas}) are present --- a ``slingshot'' cold front, a RT tongue of
low-entropy gas flowing back into the core, and subsequent sloshing of this
gas in the core with the creation of non-concentric cold fronts. The notable
difference is that the amplitude of this sloshing is wider and its period
longer (and therefore the cold fronts are fewer) than in \S\ref{secGas}, as
expected for the weaker gravitational attraction toward the cluster center.
Thus, sloshing of the cool, dense central gas and the resulting cold fronts
can develop in a wide range of cluster DM profiles, from flat to cuspy.

\subsection{Subcluster mass ratio}
\label{secR}

In this and the following section, we investigate which merger mass ratios
and impact parameters are more favorable to the formation of cold fronts in
the cluster centers.

Encounters with different mass ratios are shown in Fig.\ \ref{figR}.  All
other parameters are set to the same values as in \S\ref{secGas}, the
satellite has its own gas component and the main cluster has both a DM cusp
and a cooling flow.  Major mergers ($R=2$, top panel) trigger extremely
strong perturbations in the ICM of the main cluster. Both ``slingshot'' and
``sloshing'' cold fronts form after the first passage of the satellite, but
its second passage about 1.5 Gyr later (for our particular merger
parameters) destroys them.  There are obvious signatures of merging during
most of the process, so the cluster would not look ``relaxed'' until long
after the merger.  When the system finally settles down into equilibrium,
sharp discontinuities in the core have been erased by shock heating and
vigorous turbulence.

More modest mergers with $R=20-100$ (middle and lower panels in Fig.\ 
\ref{figR}) produce longer-lived cold fronts in the main core.  Such small
subclusters, merging with this particular impact parameter ($b=500$ kpc), do
not cause a complete detachment of the cool gas peak from the DM peak,
contrary to the $R=2$ case above and our default simulation with $R=5$.
There is a weak slingshot front, but with little (for $R=20$) or no (for
$R=100$) RT-like filament of cool gas flowing from the front inward --- the
lowest-entropy gas has never left the center.  The pressure disturbance is
still sufficient to displace the coolest gas and cause its sloshing. Thus,
during mergers with the smallest subclusters, the behavior of the gas in the
center of the main cluster initially looks almost like that in a pure-DM
subcluster run (\S\ref{secDM}). A big difference, however, is the cool gas
from the subcluster that is completely stripped during the first passage.
This gas quickly falls into the center, causing turbulence and destroying
any coherent structure there.  This occurs more quickly for less massive
subclusters --- about 0.8--1 Gyr after the core passage for the $R=20$
merger, but only 0.4 Gyr for the $R=100$ merger.  In a few gigayears, this
gas will relax into a new cool core, and the remaining pure-DM satellite may
cause sloshing during subsequent pericenter passages, as in \S\ref{secDM}.

\subsection{Impact parameter of the merger}
\label{secB}

The generation and survival of cold fronts also depends on the merger impact
parameter. In one extreme, a head-on merger with a subcluster containing gas
is likely to destroy the central cool core and generate a chaotic velocity
field which precludes the formation of any coherent structures (top panel in
Fig.\ \ref{figB}). The exception is a merger in which the subcluster has
such low gas density that it is completely stripped before the DM subcluster
passes the main core.  This case would be similar to a head-on merger with a
DM-only subcluster, which is a simple symmetric limiting case of the
encounter discussed in \S\ref{secDM}. It generates sloshing and concentric
cold fronts at smaller and smaller radii in the main core via processes
discussed in detail in that section.  The significant difference here would
be that the stripped subcluster gas eventually falls into the center and
disturbs those fronts.

Encounters with a large impact parameter are more favorable to the formation
and survival of the cold fronts in the center. Lower panel in Fig.\ 
\ref{figB} shows a simulation for $b=1$ Mpc.  The overall picture is
qualitatively similar to our default $b=500$ kpc case (Fig.\ \ref{figGas}).
However, the slingshot cold front forms at a greater radius and is more
detached from the densest gas in the core, which is never significantly
separated from the main DM peak. The low-entropy gas (not quite so
low-entropy as near the very center) still forms a flow from the middle of
the slingshot front toward the center. The angular momentum of the dense gas
behind the slingshot front is now higher (because the angular momentum of
the merger is higher), which makes the gas to flow toward the center in a
well-pronounced spiral pattern.  Close examination shows that this spiral
flow is a limiting, very lopsided case of the staggered mushroom-like
structures that we saw in Figs.\ \ref{figDMzoom} and \ref{figGas} (see
\S\ref{secspiral}).

Between the time when the slingshot cold front moves beyond $r=500$ kpc and
the return of the subcluster, there is a 1--1.5 Gyr period when the cluster
looks relaxed on a 1 Mpc scale, except for the central spiral cold front
(e.g., the 3.0 Gyr panel in Fig.\ \ref{figB}).  Around $t=4.0$ Gyr, the
subcluster makes a second passage with a smaller angular momentum, hits
close to the center and its gas core disturbs the cold fronts.  A few
gigayears later, we have a relatively relaxed cluster with a sloshing core
similar to the one in our $b=500$ kpc run (the 4.8 Gyr panel in Fig.\ 
\ref{figGas}), except its central gas is now a mixture of gases from both
subunits.

To summarize, minor mergers ($R\ga5$) with relatively large initial impact
parameters ($b\ga500$ kpc, corresponding to pericentric distances of a few
hundred kpc) can easily cause sloshing of the innermost cool, dense gas of
the main cluster around the minimum of the gravitational potential, giving
rise to ``sloshing'' cold fronts.  For mergers with a DM+gas subcluster,
there are periods (albeit relatively short) when the system appears relaxed
on large scales and the most visible features are these cold fronts.  In
general, the bigger the mass ratio and the bigger the impact parameter, the
less violent is the large-scale disturbance and the more regular the central
cold front structure (with the caveat that eventual infall of the cool gas
stripped from the subcluster disturbs it faster if the subcluster is less
massive). For DM-only subclusters, a merger with any reasonable mass ratio
and impact parameter would look relaxed.

\section{Numerical effects}
\label{secnum}

\subsection{Resolution}

\begin{figure}
\centering
\plotone{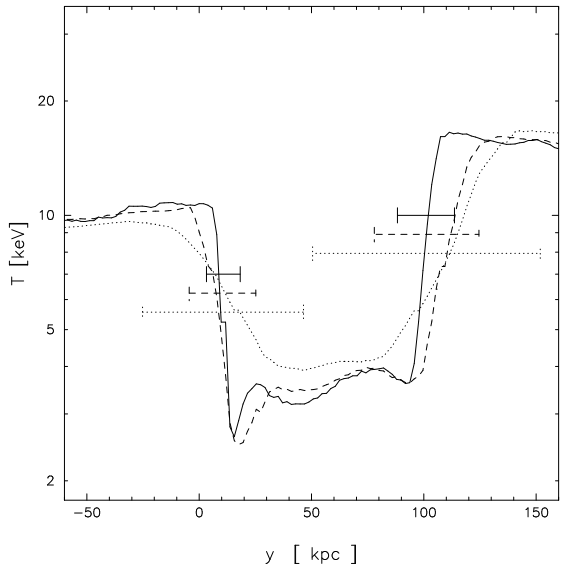}
\caption{
  Temperature across the cold fronts at $t=1.7~$Gyr in the encounter with a
  DM+gas satellite described in \S\ref{secGas}.  Dotted, dashed and solid
  correspond to the same simulation, run with $N=10^5$, $10^6$ and $10^7$
  gas particles.  Horizontal lines indicate the resolution of each
  experiment, computed as twice the SPH smoothing length at $y=7$ and 100
  kpc. The widths of both fronts are equal to the resolution in all cases.}
\label{figResol}
\end{figure}

Before proceeding to discussing the above results, we need
to address several numerical issues. First, we would like
to check whether the sharp edges that we see in the
temperature distribution are indeed discontinuities (at
least at our resolution) or just steep gradients. This can
be done by comparing the width of these features with the
code resolution.  We selected an interesting moment in our
default DM+gas subcluster run ($t=1.7$ Gyr, see the 1.67 Gyr
panel shown in Fig.\ \ref{figGas}) with two cold fronts
developed in regions of different gas densities, and reran
the whole simulation with a lower number of gas particles,
$10^6$ and $10^5$ (compared to the default $10^7$). Initial
conditions were identical.

Figure~\ref{figResol} shows the temperature profile along a
line passing approximately perpendicularly through both cold
fronts. Typically, SPH simulations can resolve structures
larger than twice the smoothing scale $h$, which in our
experiments roughly corresponds to the distance to the 64-th
neighbor. As can be seen in Fig.\ \ref{figResol}, the width
of our temperature gradients is indeed close to (or even
smaller than) $2h$\/ at the locations of the cold fronts,
shown by the horizontal bars. As the resolution improves,
both features become correspondingly narrower. Although
this does not guarantee that increasing the number of
particles still further will not eventually resolve them, it
strongly suggests that they are genuine discontinuities and
not finite gradients.

In any case, it is encouraging that both cold fronts are
present in the same places at the same time at all 3
resolutions. Furthermore, the $N=10^6$ profile is very
similar to that from our default $N=10^7$ experiment, except
for the gradient widths. This suggests that we have
achieved numerical convergence. It is also clear that
$N=10^5$ gas particles is not enough to model the central
cold fronts.

\begin{figure}
\centering
\plotone{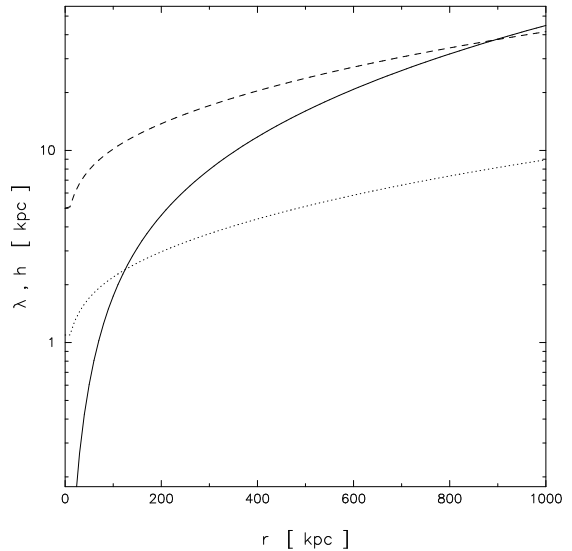}
\caption{
  Mean free path of the electrons (solid lines) for our
  $1.4\times10^{15}\Msun$ cluster with a Hernquist dark matter halo and a
  cool core, compared to the resolution of our simulations with $N=10^7$ SPH
  particles (dashed line) and a hypothetical experiment with $N=10^9$
  (dotted lines).  }
\label{figLambda}
\end{figure}

It is interesting to compare our linear resolution with the
electron mean free path due to Coulomb collisions,
\begin{equation}
\lambda_e\approx 23\,{\rm kpc} 
    \left(\frac{T}{10^8\,{\rm K}}\right)^2
    \left(\frac{n_e}{10^{-3}\,{\rm cm}^{-3}}\right)^{-1},
\label{eqLambda}
\end{equation}
where $n_e$ is the electron number density (e.g., Sarazin
1988). In the absence of magnetic fields, this is the
minimum scale at which the hydrodynamic approximation is
applicable. In reality, the ICM is magnetized and
collisionless, so the true mean free path is significantly
smaller (which was shown observationally for the cold front
in A3667, Vikhlinin et al.\ 2001). Figure \ref{figLambda}
shows $\lambda_e$ from eq.\ (\ref{eqLambda}) for our
fiducial main cluster, along with the numeric resolution for
our $10^7$ particle runs.  Throughout the interesting radial
range, our numeric resolution element is bigger, so we do
not yet have to worry about what happens in the intracluster
plasma on a microscopic level. Future increases in numeric
resolution (e.g., dotted line in Fig.\ \ref{figLambda} for
$10^9$ particles) might pose such questions.


On the other hand, Figure~\ref{figLambda} also shows that we
may be underestimating the amount of small-scale
instabilities and turbulence, because our numeric resolution
essentially imposes an artificial damping scale for
them. This may be important for the survival of cold fronts,
because as we saw in \S\ref{secGas}, they are disrupted by
turbulence (at least on those scales at which turbulence
does occur in our simulations). It is instructive to
evaluate the characteristic Reynolds numbers in our
simulations.  For an ideal gas, it is ${\rm Re}\sim M d/l$,
where $M$\/ is the Mach number of the flow, $d$ is the
characteristic linear scale and $l$\/ is the mean free path
of the gas particles. Turbulence can develop when ${\rm
Re}\ga 10$. At the crudest approximation, we may substitute
our linear resolution (Fig.\ \ref{figLambda}) for $l$. Then
looking at the later panels in Fig.\ \ref{figDMzoom} (the DM
infall case), Reynolds numbers for the flow around the cool
edge features are in the range 3--6, taking the scale of
those features as $d$. So at our resolution, turbulence
cannot develop there. In the gas subcluster run, ${\rm
Re}\sim 30$ on a 0.5 Mpc scale, and the destructive effects
of turbulence on the cold fronts can indeed be seen in Fig.\
\ref{figGas}.


\subsection{Artificial viscosity}

\begin{figure*}
\centering
\plottwo{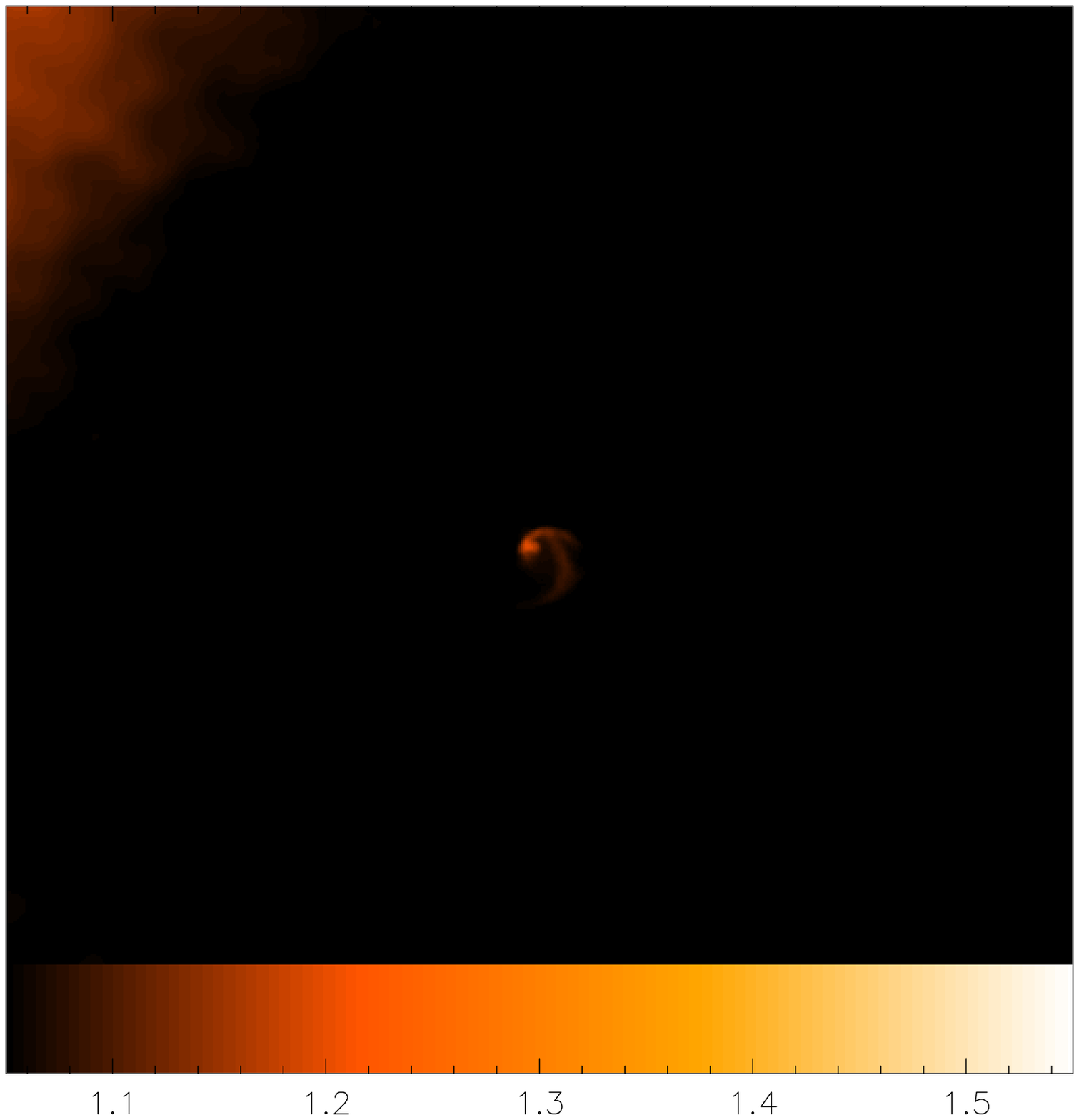}{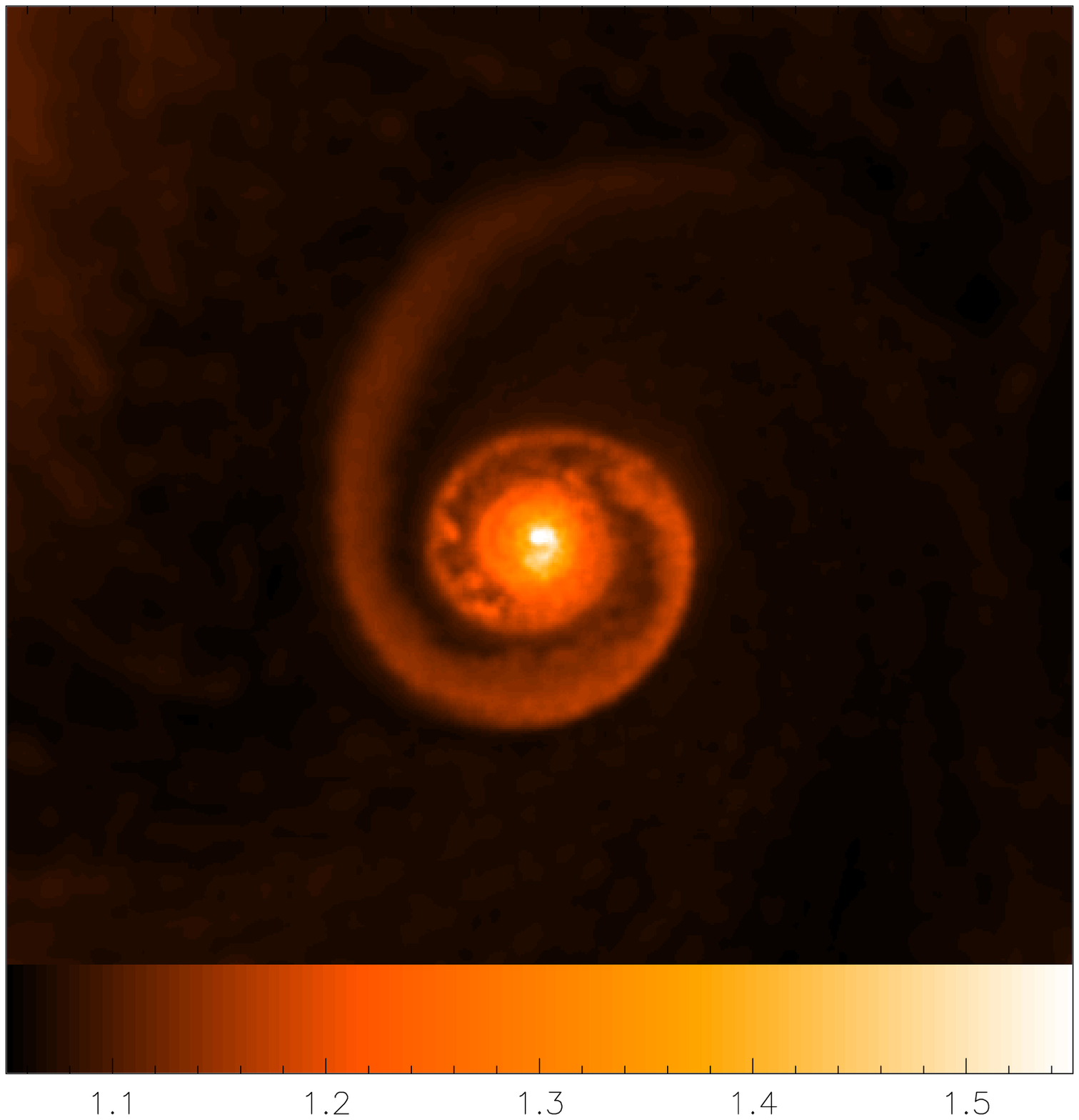}
\caption{The ratio of the gas entropy increase relative to the initial
  entropy (a numeric artifact), for DM-only run, at $t=2.0$ and 6.0 Gyr.}
\label{figDS}
\end{figure*}

Apart from the resolution, there is a more insidious numerical effect,
recently pointed out by \citet{Dolag_05}, that might affect our results. The
SPH formalism resorts to artificial viscosity in order to properly model
shock waves.  Artificial viscosity tends to erase structures in the velocity
field and suppress turbulence.  A comparison between the artificial
viscosity implemented in {\sc Gadget} with the new scheme proposed by Dolag
et al.\ (2005, see their Fig.\ 9) shows that turbulence increases in the
latter and tends to break up the fluid interfaces corresponding to our
stripping-induced cold fronts. Judging from the apparent magnitude of the
effect in \citet{Dolag_05}, sloshing-induced cold fronts should still form
in the cores if one uses the improved scheme, but their structure may be
more irregular and the survival time shorter than suggested by our results.
On the other hand, our cold fronts, both sloshing and subcluster-stripping
types, have regular shapes very similar to those observed in real clusters
(\S\ref{secObs} below), suggesting that physical viscosity in real systems
may not be negligible.  This merits a detailed investigation in future works
(and may even help to measure the viscosity in the ICM).

The artificial viscosity also increases the entropy of the gas in locations
of a steep velocity gradient. To quantify this effect in our simulations, in
Fig.\ \ref{figDS} we plot the relative entropy increase of SPH particles,
$S(t)/S(0)$, for the encounter with the pure DM satellite (\S\ref{secDM}).
In this run, the subcluster wake becomes a weak shock only after pericentric
passage, so the entropy in the central region should stay constant
throughout the whole simulation.  We can see, however, that artificial
viscosity injects a small amount of ``numerical entropy'' into the ICM at
the locations of the cold fronts, which often exhibit velocity shear.  At
$t=2$ Gyr, this results in a $\sim 20$\% increase in the entropy of some
particles.  The effect becomes stronger as the large spiral structure
develops.  By 6 Gyr, for some particles the specific entropy has increased
by 50\% from its original value.  At constant pressure (imposed by
hydrostatic equilibrium), that translates into temperatures higher, and
densities lower, by 12\% and 28\% with respect to their true values.  As we
mentioned in \S\ref{secspiral}, this may inhibit sinking of the dense gas
from the cold fronts back toward the center, and therefore make the outer
parts of the cool spiral structure more prominent. However, compared to the
steep overall gas density gradient in the cluster, this artificial density
change is small and should not affect the picture qualitatively.

\begin{figure*}
\centering
\includegraphics[width=5.4cm]{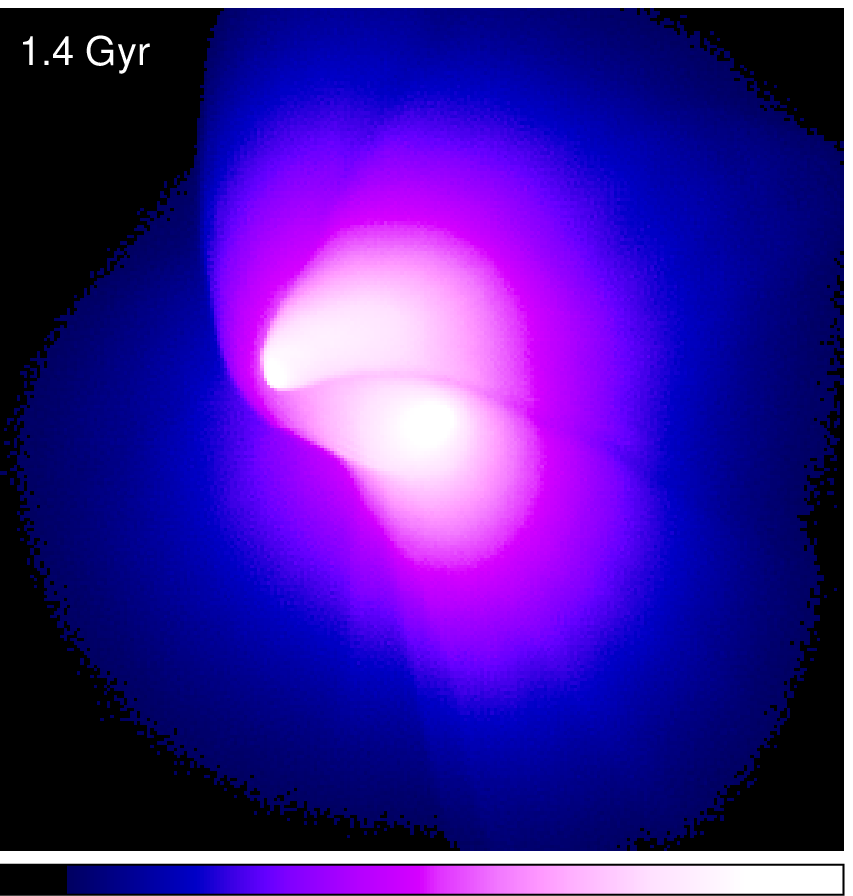}
\includegraphics[width=5.4cm]{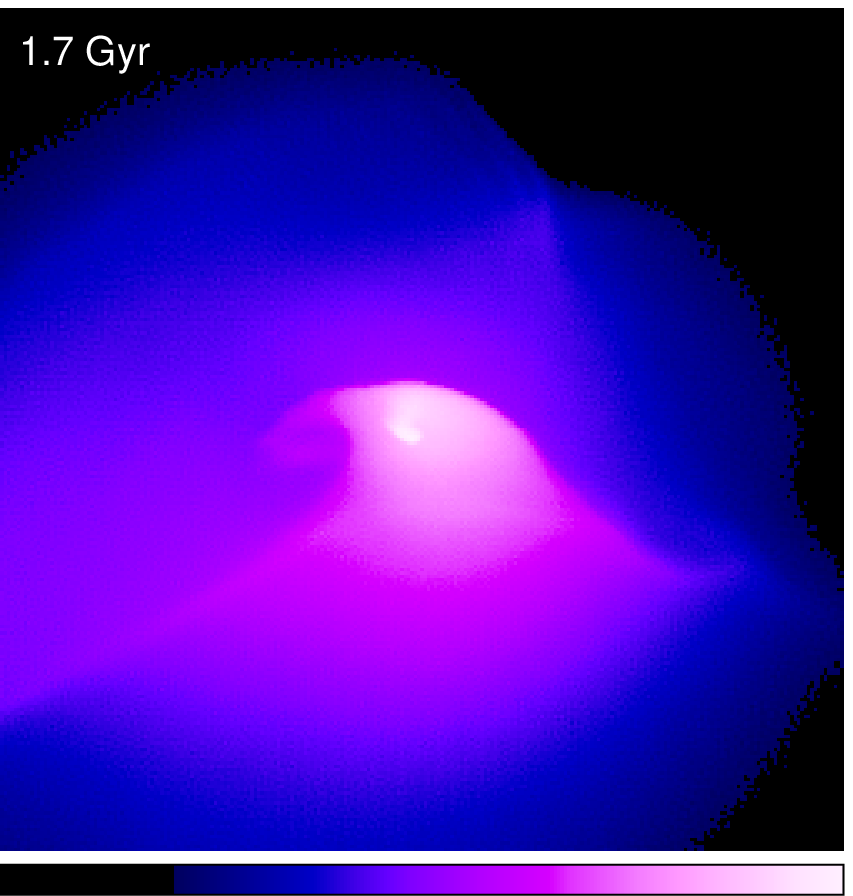}
\includegraphics[width=5.4cm]{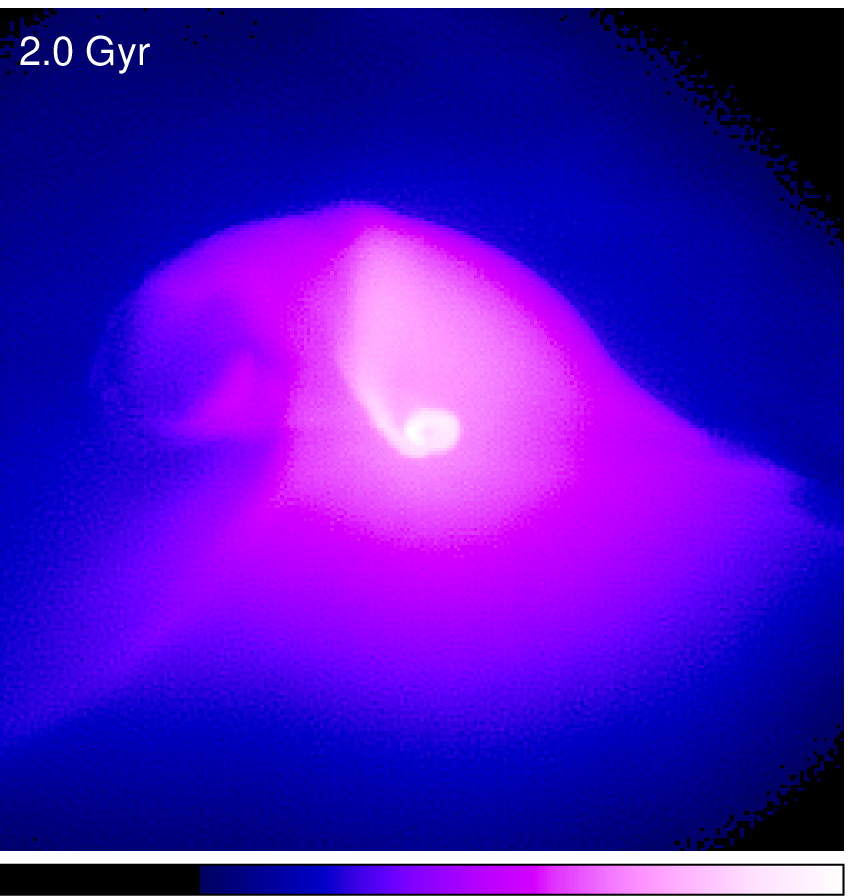}

\vspace{1mm}
\includegraphics[width=5.4cm]{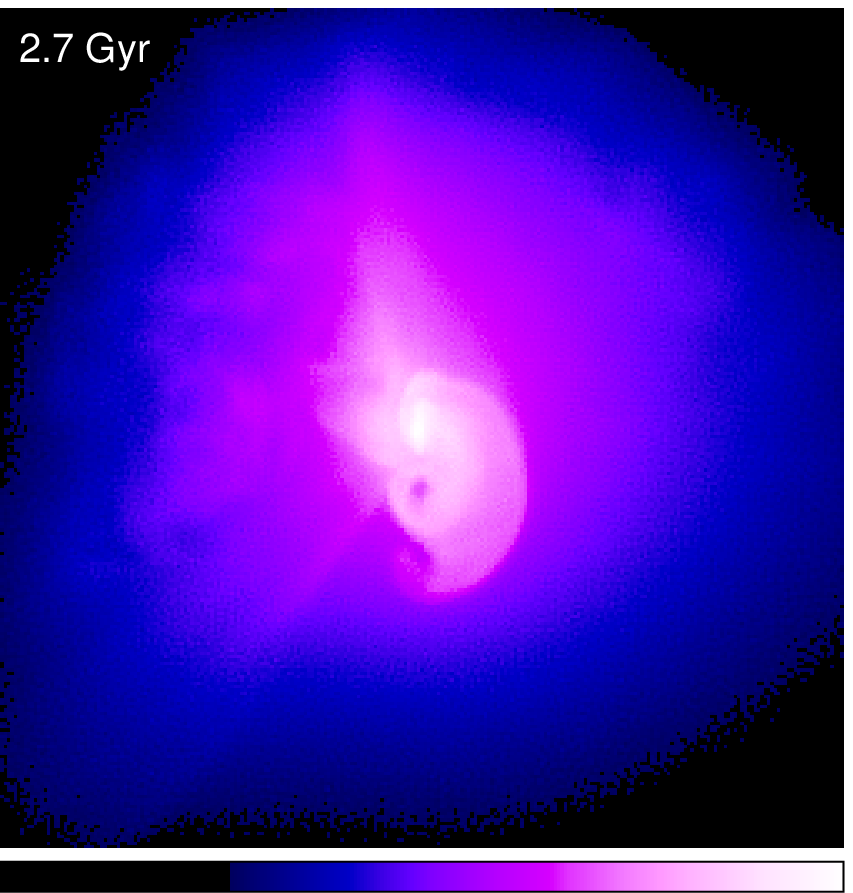}
\includegraphics[width=5.4cm]{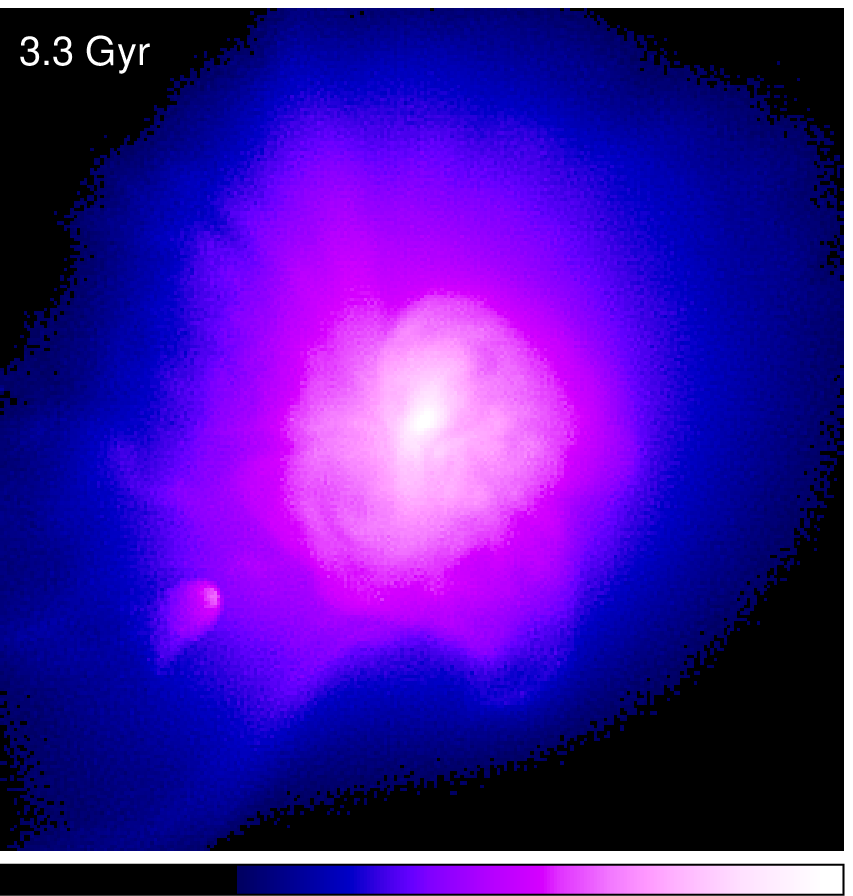}
\includegraphics[width=5.4cm]{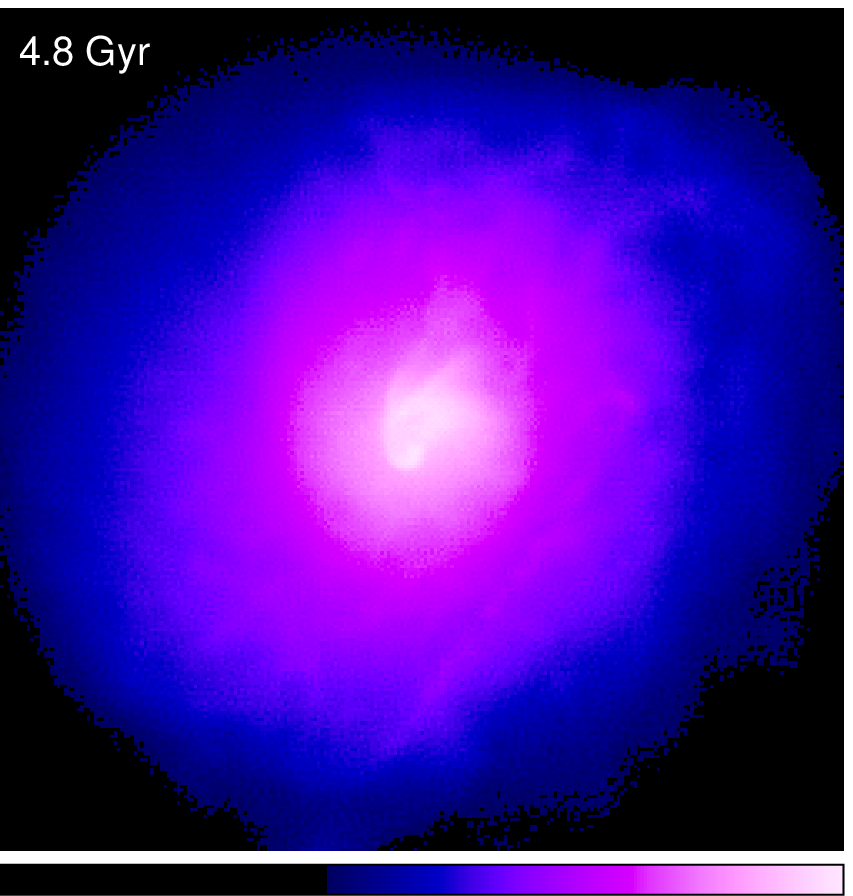}
\caption{
  Projected X-ray brightness for the merger shown in Fig.\ \ref{figGas}
  ($R=5$, $b=500~$kpc, DM+gas subcluster). The panel size is 1 Mpc.
  Brightness scale is logarithmic; color scale is adjusted in each plot to
  emphasize interesting features. The cluster is very disturbed on all
  scales; it needs 3--4 Gyr after the first core passage to start looking
  relaxed on large scales (last panel).}
\label{figLxgas}
\end{figure*}

Finally, in the first panel of Fig.\ \ref{figGas}, one may notice a narrow
hot strip along the lower boundary of the gas stripped from the subcluster.
It is seen more prominently in the corresponding X-ray image (the first
panel in Fig.\ \ref{figLxgas}, slightly shifted in time) as a narrow
low-brightness strip separating the two subclusters.  Its width is a couple
of resolution elements. This strip contains high-entropy gas from the
outskirts of the main cluster that the subcluster has dragged inside, due to
the fact that the gas cannot flow freely along the interface.  This
``viscosity'' appears to be a resolution effect. It may or may not reflect
what happens in real clusters, but it should not affect our present results
in any qualitative way.

  \section{Discussion}
  \label{secDiscussion}

\begin{figure*}
\centering
\includegraphics[width=5.4cm]{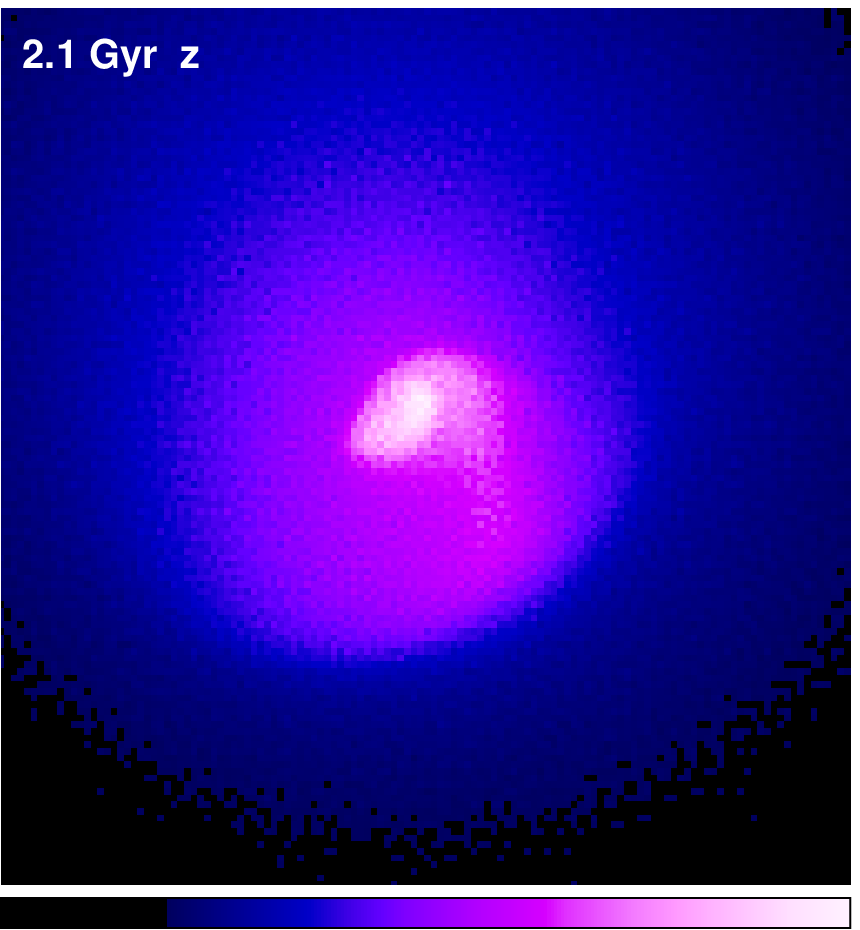}
\includegraphics[width=5.4cm]{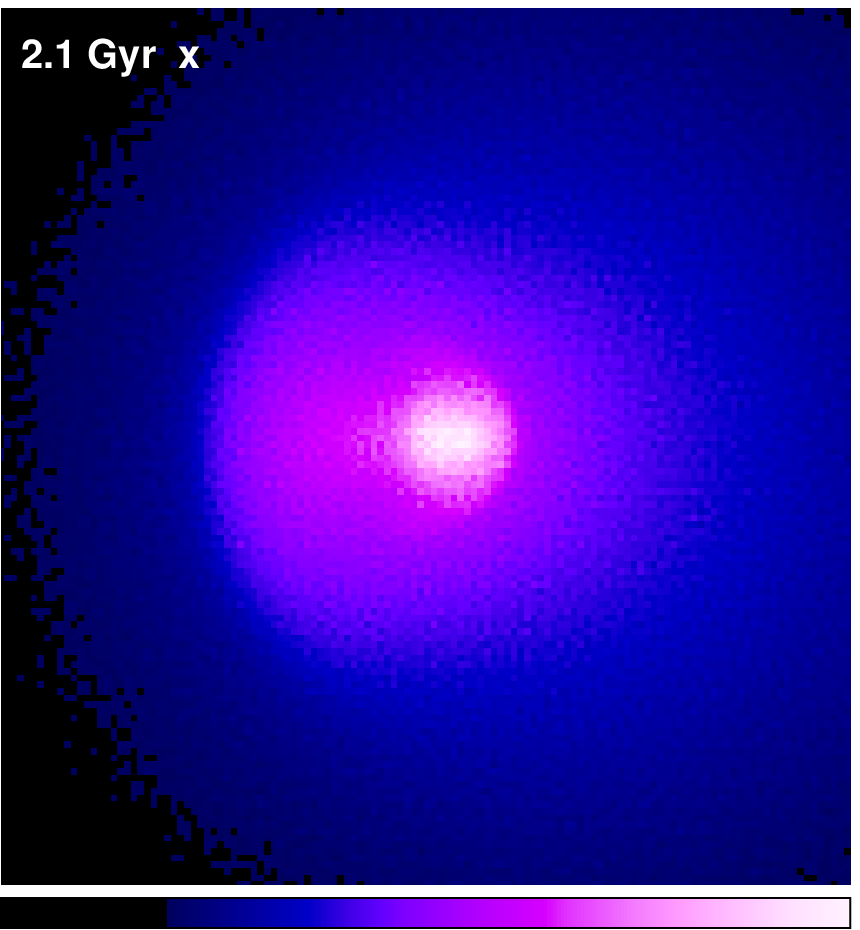}
\includegraphics[width=5.4cm]{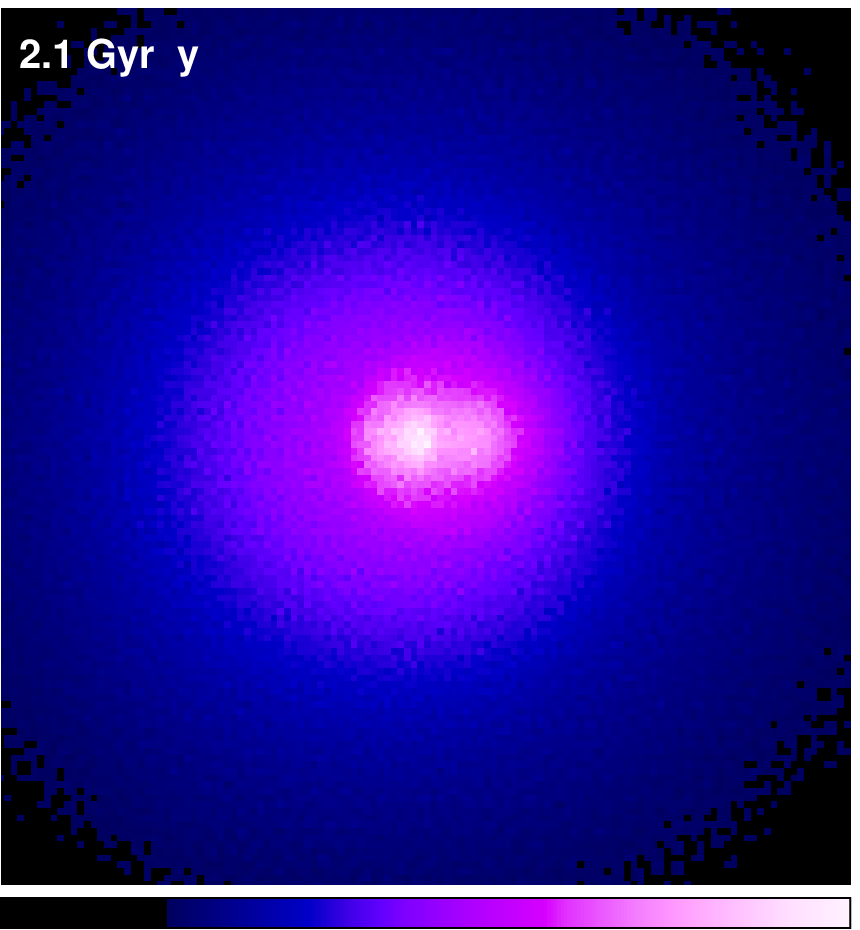}

\vspace{1mm}
\includegraphics[width=5.4cm]{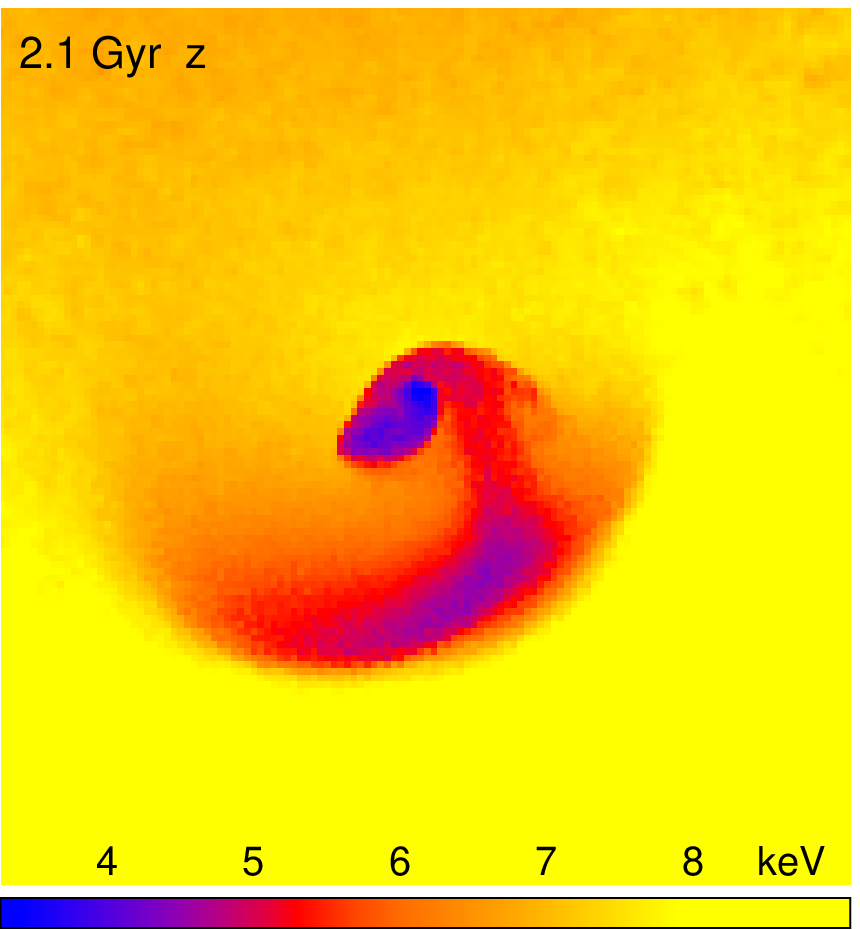}
\includegraphics[width=5.4cm]{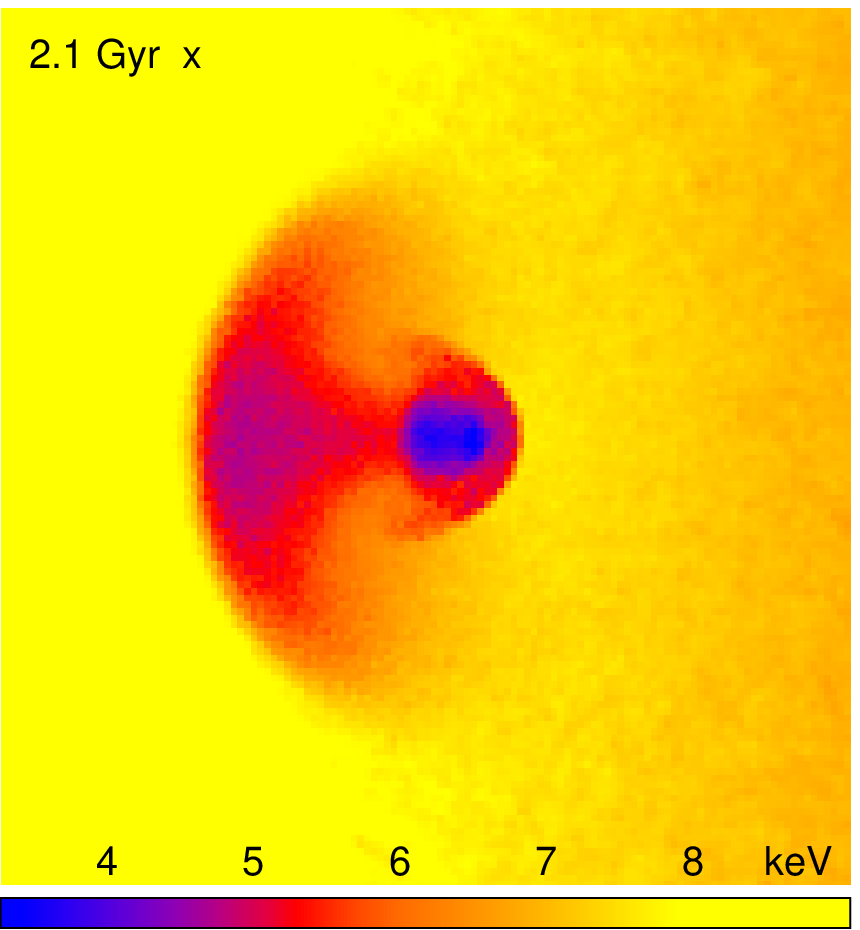}
\includegraphics[width=5.4cm]{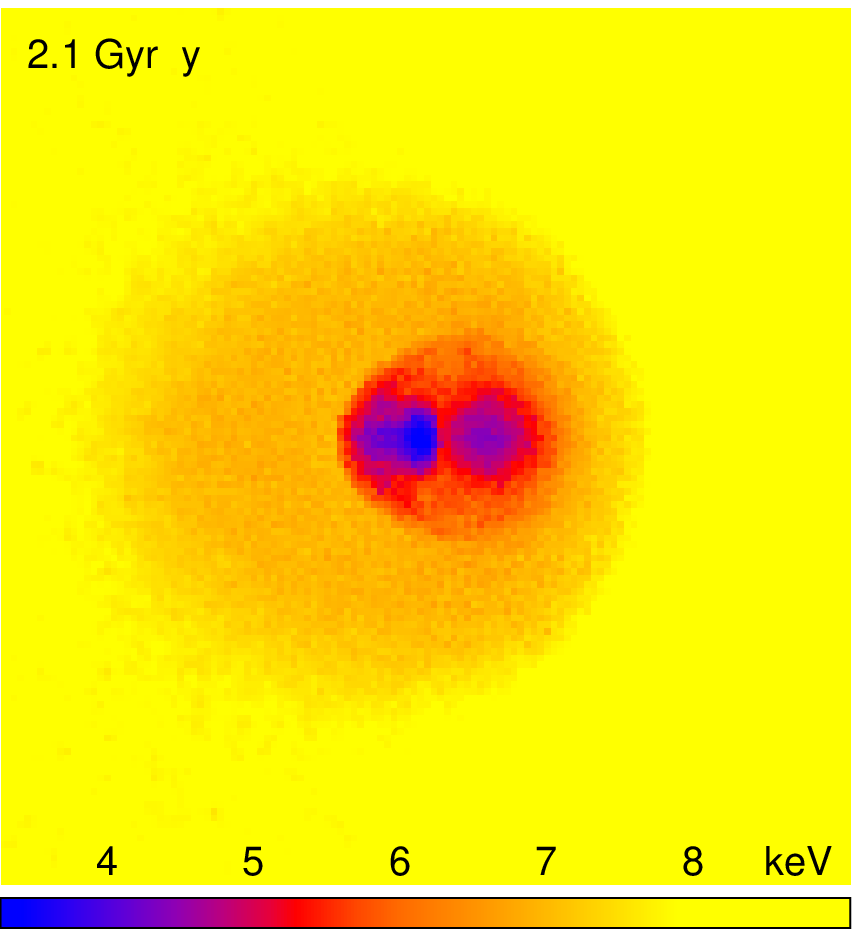}
\caption{
  X-ray images (top) and projected temperature maps (bottom) of the central
  region (0.25 Mpc panel size) for the run shown in Fig.\ \ref{figDM}
  ($R=5$, pure DM subcluster), as seen from three different
  directions. The $z$\/ projection is the merger plane projection shown in
  Fig.\ \ref{figDM}.  In $x$\/ projection, the spiral structure looks like
  the often-observed concentric arcs.}
\label{figLxproj}
\end{figure*}

\begin{figure}
\centering
\plotone{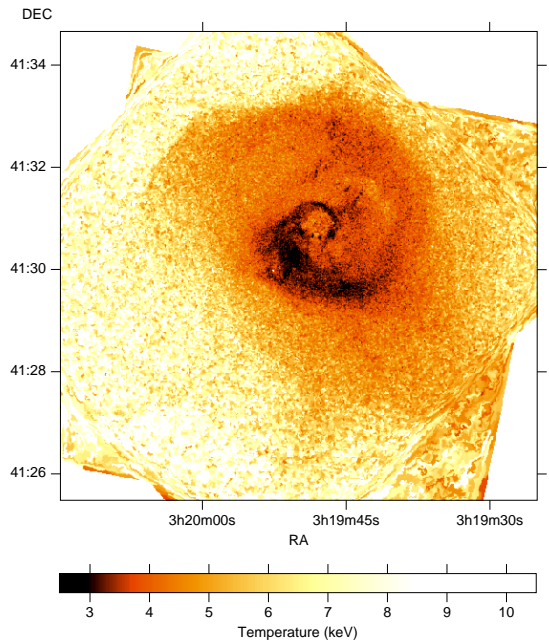}
\caption{\emph{Chandra} projected gas temperature map of the central region
  of the Perseus cluster, the most detailed map of any cluster obtained to
  date (reproduced from Fabian et al.\ 2005b). A large-scale spiral structure
  such as those in our simulations can be clearly seen. (A $\sim1'$ diameter
  cool ring at the innermost end of the spiral is a boundary of an AGN
  bubble, unrelated to sloshing.) The linear size of the map is 200 kpc.}
\label{figPer}
\end{figure}

\begin{figure*}
\centering
\includegraphics[width=5.4cm]{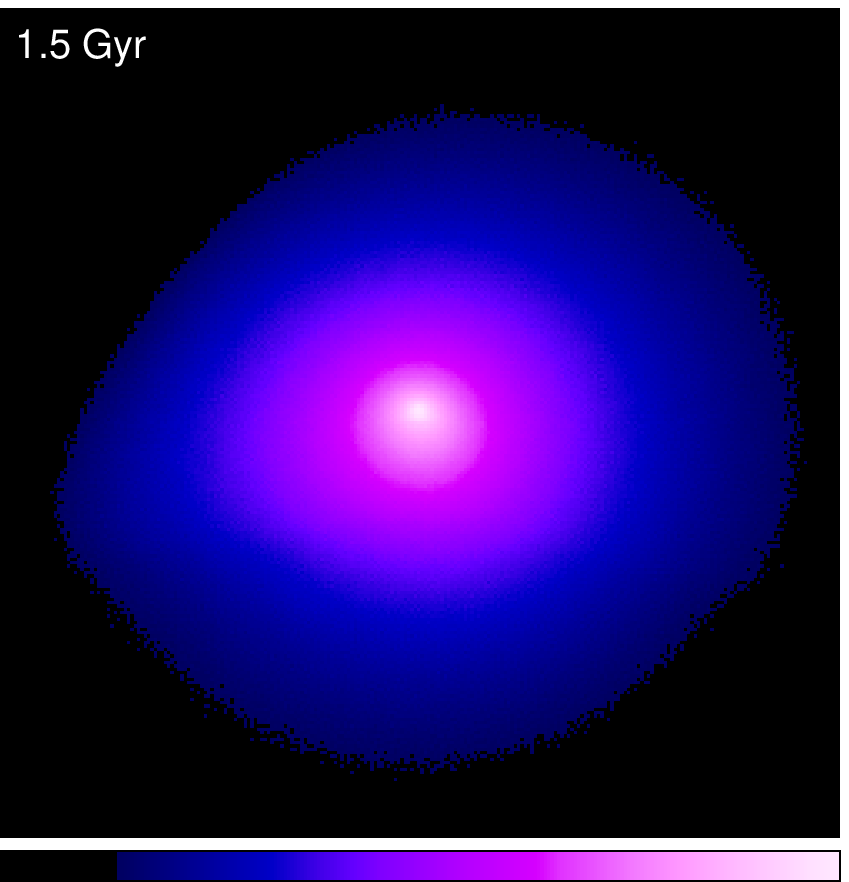}
\includegraphics[width=5.4cm]{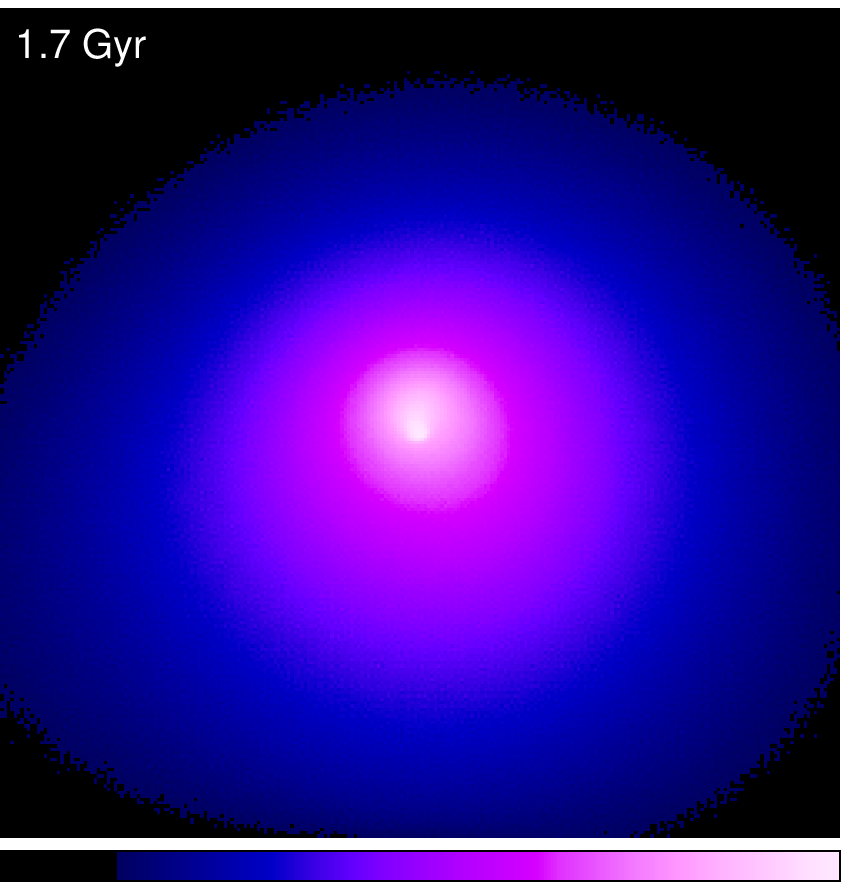}
\includegraphics[width=5.4cm]{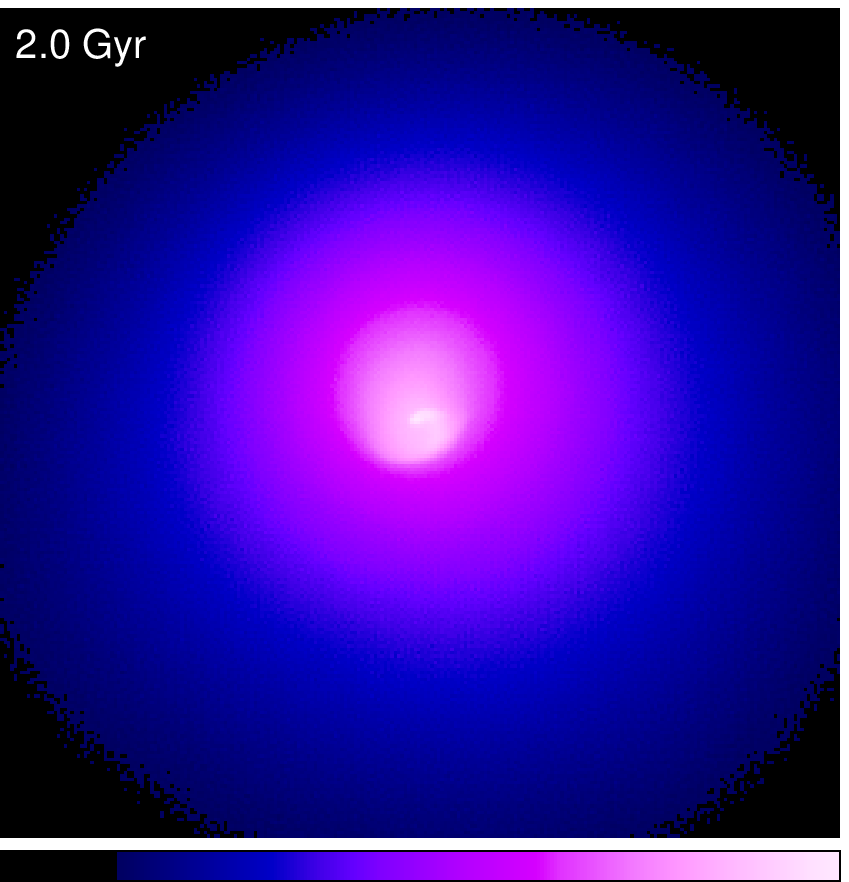}

\vspace{1mm}
\includegraphics[width=5.4cm]{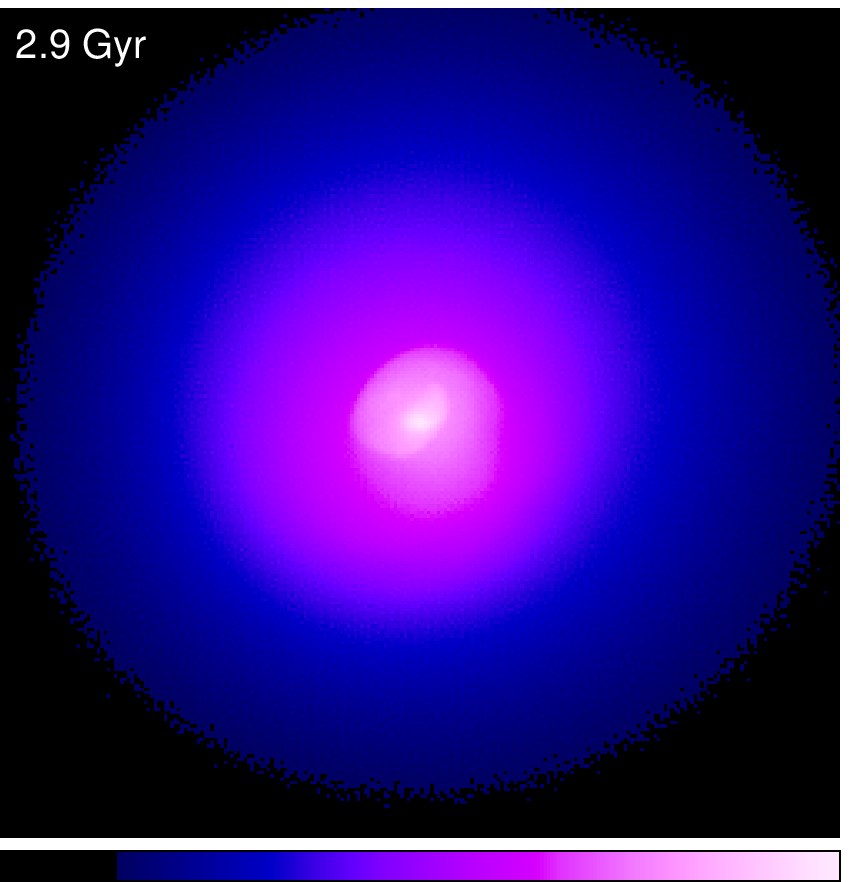}
\includegraphics[width=5.4cm]{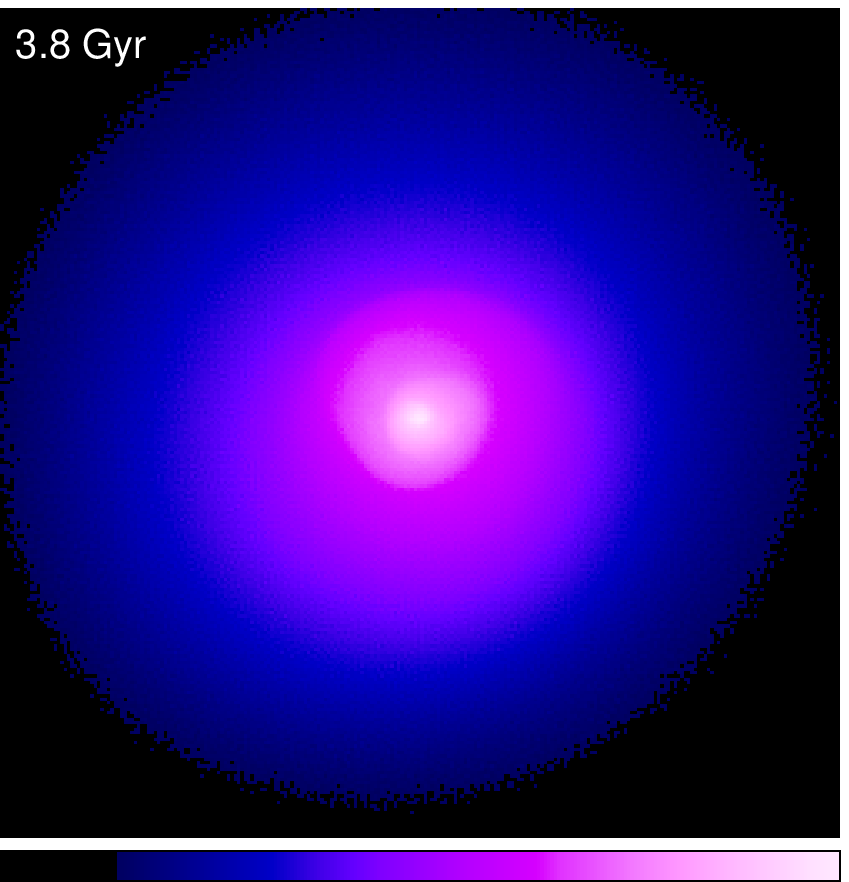}
\includegraphics[width=5.4cm]{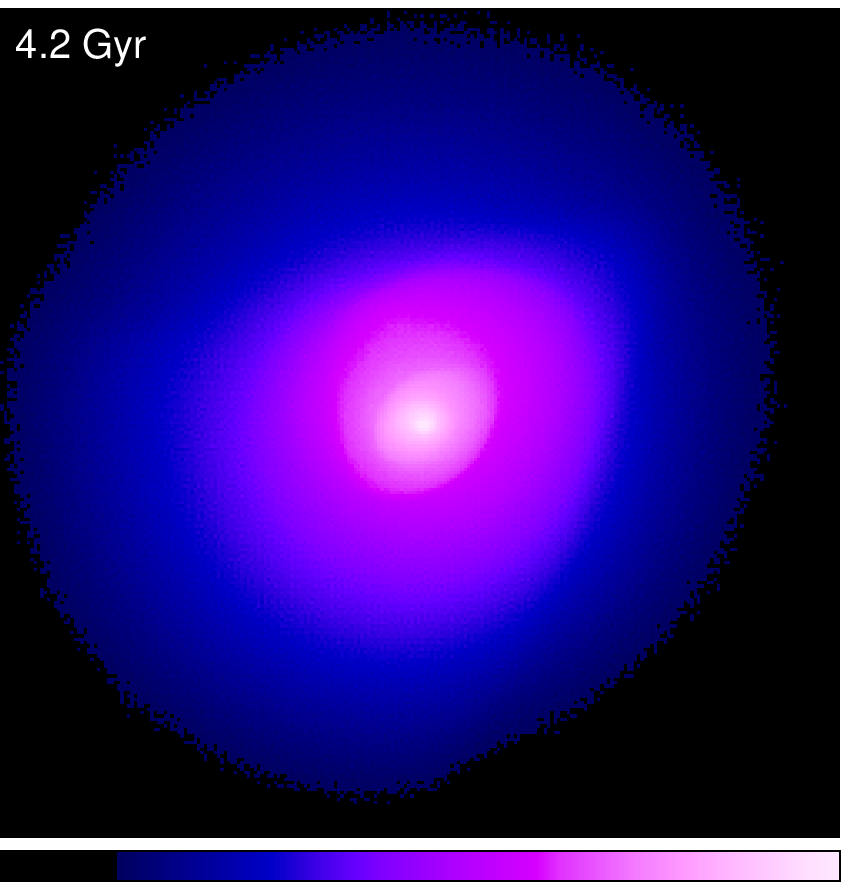}
\caption{
  Projected X-ray brightness for the merger shown in Fig.\ \ref{figDM}
  ($R=5$, pure DM subcluster). The panel size is 1 Mpc. Brightness
  scale is logarithmic and the same in all panels. With the possible
  exception of short moments of the subcluster flyby generating a conical
  wake (1.5 Gyr and 4.2 Gyr), the cluster stays very symmetric on large
  scales; the only structure is edges in the center.}
\label{figLxDM}
\end{figure*}

  \subsection{Comparison with observations}
  \label{secObs}

Our simulations have shown that mergers with a wide range of parameters, even
with the subcluster lacking any gas at all, easily set off sloshing in the
cool central gas and generate cold fronts with density and temperature jumps
of a factor of 2 and more, the amplitudes observed in real clusters.
However, observers deal with projected X-ray brightness and temperatures, so
it is interesting to see if our simulated clusters would look similar in
projection.

To match \chandra\ images and spectra, we have constructed X-ray surface
brightness and temperature maps according to the spectral weighting scheme
described in \citet{Vikhlinin_05}.  The X-ray emission of each particle is
computed as
\begin{equation}
L_i = m_i\, \rho_i\, c(T_i),
\end{equation}
where $m_i$, $\rho_i$ and $T_i$ are the mass, density and temperature of the
SPH particle and $c(T)$ is a function containing the \emph{Chandra}
sensitivity to thermal plasma spectra \citep{Vikhlinin_05}.  To obtain the
temperature that an observer fitting a spectrum from a certain cluster
region with a single-temperature model would derive, the gas temperatures
along the line of sight are weighted proportional to
\begin{equation}
w_i = L_i\, T_i^{-0.875}.
\end{equation}
X-ray luminosities and temperatures have been projected along each of the
three main axes of the simulation box and then smoothed with the
two-dimensional kernel
\begin{equation}
W(u,v)\!=\!\!\left(\!\frac{1}{2}\!-\!\frac{u\!-\!u_i}{h_i}\!\right)
         \!\!\left(\!\frac{1}{2}\!+\!\frac{u\!-\!u_i}{h_i}\!\right)
         \!\!\left(\!\frac{1}{2}\!-\!\frac{v\!-\!v_i}{h_i}\!\right)
         \!\!\left(\!\frac{1}{2}\!+\!\frac{v\!-\!v_i}{h_i}\!\right)
\end{equation}
where $u$ and $v$ are projected coordinates and the smoothing length $h_i$ is inversely proportional to the cubic root of the density of each particle, computed by the {\sc FiEstAS} algorithm \citep{AscasibarBinney05}.

\subsubsection{Small-scale X-ray and temperature images}

Figure \ref{figLxproj} shows X-ray brightness and temperature for three
projections of the central 250 kpc region of the cluster from our pure-DM
subcluster run (\S\ref{secDM}) at $t=2.1$ Gyr, a moment when the spiral
structure is prominent.  All three edges seen in the corresponding
temperature slice (Fig.\ \ref{figDMzoom}) exhibit density and temperature
jumps by a factor of 2.  In projection, a spherical discontinuity becomes a
discontinuity in the gradient of the respective quantity.  So as expected,
the features are less prominent in projection, but still clearly seen (with
the help of the quadratic dependence of the X-ray brightness on gas
density).  The spiral edge pattern projected along the $x$ axis looks like
several concentric arcs, while the $y$\/ projection is less favorable and
barely reveals only the two central discontinuities.

The spiral-like structures such as that in the $z$\/ projection in Fig.\ 
\ref{figLxproj} look very similar to those observed in X-ray images of real
clusters, such as A2029 and especially Ophiuchus (Fig.\ \ref{figObs}).  The
2.1 Gyr panel of Fig.\ \ref{figDMzoom} is probably the best illustration of
the gas flows inside the Ophiuchus core.  Sufficiently detailed temperature
maps for real clusters are much harder to come by, but a 1 Ms \chandra\ 
observation of Perseus, the best-exposed cooling flow cluster to date,
reveals just such a spiral temperature structure, with several edges at
different scales (Fig.\ \ref{figPer}, reproduced from Fabian et al.\ 2005b; a
larger-scale map from \xmm\ can be found in Churazov et al.\ 2003).  The
$\sim1'$ diameter black ring in the very center is a cool boundary of the
AGN bubble unrelated to our subject, but the structure on larger scales
bears a remarkable resemblance to that in Fig.\ \ref{figLxproj}.  A
temperature map for the core of another cooling flow cluster, Centaurus
(Fabian et al.\ 2005a), looks like the $x$\/ projection in Fig.\ 
\ref{figLxproj}.

\subsubsection{Large-scale X-ray images}

Since the primary motivation for this study is to explain cold fronts in
relaxed clusters, now we will check how relaxed our simulated clusters look
on larger scales.  Figure \ref{figLxDM} shows 1 Mpc X-ray images for several
interesting moments of our merger with a DM-only subcluster (\S\ref{secDM}
and Fig.\ \ref{figDM}). Most of the time, the only disturbances seen in the
images are the central cold fronts. For several brief periods, when the DM
satellite crosses the cluster, it generates a subtle conical wake (first and
last panels), but the subcluster spends most of the time in the outskirts.
This simulation looks very much like the most relaxed cooling-flow clusters
in the real world, such as A2029 (compare its image in Fig.\ \ref{figObs}
and the 2.0 Gyr panel in Fig.\ \ref{figLxDM}) and A1795.

Figure \ref{figLxgas} shows a merger with the DM+gas subcluster
(\S\ref{secGas}, Fig.\ \ref{figGas}).  At early stages, it generates a
wealth of prominent cold fronts, including a ram pressure stripping front in
the subcluster, a large ``slingshot'' front 0.3--0.6 Gyr after the core
passage, and sloshing fronts in the center around the same and later time.
However, it also generates a strong general disturbance and asymmetry. At
all stages, except long after the merger (the last panel), such disturbance
will be obvious in any well-exposed X-ray observation. As discussed in
\S\ref{secB}, if the satellite is less massive and/or has a greater impact
parameter, there may be relatively short periods (of the order of 1 Gyr) when
disturbances in the $r<500$ kpc region (other than the central cold fronts)
are weak and the cluster may pass as ``relaxed''. However, we conclude that
sloshing in the cores of the relaxed clusters is most probably caused by
near-center passages of pure-DM subhalos preserved within the cluster.

\subsection{Other observable effects}
\label{secabund}

\subsubsection{Abundance discontinuities}

The mergers and processes modeled here to explain the central cold fronts
have other observable effects.  We expect discontinuities in the metal
abundance to accompany the temperature discontinuities, even if the
satellite did not contribute gas with a different abundance, or any gas at
all. They would result from the centrally peaked initial abundance profiles,
observed in most, if not all, cooling-flow clusters (e.g., Vikhlinin et al.\ 
2005).  Sloshing brings into contact the gases from different initial radii
and should create abundance jumps just as it creates the temperature jumps
(\S\ref{secOrigin} and Fig.\ \ref{figRing}). Such abundance jumps at the
cold fronts are indeed seen in Centaurus (Fabian et al.\ 2005a) and Perseus
(Sanders et al.\ 2004), although Dupke \& White (2003) reported no
significant abundance discontinuity in A496 from a less statistically
accurate measurement.

\subsubsection{cD oscillations}
\label{seccd}

The centers of all cooling flow clusters contain giant cD galaxies, which
often exhibit peculiar line-of-sight velocities up to a few hundred \kms\ 
relative to the average velocity of other cluster galaxies (e.g., Oegerle \&
Hill 2001). If the cD galaxy sits exactly at the cusp of the DM distribution
(as is most likely), we expect it to start oscillating along with the DM
peak after each subcluster flyby, as shown in Fig.\ \ref{figtraj}.  Gas
sloshing and cD peculiar velocities can thus be caused by the same minor
mergers.  Indeed, the observed cD peculiar velocities are of the order of
what we see in our simulations (e.g., the velocity of the DM peak during the
subcluster flyby in Fig.\ \ref{figtraj} is around 200 \kms).

A peculiar velocity of the cD galaxy, and perhaps a
disturbed shape of its extended stellar envelope, could be
tell-tale signs of a close encounter with a massive
subcluster. Such signatures may be expected in cold fronts
caused by gasless satellites. At the same time, as we have
seen above, cold fronts with similar amplitudes may also be
caused by hydrodynamical disturbances (i.e. a shock wave)
from a gas subcluster orbiting farther away from the center,
which may leave the cD galaxy undisturbed. In principle,
this difference may be used to determine the cause of the
central disturbance in a particular cluster. A quantitative
analysis is beyond the scope of the present paper.

\subsection{Minor mergers or AGN explosions?}
\label{secagn}

Our simulations have shown that it is very easy to set off sloshing of the
cool gas in the cluster center. In principle, it could be caused by any
disturbance other than a minor merger. For example, an AGN blowing a bubble
in the dense gas, such as those seen in many cooling flow clusters (e.g.,
McNamara et al.\ 2000) may set off gas sloshing, provided the explosion was
offset from the density peak, as seen in simulations by Quilis et al.\ 
(2001). It would be interesting to model this process in more detail.
However, we can put forward two arguments against AGN explosions being a
prevalent mechanism for the central cold fronts.  First, it is not clear how
an AGN could provide angular momentum to the gas, which is necessary to
produce the often-observed spiral fronts.  Second, the ubiquitous cD
peculiar velocities mentioned above indicate that minor mergers such as
those simulated here occur often, and should generate long-lived sloshing
via the mechanism presented here, regardless of whether there are other
mechanisms at work.

Note that in some clusters showing X-ray edges near the center, those edges
may be an altogether different phenomenon. In the Hydra-A cluster, the edge
appears to be a weak shock propagating in front of a large AGN-blown bubble
(Nulsen et al.\ 2005). Such edges look somewhat differently from the
``sloshing'' edges considered here, spanning a larger sector (in Hydra-A, it
can be traced amost all the way around the cluster core). In addition, very
subtle brightness edges or ``ripples'' observed in the core of the Perseus
cluster have been attributed to sound waves from the central AGN explosions
(Fabian et al.\ 2005).

\begin{figure*}
\centering
\plotone{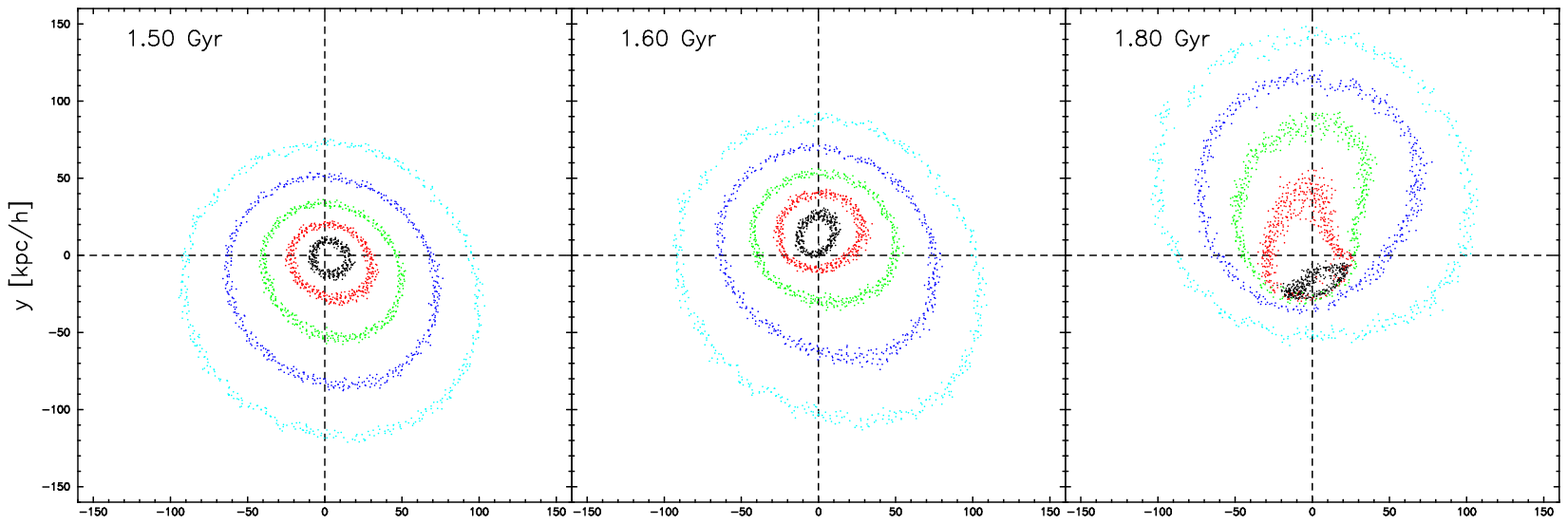}
\plotone{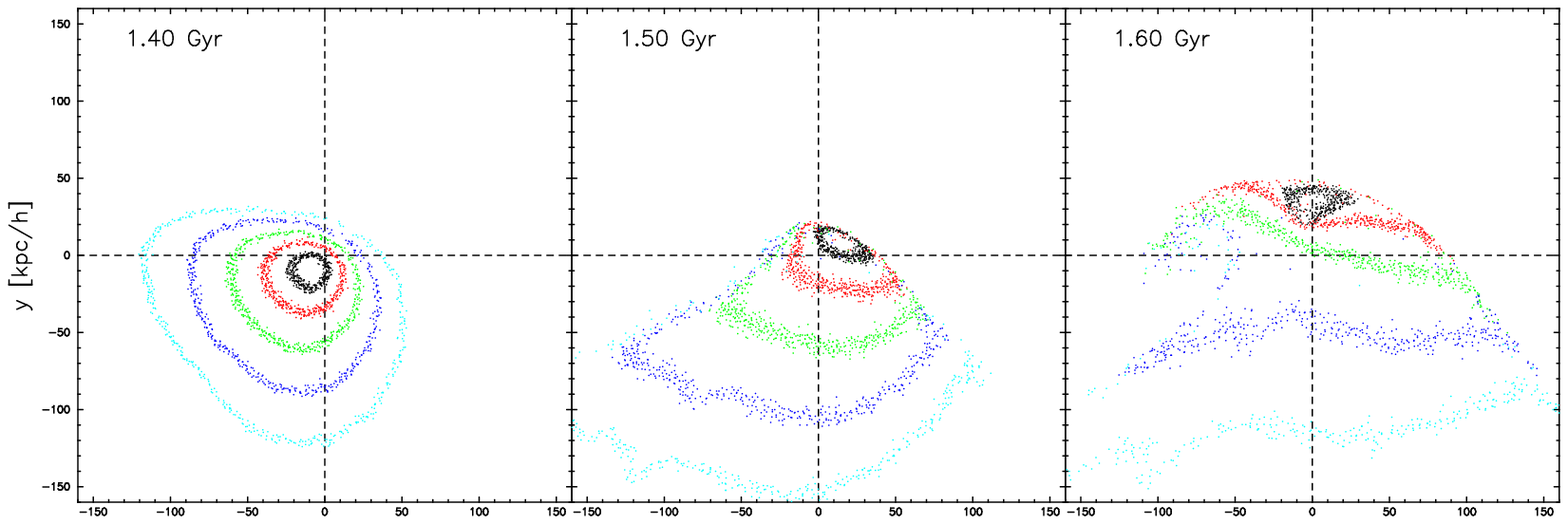}

\caption{Evolution of initially concentric SPH particles during the
  encounter with a DM-only satellite as in Fig.\ \ref{figDM} (top) and with
  gaseous satellite as in Fig.\ \ref{figGas} (bottom), showing the emergence
  of a sharp density discontinuity. The cross shows the DM peak.}
\label{figRing}
\end{figure*}

\subsection{Physical origin of contact discontinuities}
\label{secOrigin}

Our simulations showed that gas sloshing in the cluster center generates
sharp contact discontinuities in the gas. We also saw, along with authors of
many earlier simulation works, that motion of a gas subcluster creates a
discontinuity at its front.  Here we would like to address a question which
was not clearly explained in the literature --- how exactly an initially
continuous gas distribution becomes a sharp contact discontinuity.  Ram
pressure stripping is usually quoted as the agent (e.g., Markevitch et al.\ 
2000; Vikhlinin et al.\ 2001).  It was shown that inside a cold front in
A3667, the azimuthal distribution of the gas pressure closely follows the
pressure of the hotter gas flowing around the front, if one models the
velocity field around the spherical front and uses the Bernoulli equation
(Vikhlinin \& Markevitch 2002).
This indicates that there is pressure equilibrium across the front and therefore its shape is stationary or at least evolves slowly (as opposed, for example, to an expanding bubble).
It is easy to understand how stripping by such a tangential gas
flow around an infalling dense gas subcluster may remove the outer layers of
the subcluster's gas until it reaches the radius where the pressure in the
cold gas equals the pressure outside.  However, at the forward tip of the
front (the stagnation point), there is no tangential gas flow and no such
stripping, yet the fronts are just as sharp there.

When the subcluster has a gaseous component, ram pressure is exerted by the
shocked ICM gas and then by the gas stripped from the subcluster, so one could
argue that there are discontinuities in the ICM to begin with.
However, in our DM-only
subcluster run, there is no shock and no other gas, the gas density and
velocity distributions are perfectly continuous until the sloshing starts.

The moment when the first contact discontinuities emerge in the main cluster
core is shown in Fig.\ \ref{figRing}, where top panels show our simulations
with the DM-only subcluster and bottom panels show the one with the gas
subcluster.  The figures show the positions of individual SPH gas particles
extracted from initially concentric annuli around the main cluster center,
in a slice along the merger plane.  It is instructive to compare this plot
with the figures showing the velocity field, Figs.\ \ref{figJumpDM} and
\ref{figDMzoom} for the DM-only subcluster and Fig.\ \ref{figGas} for the
gas subcluster.  For the DM-only run, 1.55 Gyr is the approximate moment
when the main core reverses the direction of motion w.r.t.\ the surrounding
gas and starts experiencing ram pressure at its downward side.  The net
force from the ram pressure is the same for each cubic centimeter of the gas
in the core (near the symmetry axis and assuming subsonic motions).
Therefore, denser gas experiences a smaller resulting acceleration. This
quickly produces a velocity gradient inside the core along the direction of
the force.  The lower-density, outer layer of the core gas (red and green
dots) is then squeezed to the sides from the region where the ambient flow
eventually meets the dense gas for which the density-proportional gravity or
inertial force prevails over the area-proportional ram pressure force (black
dots; 1.80 Gyr panel).

The same process occurs in the run shown in lower panels of Fig.\ 
\ref{figRing}.  At 1.4 Gyr, the core starts experiencing ram pressure (much
stronger than in the case above), and by the next snapshot the velocity
gradient along the ram pressure force has brought the dense gas in contact
with the outer gas.  By the next snapshot, the process is repeated on a
smaller scale, when the densest gas turns around and again starts moving
toward the center against an ambient flow.

One may note that in both cases, the first discontinuity arises right after
the dense gas overshoots the DM peak. Although this increases the
temperature contrast due to adiabatic expansion of the dense gas as it moves
to a lower-pressure environment, it is not a necessary condition and is
largely a coincidence.  Indeed, sharp contact discontinuities form in the
infalling subcluster (e.g., the 1.39 Gyr panel in Fig.\ \ref{figGas}) and in
our sloshing core (e.g., the 1.67 Gyr panel) without overshooting the DM
core.
We conclude that whenever a smooth gas density peak
encounters a flow of ambient gas (caused by subcluster
infall, shock passage, sloshing, etc.), a contact
discontinuity quickly forms by ``squeezing out'' the gas
layers near the future stagnation point that are not in
pressure equilibrium with the flow.
Away from the axis of
symmetry of the cold front, such gas is stripped by the
shear flow.

  \subsection{Comparison with previous works}
  \label{secprev}

A number of earlier hydrodynamic simulations have been devoted to the origin
of cold fronts in clusters since their discovery in real systems.  Cold
fronts have been seen in several cosmological experiments, always as a
consequence of merging activity \citep[e.g.][]{Bialek02,NagaiKravtsov03,Mathis05}.
The fronts seen in those
simulations are analogous to our large-scale discontinuities separating the
ICM of the two merging clusters --- the ram pressure stripping front in the
infalling subcluster, and the ``slingshot'' front in the main cluster after
the first core passage (Fig.\ \ref{figGas}).  The small-scale
sloshing-induced cold fronts described in this work have never been reported
in a cosmological setting.  This is not surprising given that, on one hand,
the resolution of these works is substantially lower than the experiments
presented here.  On the other hand, in realistic cosmological simulations,
it is often difficult to disentangle the effects of concurrent physical
processes.  For this reason, idealized simulations with controlled initial
conditions are very helpful in deriving a physical interpretation of the
observed phenomenology.

Several earlier idealized simulations obtained results relevant for the
phenomenon considered here.  Heinz et al.\ (2003) used 2D simulations to
model a passage of a shock wave through an isothermal $\beta$-model gas core
in a stationary gravitational potential. Their core developed a cold front
and a circulation pattern where the lower-entropy gas was flowing from the
center toward the front and was then redirected to the sides along the front
surface.  We do see a similar circulation pattern forming in our simulations
whenever the cluster core is subjected to an ambient flow (e.g., Fig.\ 
\ref{figDMzoom}).  In our case, however, the initial gas core is not flat
but resembles a cooling flow with a steep entropy decline toward the center,
and the gas is supported by a cuspy DM profile.  This produces additional
reverse flows of the lowest-entropy gas from the front toward the center
along the filaments forming via RT instability.  This was absent in the
Heinz et al.\ (2003) setup.

Churazov et al.\ (2003) and Fujita et al.\ (2004) used 2D simulations to
show that a weak shock or acoustic wave propagating toward a cooling flow
can displace the cool gas from its equilibrium in the gravitational
potential well. According to Churazov et al., the cool gas then starts
oscillating in the potential well with different periods for the gas at
different radii, which produces a series of cold fronts at different radii
at the opposite sides of the core. The cold fronts are locations of the
apocenters of oscillations where the gas spends most of the time. This is
approximately what we see, except that the disturbance in our setup is
created by a subcluster flyby. We also find that cold fronts are not the end
points of the gas oscillations (which was also proposed in the original
paper by Markevitch et al.\ 2001), but rather structures continuously moving
outward through a combination of gas sloshing and circulation inside the
front (Figs.\ \ref{figDMzoom} and \ref{figR} and \S\ref{secspiral}).

Finally, the work that used the setup closest to ours is
\citet{TittleyHenriksen05}, who carried out three-dimensional numerical
simulations of off-center mergers with mass ratios 10 and 30. The mergers
were selected from a cosmological run.  They propose that the central cold
fronts arise as a result of oscillations of the main cluster's DM peak
initiated by the passage of the substructure.  In their scenario, the
central gas is dragged along by the DM peak and cold fronts form due to
compression of the gas on the forward side. They did not actually see the
cold fronts because of insufficient resolution ($N\sim10^4-10^5$ particles,
compared to our $10^7$).
In our DM-only subcluster run, the initial decoupling of the cool gas from
the DM peak is indeed caused by a swing of that peak during the subcluster
flyby (\S\ref{secdecoup}). However, the subsequent gas sloshing, which
generates cold fronts, proceeds with a period much shorter than oscillations
of the DM core (\S\ref{secslosh}). When the infalling subcluster has gas, we
find that the initial gas-DM displacement is also mostly a hydrodynamic
effect (\S\ref{secGas}).

  \section{Summary}
  \label{secsum}

We used high-resolution hydrodynamic simulations of idealized cluster
mergers in order to determine if and how they can cause the ubiquitous cold
fronts observed by \chandra\ in the centers of many ``relaxed'' clusters
with cooling flows. Several numerical experiments have been carried out,
varying the mass ratio of the merger, the initial impact parameter, and the
shape of the central gravitational potential.  We also tried a merger with a
dark matter subcluster devoid of any gas.  Our initial gas density and
temperature profiles match closely those observed in real cooling-flow
clusters, such as A2029. As a test, we also tried a cluster with an
isothermal gas core.

Our results show that mergers with small subclusters easily set off sloshing
of the cool gas in the main cluster central potential minimum, which gives
rise to cold fronts very similar to those observed in real clusters.  The
necessary conditions for such sloshing are the presence of an initial steep
entropy drop in the center (such as that in cooling flows) and that the
merger does not completely destroy the cool core in a head-on impact. The
presense of a central DM cusp is not necessary, although it does make the
cold front structure look more realistic.

It is difficult to set
off central gas sloshing without causing an obvious global disturbance in
the cluster. When the infalling subcluster has gas, we observe shocks, a
cold front in the subcluster, gas stripped from the subcluster, and a
large-scale ``slingshot'' cold front propagating from the center of the main
cluster and separating the ICM of the two halos. If the main cluster had a
cooling flow, we additionally see sloshing in the center.  The cool gas
stripped from the subcluster eventually falls into the center of the main
cluster and disturbs any coherent motions there.  We find that minor mergers
with mass ratios $R\ga 5$ and large impact parameters ($b=0.5-1$ Mpc,
corresponding to distance during core passage of $100-400$kpc) may produce
relaxed-looking clusters with the visible disturbance limited to the central
sloshing, but only for brief ($\sim 1$ Gyr) periods or very late in the
merger.

However, if the infalling subcluster did not have any gas during core
passage (e.g., it was completely stripped at an early stage), the only
noticeable disturbance during the merger is the central sloshing. The
resulting pattern of cold fronts can survive for several gigayears.  This
type of mergers appears to provide the best fit to the most-relaxed observed
clusters with central cold fronts, such as A2029 and A1795.

Gas sloshing is set off when the gas peak is displaced from the DM peak and
starts falling back.  In the main core, this occurs as a result of a ``ram
pressure slingshot'', when the core gas is first compressed and displaced by
ram pressure, and then it suddenly diminishes. When the infalling subcluster
has gas, ram pressure is caused mainly by the passage of the shock. In the
DM-only subhalo case, it is caused mostly by the orbital swing of the main
DM peak during the subhalo flyby. A gravitational wake created by the DM
subcluster in the gas of the main cluster also plays a role in the latter
case.  In a process similar to Rayleigh-Taylor instability, the densest
fraction of the displaced gas quickly turns around and starts falling back
toward the potential minimum.  It encounters the less-dense gas that is
still flowing in the opposite direction, and forms a mushroom-shaped cold
front. The mushroom head overshoots the potential minimum, after which the
densest gas turns around again and the picture repeats itself at
progressively smaller linear scales.  This creates the often-observed
pattern of near-concentric density edges at different radii on the opposite
sides from the center.  If the subcluster had a nonzero impact parameter,
these edges are not exactly concentric but instead form a spiral pattern,
also seen in real clusters.  This spiral is just a superposition of the
independent edges and does not represent any coherent spiraling motion, at
least not initially.

We find that although the lowest-entropy gas indeed sloshes back and forth
in the potential minimum, each cold front, once formed, propagates outward
from the center and does not ``turn around'' or ``straighten out''. There
appears to be a circulation pattern in which the lowest-entropy gas
initially forming this cold front, turns around and sinks back towards the
center, while being replaced at the front by higher-entropy gas. The caveat
here is that details of the long-term evolution of the fronts in our
simulations may be affected by a numerical artifact, namely, the spurious
generation of entropy by the artificial viscosity inherent to the SPH
scheme.

On the other hand, our SPH simulations can trace the origin of gas particles,
enabling us to see exactly how the initially continuous gas density and
velocity field gives rise to a contact discontinuity. It arises naturally
when the gas density peak starts moving and experiencing ram pressure from
the ambient ICM gas, which creates an acceleration gradient
along the direction of the flow.  The highest-density gas, for
which the density-proportional gravity or inertial force prevails over the
area-proportional ram pressure force, then squeezes the lower-density gas,
more easily affected by ram pressure, to the sides and eventually comes into
contact with the ambient medium.

We have shown that minor mergers easily create sloshing and cold fronts in
the cooling flow clusters. In principle, any gas disturbance may set off
sloshing, e.g., AGN explosions, and it would be interesting to model such a
process in detail.  However, there are arguments against AGN outbursts being
a prevalent mechanism.  Most notably, the ubiquitous cD peculiar velocities
indicate that minor mergers, such as those simulated here, do occur and
should generate sloshing via the mechanism presented here.

In a forthcoming paper, we will use these simulations to address the effects
of the central gas sloshing on cooling flows and on the total mass estimates
derived under the assumption of hydrostatic equilibrium.


\acknowledgments

We are grateful to L. Hernquist, P. Nulsen, A. Vikhlinin, P. Ortiz, and C.
Jones for useful discussions. YA thanks Harvard-Smithsonian Center for
Astrophysics for hospitality; all simulations presented in this work have
been carried out at the CfA's Institute for Theory and Computation.  The
work was supported by NASA grants G02-3164X and G04-5152X and NASA contract
NAS8-39073.

 \bibliographystyle{apj}
 \bibliography{/home/yago/old/docs/LaTeX/BibTeX/DATABASE,/home/yago/old/docs/LaTeX/BibTeX/PREPRINTS}

\begin{thebibliography}{15}
\expandafter\ifx\csname natexlab\endcsname\relax\def\natexlab#1{#1}\fi

\bibitem[{{Acreman} {et~al.}(2003){Acreman}, {Stevens}, {Ponman}, \&
  {Sakelliou}}]{Acreman03}
{Acreman}, D.~M., {Stevens}, I.~R., {Ponman}, T.~J., \& {Sakelliou}, I. 2003,
  \mnras, 341, 1333

\bibitem[{{Asai} {et~al.}(2004){Asai}, {Fukuda}, \& {Matsumoto}}]{Asai04}
{Asai}, N., {Fukuda}, N., \& {Matsumoto}, R. 2004, \apjl, 606, L105

\bibitem[{{Ascasibar} \& {Binney}(2005)}]{AscasibarBinney05}
{Ascasibar}, Y. \& {Binney}, J. 2005, \mnras, 356, 872

\bibitem[{{Bialek} {et~al.}(2002){Bialek}, {Evrard}, \& {Mohr}}]{Bialek02}
{Bialek}, J.~J., {Evrard}, A.~E., \& {Mohr}, J.~J. 2002, \apjl, 578, L9

\bibitem[Buote and Tsai(1996)]{1996ApJ...458...27B}Buote, D.A.~and Tsai, 
J.C., 1996, \apj, 458, 27

\bibitem[Churazov \emph{et al.}(2003)]{Churazov03}
Churazov, E., Forman, W., Jones, C., and B{\"o}hringer, H. 2003, ApJ 590,
225
 
\bibitem[\protect\citeauthoryear{{Clarke}, {Blanton} \& {Sarazin}}{{Clarke}
  et~al.}{2004}]{Clarke04}
{Clarke} T.~E.,  {Blanton} E.~L.,    {Sarazin} C.~L.,  2004, \apj, 616, 178

\bibitem[{{Dolag} {et~al.}(2005){Dolag}, {Vazza}, {Brunetti}, \&
  {Tormen}}]{Dolag_05}
{Dolag}, K., {Vazza}, F., {Brunetti}, G., \& {Tormen}, G. 2005,
  {\tt(astro-ph/0507480)}

\bibitem[\protect\citeauthoryear{{Dupke} \& {White}}{{Dupke} \&
  {White}}{2003}]{Dupke03}
{Dupke} R.,  {White} R.~E.,  2003, \apjl, 583, L13

\bibitem[\protect\citeauthoryear{{Fabian}, {Sanders}, {Taylor} \&
  {Allen}}{{Fabian} et~al.}{2005a}]{Fabian05}
{Fabian} A.~C.,  {Sanders} J.~S.,  {Taylor} G.~B.,    {Allen} S.~W.,  2005,
  \mnras, 360, L20

\bibitem[\protect\citeauthoryear{{Fabian}, {Sanders}, {Taylor}, {Allen},
  {Crawford}, {Johnstone} \& {Iwasawa}}{{Fabian} et~al.}{2005b}]{Fabian_05}
{Fabian} A.~C.,  {Sanders} J.~S.,  {Taylor} G.~B.,  {Allen} S.~W.,  {Crawford}
  C.~S.,  {Johnstone} R.~M.,    {Iwasawa} K.,  2005, astro-ph/0510476

\bibitem[\protect\citeauthoryear{{Fujita}, {Matsumoto} \& {Wada}}{{Fujita}
  et~al.}{2004}]{Fujita04}
{Fujita} Y.,  {Matsumoto} T.,    {Wada} K.,  2004, \apjl, 612, L9

\bibitem[Hallman and Markevitch(2004)]{Hallman04}
Hallman, E.J.~and Markevitch, M. 2004, \apjl, 610, L81

\bibitem[{{Heinz} {et~al.}(2003){Heinz}, {Churazov}, {Forman}, {Jones}, \&
  {Briel}}]{Heinz03}
{Heinz}, S., {Churazov}, E., {Forman}, W., {Jones}, C., \& {Briel}, U.~G. 2003,
  \mnras, 346, 13

\bibitem[{{Hernquist}(1990)}]{Hernquist90}
{Hernquist}, L. 1990, \apj, 356, 359

\bibitem[Machacek \emph{et al.}(2005)]{2005ApJ...621..663M}
Machacek, M., 
Dosaj, A., Forman, W., Jones, C., Markevitch, M., Vikhlinin, A., Warmflash, 
A., and Kraft, R., 2005, \apj, 621, 663

\bibitem[\protect\citeauthoryear{{Markevitch}, {Gonzalez}, {David},
  {Vikhlinin}, {Murray}, {Forman}, {Jones} \& {Tucker}}{{Markevitch}
  et~al.}{2002}]{Markevitch02}
{Markevitch} M.,  {Gonzalez} A.~H.,  {David} L.,  {Vikhlinin} A.,  {Murray} S.,
   {Forman} W.,  {Jones} C.,    {Tucker} W.,  2002, \apjl, 567, L27

\bibitem[\protect\citeauthoryear{{Markevitch}, {Vikhlinin} \&
  {Forman}}{{Markevitch} et~al.}{2003}]{Markevitch_03}
{Markevitch} M.,  {Vikhlinin} A.,    {Forman} W.~R.,  2003, in ASP Conf. Proc.,
  Vol. 301, 37 (astro-ph/0208208)

\bibitem[\protect\citeauthoryear{{Markevitch}, {Vikhlinin} \&
  {Mazzotta}}{{Markevitch} et~al.}{2001}]{Markevitch01}
{Markevitch} M.,  {Vikhlinin} A.,    {Mazzotta} P.,  2001, \apjl, 562, L153

\bibitem[\protect\citeauthoryear{{Markevitch et~al.}}{{Markevitch
  et~al.}}{2000}]{Markevitch00sh}
{Markevitch et~al.} 2000, \apj, 541, 542

\bibitem[{{Mathis} {et~al.}(2005){Mathis}, {Lavaux}, {Diego}, \&
  {Silk}}]{Mathis05}
{Mathis}, H., {Lavaux}, G., {Diego}, J.~M., \& {Silk}, J. 2005, \mnras, 357,
  801

\bibitem[Mazzotta et al.(2001)]{2001ApJ...555..205M} Mazzotta, P., 
Markevitch, M., Vikhlinin, A., Forman, W.~R., David, L.~P., \& 
VanSpeybroeck, L.\ 2001, \apj, 555, 205 

\bibitem[McNamara \emph{et al.}(2000)]{Mcnamara00}
McNamara, B.R., Wise, M., Nulsen, P.E.J., David, L.P., Sarazin, C.L., Bautz,
M., Markevitch, M., Vikhlinin, A., Forman, W.R., Jones, C., and Harris,
D.E. 2000, \apj, 534, L135

\bibitem[{{Nagai} \& {Kravtsov}(2003)}]{NagaiKravtsov03}
{Nagai}, D. \& {Kravtsov}, A.~V. 2003, \apj, 587, 514

\bibitem[Nulsen \emph{et al.}(2005)]{Nulsen05}
Nulsen, P. E. J., McNamara, B. R., Wise, M. W., \& David, L. P. 2005, \apj, 
628, 629

\bibitem[Oegerle \& Hill(2001)]{Oegerle01}
Oegerle, W.R. \& Hill, J.M. 2001, \aj, 122, 2858
 
\bibitem[Quilis, Bower, \& Balogh(2001)]{Quilis01}
Quilis, V., Bower, R. G., and Balogh, M. L. 2001, \mnras, 328, 1091

\bibitem[{{Ricker} \& {Sarazin}(2001)}]{RickerSarazin01}
{Ricker}, P.~M. \& {Sarazin}, C.~L. 2001, \apj, 561, 621

\bibitem[{{Sakelliou}(2000)}]{Sakelliou00}
{Sakelliou}, I. 2000, \mnras, 318, 1164

\bibitem[\protect\citeauthoryear{{Sanders}, {Fabian}, {Allen} \&
  {Schmidt}}{{Sanders} et~al.}{2004}]{Sanders04}
{Sanders} J.~S.,  {Fabian} A.~C.,  {Allen} S.~W.,    {Schmidt} R.~W.,  2004,
  \mnras, 349, 952

\bibitem[\protect\citeauthoryear{{Sanders}, {Fabian} \& {Taylor}}{{Sanders}
  et~al.}{2005}]{Sanders05}
{Sanders} J.~S.,  {Fabian} A.~C.,    {Taylor} G.~B.,  2005, \mnras, 356, 1022

\bibitem[Sarazin(1988)]{Sarazin}
Sarazin, C. L. 1988, X-Ray Emission in Cluster of Galaxies (Cambridge:
Cambridge Univ. Press)

\bibitem[{{Springel}(2005)}]{Gadget2}
{Springel}, V. 2005, \mnras, 364, 1105

\bibitem[{{Takizawa}(2005)}]{Takizawa05}
{Takizawa}, M. 2005, \apj, 629, 791

\bibitem[{{Tittley} \& {Henriksen}(2005)}]{TittleyHenriksen05}
{Tittley}, E.~R. \& {Henriksen}, M. 2005, \apj, 618, 227

\bibitem[\protect\citeauthoryear{{Vikhlinin}}{{Vikhlinin}}{2005}]{Vikhlinin_05}
Vikhlinin, A.,  2005, \apj, in press (astro-ph/0504098)

\bibitem[Vikhlinin \& Markevitch(2002)]{2002AstL...28..495V}
Vikhlinin, A. \& Markevitch, M. 2002, Astronomy Letters, 28, 495

\bibitem[\protect\citeauthoryear{{Vikhlinin}, {Markevitch} \&
  {Murray}}{{Vikhlinin} et~al.}{2001}]{Vikhlinin01a}
Vikhlinin, A.,  Markevitch, M., \& Murray, S.~S.,  2001, \apj, 551, 160

\bibitem[Vikhlinin et al.(2005)]{Vikhlinin2005}
Vikhlinin, A., Markevitch, M., Murray, S.~S., Jones, C., 
Forman, W., \& Van Speybroeck, L.\ 2005, \apj, 628, 655 

\end{thebibliography}



\end{document}